\documentclass{aastex61}

\submitjournal{ApJS}

\shorttitle{A large-scale survey toward RMC}
\shortauthors{Li et al.}

\begin{document}

\title{A large-scale survey of CO and its isotopologues toward the Rosette molecular cloud}

\correspondingauthor{Hongchi Wang}
\email{hcwang@pmo.ac.cn}

\author{Chong Li}
\affil{Purple Mountain Observatory and Key Laboratory of Radio Astronomy, Chinese Academy of Sciences, 2 West Beijing Road, Nanjing 210008, China}
\affil{University of Chinese Academy of Sciences, 19A Yuquan Road, Shijingshan District, Beijing 100049, China}

\author{Hongchi Wang}
\affil{Purple Mountain Observatory and Key Laboratory of Radio Astronomy, Chinese Academy of Sciences, 2 West Beijing Road, Nanjing 210008, China}

\author{Miaomiao Zhang}
\affil{Purple Mountain Observatory and Key Laboratory of Radio Astronomy, Chinese Academy of Sciences, 2 West Beijing Road, Nanjing 210008, China}

\author{Yuehui Ma}
\affil{Purple Mountain Observatory and Key Laboratory of Radio Astronomy, Chinese Academy of Sciences, 2 West Beijing Road, Nanjing 210008, China}
\affil{University of Chinese Academy of Sciences, 19A Yuquan Road, Shijingshan District, Beijing 100049, China}

\author{Min Fang}
\affil{Department of Astronomy, University of Arizona, 933 North Cherry Avenue, Tucson, AZ 85721, USA}

\author{Ji Yang}
\affil{Purple Mountain Observatory and Key Laboratory of Radio Astronomy, Chinese Academy of Sciences, 2 West Beijing Road, Nanjing 210008, China}

\begin{abstract}

Using the PMO-13.7m millimeter telescope at Delingha in China, we have conducted a large-scale simultaneous survey of $^{12}$CO, $^{13}$CO, and C$^{18}$O J=1-0 emission toward the Rosette molecular cloud (RMC) region with a sky coverage of $3.5\arcdeg \times 2.5\arcdeg$. The majority of the emission in the region comes from the RMC complex with velocities lying in the range from -2 km s$^{-1}$ to 20.5 km s$^{-1}$. Beyond this velocity range, 73 molecular clumps are identified with kinematic distances from 2.4 kpc to 11 kpc. Based on the spatial and velocity distribution, nine individual clouds, C1-C9, have been identified for the RMC complex. It appears that the C3 cloud is different from other clouds in the RMC complex in view of its characteristic velocity, excitation temperature, and velocity dispersion. Most of the young stellar clusters in the region are located in positions of both high column density and high excitation temperature. Seven new molecular filaments are discovered in the RMC complex. Evidence for cloud-cloud collision is found in the region of young stellar clusters REFL9 and PouF, showing that these young stellar clusters probably result from a cloud-cloud collision. The abundance ratios of $^{13}$CO to C$^{18}$O in the region have a mean value of 13.7 which is 2.5 times larger than the solar system value, showing that UV photons from the nearby OB clusters have strong influence on the chemistry of clouds in the RMC complex.

\end{abstract}




\section{Introduction} \label{sec:intro}

Stars form in cold and dense molecular clouds. Large scale surveys of molecular clouds in the Milky Way revealed that the major part of molecular clouds in the Milky Way is accumulated into cloud complexes which are called Giant Molecular Clouds (GMCs) \citep{1987ApJ...322..706D,2001ApJ...547..792D}. GMCs are found to be in virial equilibrium state and bound by gravity while the constituent clouds of GMCs and isolated molecular clouds with M $< 10^3$ M$_\sun$ are not in self-gravitational equilibrium \citep{1987ApJ...319..730S, 2001ApJ...551..852H}. The mass function of molecular clouds follows a power law with an index of around -1.7 \citep{1997ApJ...476..166W, 2005PASP..117.1403R} and there exists a scaling relation between the line-width and the size of molecular clouds \citep{1981MNRAS.194..809L, 1987ApJ...319..730S}. It has long been realized that GMCs have complex and hierarchical structure which can be divided into substructures of clouds, clumps, and cores \citep{1999ASIC..540....3B}. With high spatial resolution and sensitivity in the sub-millimeter regime, Herschel observations reveal the ubiquitous presence of filaments in molecular clouds \citep{2014prpl.conf...27A} and have found that star formation in molecular clouds occurs mainly at the junctions of filaments \citep{2012A&A...540L..11S}. Turbulence has been proposed to be responsible for the origin of the hierarchical and filamentary structure of molecular clouds \citep{1994ApJ...423..681V, 2001ApJ...553..227P}. Stellar feedbacks, in particular that from young OB clusters, have strong influence on the evolution of molecular clouds through stellar winds, outflows, and radiation \citep{2002ApJ...566..302M, 2017ApJ...850..112R}. Recently, both numerical simulations and observations have shown that cloud-cloud collision may play an important role in the dynamics of molecular clouds and may be an important mechanism for triggered star formation \citep{2009ApJ...696L.115F, 2010MNRAS.405.1431A, 2017ApJ...835L..14G}. However, it remains unclear what are roles of turbulence, stellar feedbacks, and cloud-cloud collision in the formation, evolution, and destruction of molecular clouds. Observations of GMCs in varying environments with large spatial dynamic range is essential to address this important question.

The Rosette molecular cloud (RMC) is an ideal target to analyse the internal structure of GMCs and to characterize the influence of stellar feedbacks. It is associated with an optical emission nebula, the Rosette Nebula. The Rosette Nebula is a well known \ion{H}{2} region which is driven by the NGC 2244 OB cluster. This OB cluster contains six massive stars of O spectral type that have a total luminosity of $\sim$ 10$^6$ L$_{\odot}$ \citep{1985A&A...144..171C}. The expanding \ion{H}{2} region interacts with the surrounding RMC and the photons from the NGC 2244 OB cluster produce a photon dominated region (PDR) at the interface between the \ion{H}{2}-region and RMC \citep{1998A&A...338..262S}. The distance to the young NGC 2244 OB cluster has been estimated using stellar photometry with results ranging from 1.4-1.7 kpc \citep{1981PASJ...33..149O,2002AJ....123..892P}. Using optical spectroscopy \cite{2000A&A...358..553H} determined the orbital and fundamental stellar parameters for each of the components of the V578 Mon binary, which is a member of the NGC 2244 OB cluster, and derived a distance of 1.39 $\pm$ 0.1 kpc for the NGC 2244 OB cluster. As in \cite{2009MNRAS.395.1805D}, we adopted a distance of 1.4 kpc for RMC in this work.

The RMC has been surveyed in multi-wavelength. \cite{1997ApJ...477..176P} imaged a region of ~0.7 deg$^2$ toward the RMC in near-infrared (JHK) and detected seven young embedded clusters, PL01-PL07. The FLAMINGOS survey conducted by \cite{2008ApJ...672..861R} confirmed the existence of PL01-P07 and detected four more young clusters, REFL08-REFL10 and the NGC 2237 cluster. The Spitzer telescope survey of the RMC with IRAC and MIPS, covering a region of 1\degr $\times$ 1.5\degr, identified a total of 751 young stellar objects with infrared excess down to a mass limit of 0.4 M$_\sun$. The observations confirmed the seven clusters of \cite{1997ApJ...477..176P} and the existence of clusters REFL08 and REFL09. Furthermore, two new small clusters, PouC and PouD, were detected near clusters PL02 and PL03, respectively \citep{2008MNRAS.384.1249P}. Herschel observations with the PACS and SPIRE instruments in the wavelength range of 70-520 $\mu$m have revealed protostars, gas clumps, filaments, and dust temperature distribution in the RMC \citep{2010A&A...518L..84H, 2010A&A...518L..91D, 2010A&A...518L..83S, 2012A&A...540L..11S}.

The first large scale CO J = 1-0 emission survey of the RMC was made with the 1.2 m telescope at Columbia University in New York \citep{1980ApJ...241..676B}. The survey revealed that the RMC is extended along the galactic plane with a maximum extent of $\sim$ 100 pc and it contains emission maxima IRS and A-J. They found that the RMC possesses an overall velocity gradient of 0.20 km s$^{-1}$ pc$^{-1}$. \citet{1986ApJ...300L..89B} mapped the RMC in J=1-0 lines of CO and $^{13}$CO with the 7 m telescope at Bell Laboratories and found that molecular clumps in the RMC are embedded in low volume density interclump molecular gas. By analyzing the 7 m telescope survey data, \citet{1995ApJ...451..252W} found an overall velocity gradient of 0.08 km s$^{-1}$ pc$^{-1}$ for the RMC and interpreted this velocity gradient as the rotation of the RMC. Using the $^{12}$CO and C$^{13}$O J = 1-0 survey data obtained with the 14 m telescope of the Five College Radio Astronomy Observatory, \citet{2006ApJ...643..956H} examined the role of turbulent fragmentation in regulating the efficiency of star formation and it was found that the effect of turbulent fragmentation must be limited and nonexclusive in the RMC. \citet{2009MNRAS.395.1805D} conducted a large-scale survey of the $^{12}$CO J=1-0 emission covering 4.8 deg$^2$ with the James Clerk Maxwell Telescope (JCMT) and it was shown that the dominant bulk molecular gas motion in the region is expansion away from the O stars in NGC 2244.

In the present work we present a new large-scale (3.5\arcdeg $\times$ 2.5\arcdeg) survey of $^{12}$CO, $^{13}$CO, and C$^{18}$O J=1-0 emission toward the RMC region. The survey is described in Section 2 and the results are presented in Section 3. We discuss our results in Section 4 and present the summary in Section 5.

\section{Observations}

The RMC region has been observed as part of the Milky Way Imaging Scroll Painting (MWISP \footnote{\url{http://www.radioast.nsdc.cn/mwisp.php
}}) project which is dedicated to the large-scale survey of molecular gas along the northern Galactic plane. The simultaneous observations of $^{12}$CO, $^{13}$CO and C$^{18}$O J=1-0 emission toward the RMC region were carried out from 2010 October to 2012 March using the PMO-13.7m millimeter telescope at Delingha in China. A superconducting spectroscopic array receiver (SSAR) containing 3 $\times$ 3 beams was used as the front-end. A specific local oscillator (LO) frequency was carefully selected so that the upper sideband is centered at the $^{12}$CO J=1-0 line while the lower sideband covers the $^{13}$CO and C$^{18}$O J=1-0 lines \citep{2012ITTST...2..593S}. For each sideband, a Fast Fourier Transform Spectrometer (FFTS) containing 16384 channels with a bandwidth of 1 GHz was used as the back-end. The effective spectral resolution of each FFTS is 61.0 KHz, corresponding to a velocity resolution of 0.16 km s$^{-1}$ at the 115 Ghz frequency of the $^{12}$CO J-1-0 line. The observations cover the sky region of galactic longitude of 204.75$^{\circ}$ $<$ L $<$ 208.25$^{\circ}$ and galactic latitude of -3.25$^{\circ}$ $<$ B $<$ -0.75$^{\circ}$ (3.5$^{\circ}$ $\times$ 2.5$^{\circ}$) and was conducted in position-switch on-the-fly (OTF) mode with a scanning rate of 75$\arcsec$ per second and a dump time of 0.2 s. The survey area is split into cells of the size of 30$\arcmin$ $\times$ 30$\arcmin$. Each cell was scanned at least twice, once along the Galactic longitude and once along the Galactic latitude, to reduce scanning effects. The pointing of the telescope has an rms accuracy of about 5\arcsec and the beam widths are about 55\arcsec and 52\arcsec at 110 GHz and 115 GHz, respectively.

The Data are processed using the CLASS package of the GILDAS \footnote{\url{http://www.iram.fr/IRAMFR/GILDAS}} software. The raw data are re-grided and converted to FITS files. All FITS files related to the same survey cells are then combined to produce the final FITS data cubes. The spatial pixel of the FITS data cube has a size of 30$\arcsec$ $\times$ 30$\arcsec$. The antenna temperature (T$_A$) is converted to the main beam temperature with the relation T$_{mb}$ = T$_A$/B$_{eff}$, where the beam efficiency B$_{eff}$ is 0.46 at 115 GHz and 0.49 at 110 GHz according to the status report of the PMO-13.7m telescope. The calibration accuracy is estimated to be within 10\%. The typical system temperature during the observation is about 350 K for the upper sideband and 250 K for the lower sideband. The sensitivity of our observation is estimated to be around 0.5 K for the $^{12}$CO J=1-0 emission and around 0.3 K for the $^{13}$CO and C$^{18}$O J=1-0 emission. Throughout this paper, all velocities are given with respect to the local standard of rest (LSR).

\section{Results}

\subsection{Overall distribution of the RMC complex}

\begin{figure}[h]
\centering
\includegraphics[width=0.3\textwidth,angle=270]{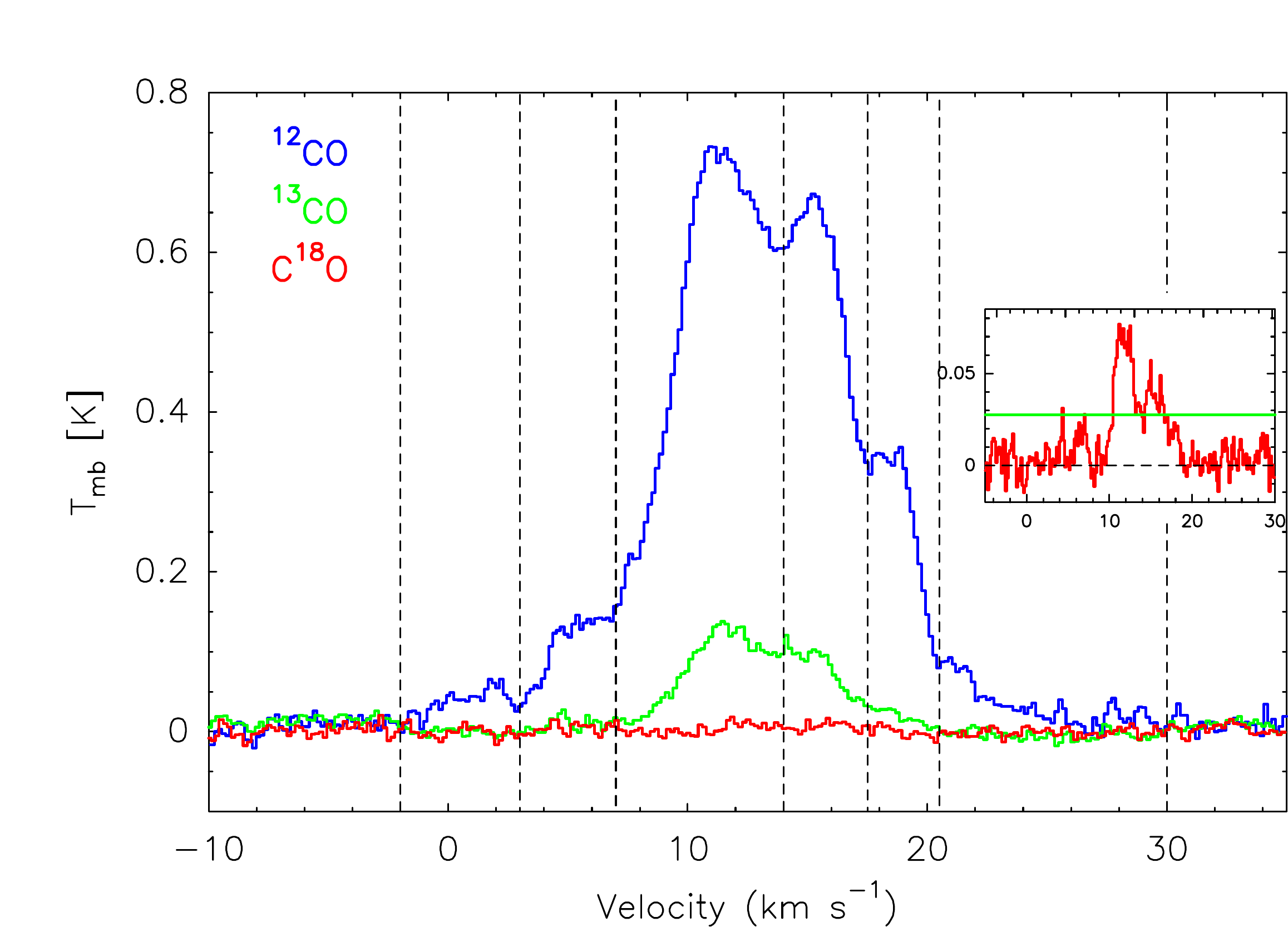}
\caption{Average spectra of the 3.5$^{\circ}$ $\times$ 2.5$^{\circ}$ region of the RMC complex with the blue, green, and red indicating the $^{12}$CO, $^{13}$CO and C$^{18}$O emission, respectively. The velocity range of the emission as seen in the $^{12}$CO spectrum is roughly divided into six sections (-2 to 3, 3 to 7, 7 to 14, 14 to 17.5, 17.5 to 20.5 and 20.5 to 30 km s$^{-1}$), which are indicated with vertical dashed lines. The inset shows the average of the C$^{18}$O spectra that have at least four contiguous channels with C$^{18}$O emission above 3$\sigma$. The green line shows the 3$\sigma$ noise level of the average spectrum. From this average spectrum, it can be seen that C$^{18}$O emission is detected in the velocity range from 11 to 17 km s$^{-1}$.}
\label{fig1}
\end{figure}

Figure \ref{fig1} shows the average spectra of the CO, $^{13}$CO, and C$^{18}$O J = 1$-$0 emission toward the 3.5$^{\circ}$ $\times$ 2.5$^{\circ}$ region of the RMC complex. Among the spectra, the $^{12}$CO emission shows the highest brightness temperature while C$^{18}$O has the lowest. As shown in Figure \ref{fig1}, the $^{12}$CO average spectrum can be divided into six velocity sections, i.e., -2 km s$^{-1}$ to 3 km s$^{-1}$, 3 km s$^{-1}$ to 7 km s$^{-1}$, 7 km s$^{-1}$ to 14 km s$^{-1}$, 14 km s$^{-1}$ to 17.5 km s$^{-1}$, 17.5 km s$^{-1}$ to 20.5 km s$^{-1}$, and 20.5 km s$^{-1}$ to 30 km s$^{-1}$. Most of the RMC complex have velocities ranging from 7 km s$^{-1}$ to 20.5 km s$^{-1}$. The average $^{13}$CO spectrum shows emission in the velocity range of 3 km s$^{-1}$ to 26 km s$^{-1}$, with major emission of velocities from 7 km s$^{-1}$ to 20 km s$^{-1}$. Comparatively, the average C$^{18}$O spectrum of this region is weak. The inset in Figure \ref{fig1} shows the average of the C$^{18}$O spectra that have at least four contiguous channels with C$^{18}$O emission above 3$\sigma$. The green line shows the 3$\sigma$ noise level of the average spectrum. From this average spectrum, it can be seen that C$^{18}$O emission is detected in the velocity range from 11 to 17 km s$^{-1}$.

\begin{figure}[h]
\centering
\includegraphics[width=0.6\textwidth]{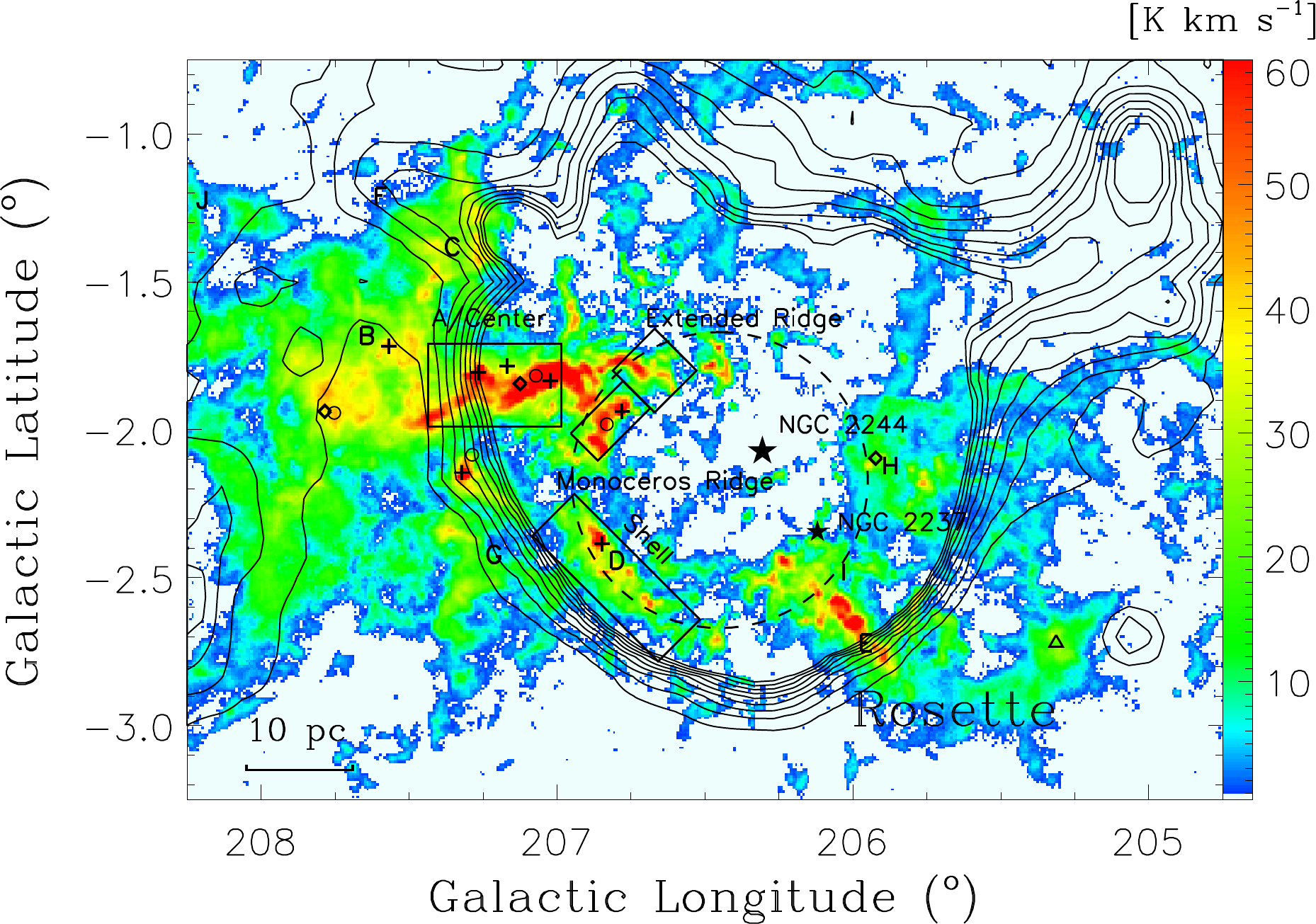}
\caption{Map of $^{12}$CO emission intensity integrated from -2 km s$^{-1}$ to 30 km s$^{-1}$. The contours indicate the 21 cm radio continuum emission from \citep{1997A&AS..126..413R}. The minimal level of the contours is 0.02 times the emission peak and the interval is 0.01 times the peak. The inner dashed circle indicates the approximate edge of the Rosette Nebula and the asterisk marks the central position of NGC2244. The pluses, diamonds, circles, and triangles indicate the embedded clusters from \citet{1997ApJ...477..176P}, \citet{2008ApJ...672..861R}, \citet{2008MNRAS.384.1249P}, and \citet{2013A&A...557A..29C}, respectively. Letters A-J indicate the cloud emission maxima identified by \citet{1980ApJ...241..676B}.}
\label{fig2}
\end{figure}

\begin{figure}[h]
\centering
\includegraphics[width=0.6\textwidth]{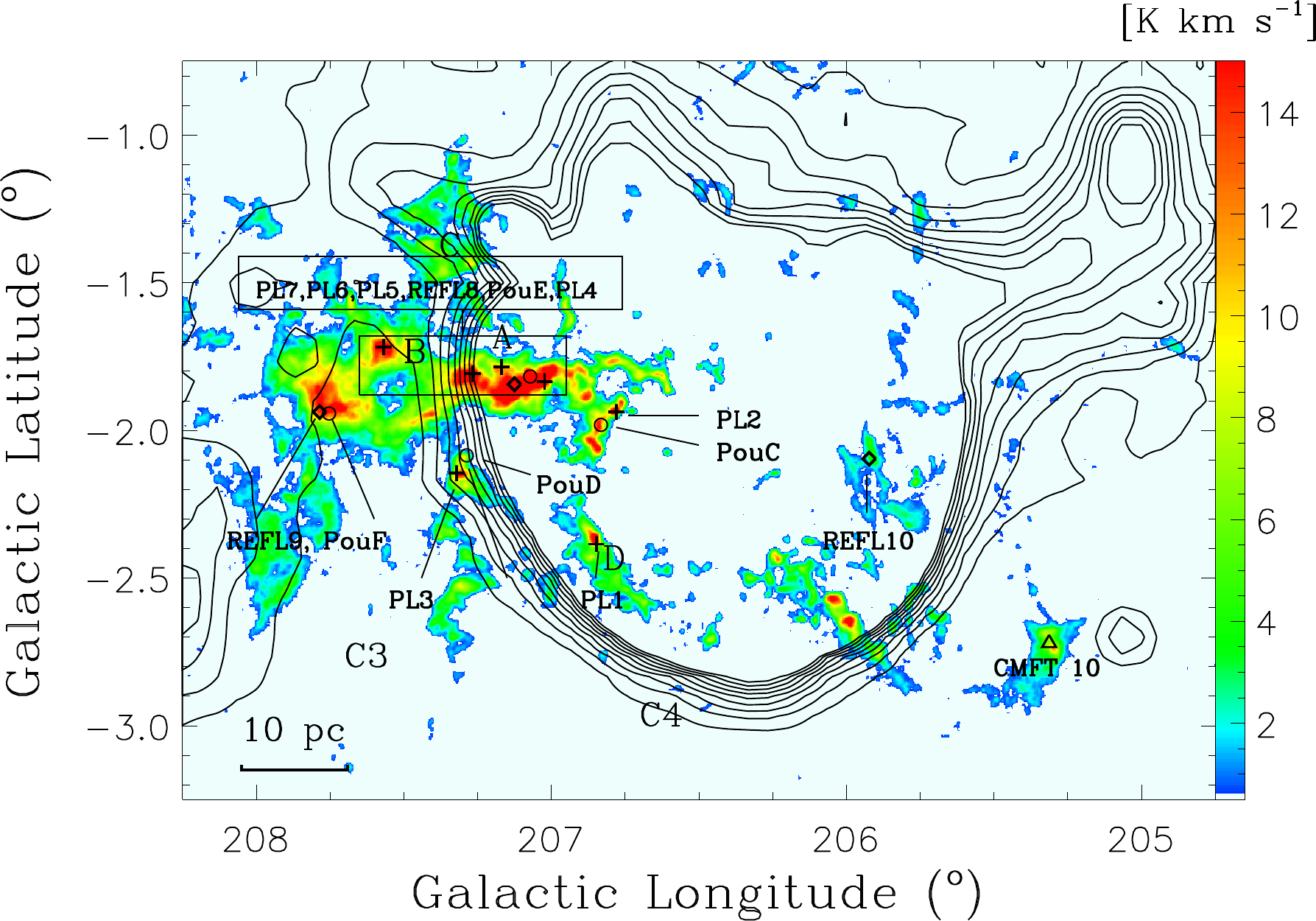}
\caption{Map of $^{13}$CO emission intensity integrated from 3 km s$^{-1}$ to 26 km s$^{-1}$. The contours are the 21 cm radio continuum emission as in Figure \ref{fig2}. The embedded clusters from \citet{1997ApJ...477..176P}, \citet{2008ApJ...672..861R}, \citet{2008MNRAS.384.1249P}, and \citet{2013A&A...557A..29C} are labeled.}
\label{fig3}
\end{figure}

\begin{figure}[h]
\centering
\includegraphics[width=0.6\textwidth]{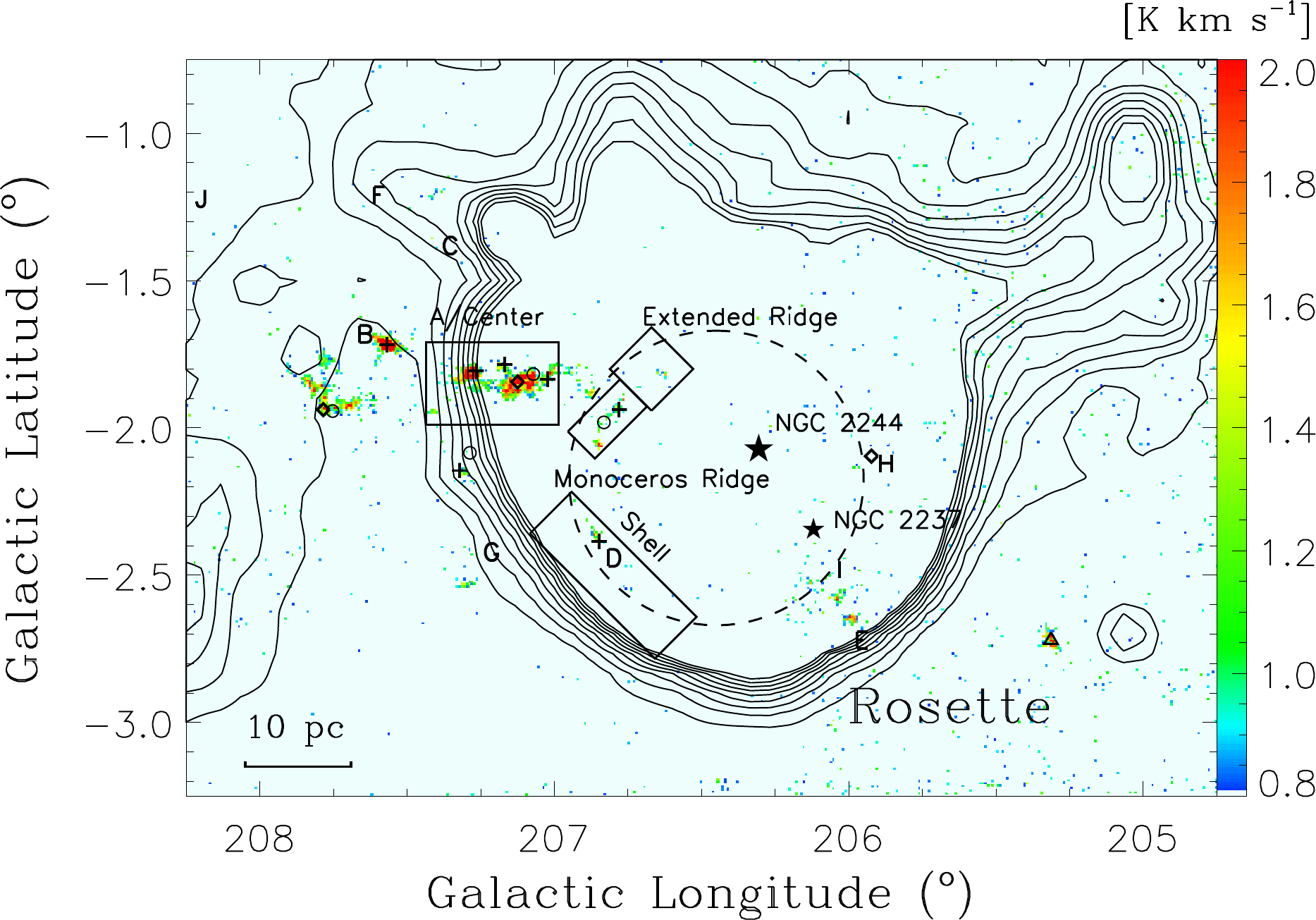}
\caption{Map of C$^{18}$O emission intensity integrated from 3 km s$^{-1}$ to 19 km s$^{-1}$. All the others are the same as in Figure \ref{fig3}.}
\label{fig4}
\end{figure}

The integrated intensity maps of the $^{12}$CO, $^{13}$CO, and C$^{18}$O J = 1$-$0 emission of the RMC region are displayed in Figures \ref{fig2}, \ref{fig3}, and \ref{fig4}. The contours in these figures indicate the 21 cm radio continuum emission from \citet{1997A&AS..126..413R}. The black dashed circle indicates the approximate edge of the Rosette Nebula and
the asterisk indicates the central position of NGC2244 from Table 3 of \cite{2008hsf1.book..928R}. Previous attempts to find molecular clouds within the Rosette Nebula have failed \citep{1969AJ.....74..985T,1970ApJ...160..485Z,1975AJ.....80..101M}. The $^{12}$CO emission map (Figure \ref{fig2}) shows that the RMC complex is a typical region where massive stars are interacting with the environmental molecular clouds. The stellar winds and radiation from the massive stars in NGC 2244 have apparently destroyed molecular clouds within the central part of the Rosette Nebula and have excavated a circle-like cavity which is similar to the S287 \ion{H}{2} region in \cite{2016A&A...588A.104G} and the infrared bubbles in \citet{2010ApJ...709..791B}. An rim with enhanced CO and $^{13}$CO emission, which includes the Extended Ridge, Monoceros Ridge and the Shell \citep{2008hsf1.book..928R}, can be seen around the OB cluster of NGC 2244 in Figure \ref{fig2}. Three young clusters, PL1, PL2 \citep{1997ApJ...477..176P}, and PouC \citep{2008MNRAS.384.1249P}, are located in the rim region. From Figure \ref{fig2} we can see that the major part of molecular clouds in the region is distributed to the east of the Rosette Nebula. However, we note that a remarkable amount of  molecular clouds can be seen to the west and southwest of the Rosette Nebula.

Compared to $^{12}$CO emission which is usually optically thick, $^{13}$CO emission is often optically thin and traces denser regions of molecular clouds than $^{12}$CO emission. As shown in Figure \ref{fig3}, the interface between the Rosette Nebula and molecular  clouds exhibits narrow shell-like structure. The eastern part of this shell structure has strong $^{13}$CO emission which shows that high column density molecular clouds exist in this part of the interaction interface. In Figure \ref{fig3} the Center region \citep{2008hsf1.book..928R} has the highest $^{13}$CO emission and it hosts 5 embedded clusters.

C$^{18}$O emission traces the molecular clouds with the highest column density compared with $^{12}$CO and $^{13}$CO emission. From Figure \ref{fig4} we can see that the C$^{18}$O emission in the RMC complex is concentrated mainly to the Center and the B clouds \citep{1980ApJ...241..676B,2008hsf1.book..928R} while in other regions, including the interface between the Rosette Nebula and the molecular clouds, the C$^{18}$O emission is very weak.

\subsection{Identification of individual clouds}
In this Section we make identification of individual clouds toward the region we surveyed on the basis of $^{12}$CO emission distribution in the position-position-velocity space. From the velocity distribution of CO emission as seen in Figure \ref{fig1}, we divide the molecular clouds in the observed region into two categories. The first category possesses velocities in the range from -2 to 20.5 km s$^{-1}$, which we refer to as the RMC complex. The second category  has velocities in the range from 20.5 up to 58 km s$^{-1}$ which lie far behind the Rosette Nebula according to their kinematic distances, and therefore we refer to them as the RMC background molecular clouds.

\subsubsection{Identification of individual clouds of the RMC complex}

\begin{figure}[h]
\centering
\includegraphics[width=0.5\textwidth]{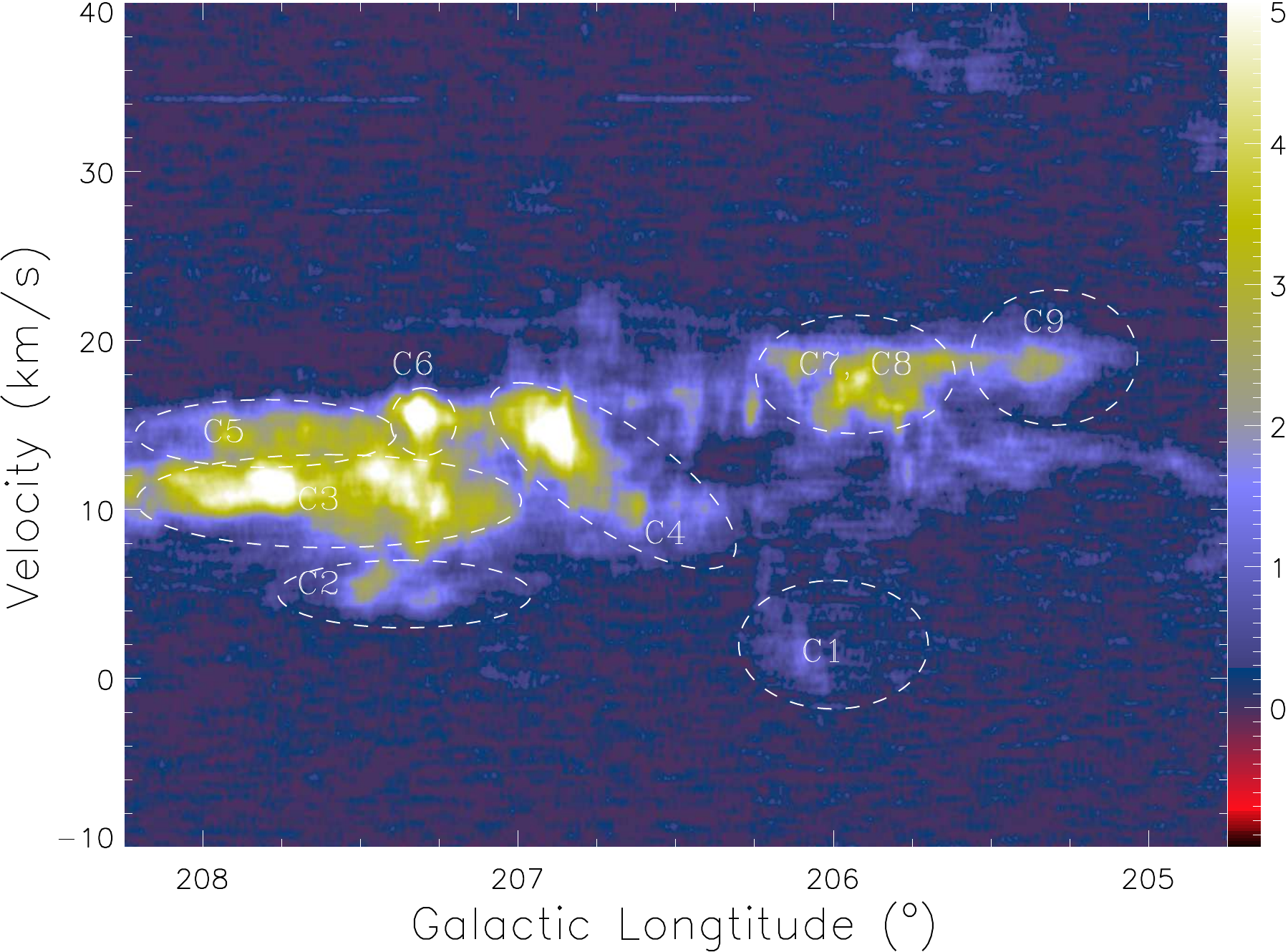}
\caption{Position-velocity map of $^{12}$CO emission along the galactic longitude from 204.75$^{\circ}$ to 208.25$^{\circ}$. The individual clouds identified in this work, clouds C1-C9, are labeled.}
\label{fig5}
\end{figure}

The Galactic longitude-velocity map of the $^{12}$CO emission of the RMC complex is shown in Figure \ref{fig5}. According to the velocity and space distribution, nine distinct molecular clouds can be identified in the RMC complex, which we designate as C1-C9. The integrated intensity maps of the $^{12}$CO and $^{13}$CO emission and the color-coded velocity distribution maps of these nine clouds are presented in Figures \ref{fig20}-\ref{fig24} in Appendix.

\begin{figure}[h]
\centering
\includegraphics[scale=.3]{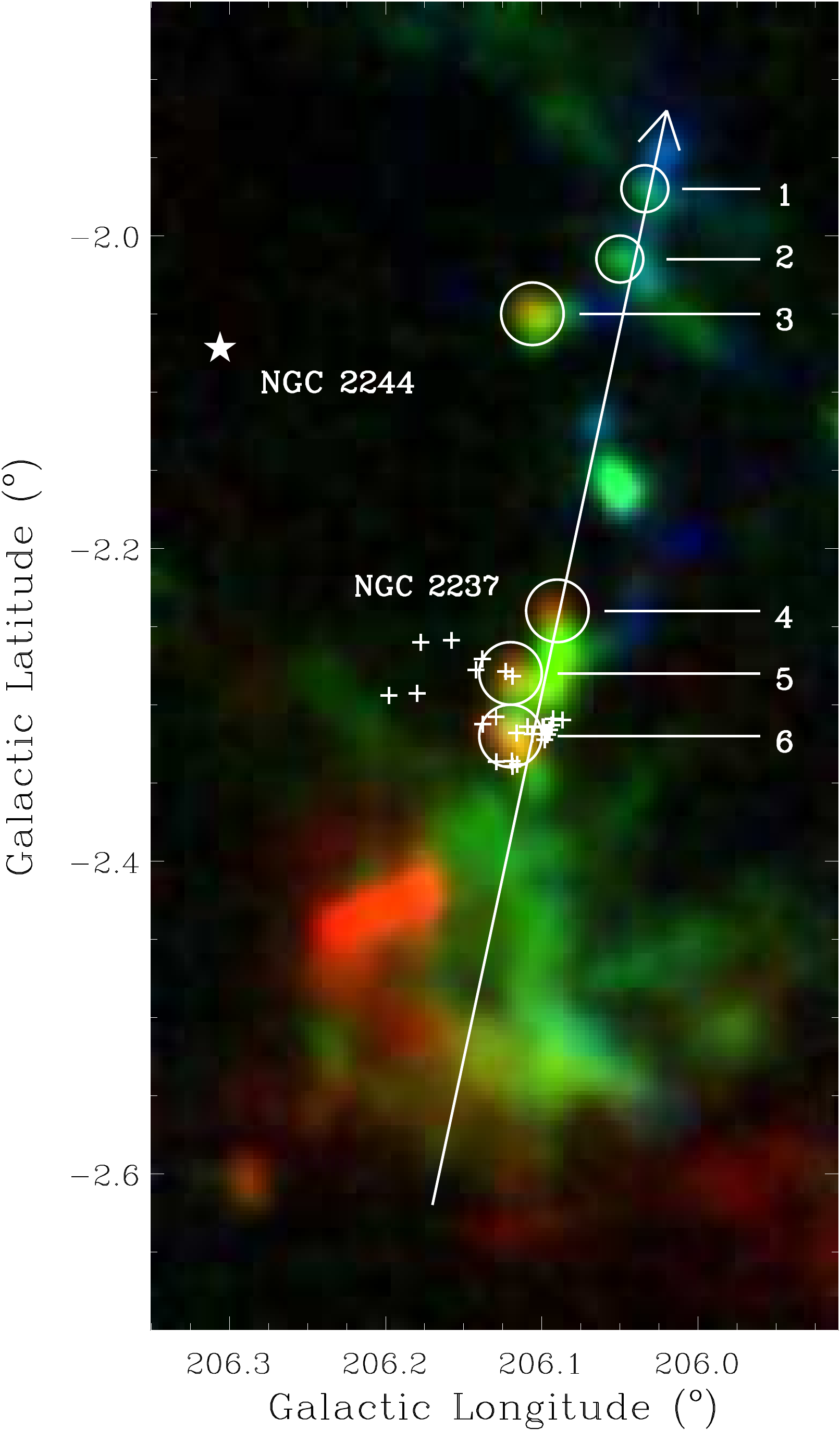}
\includegraphics[scale=.3]{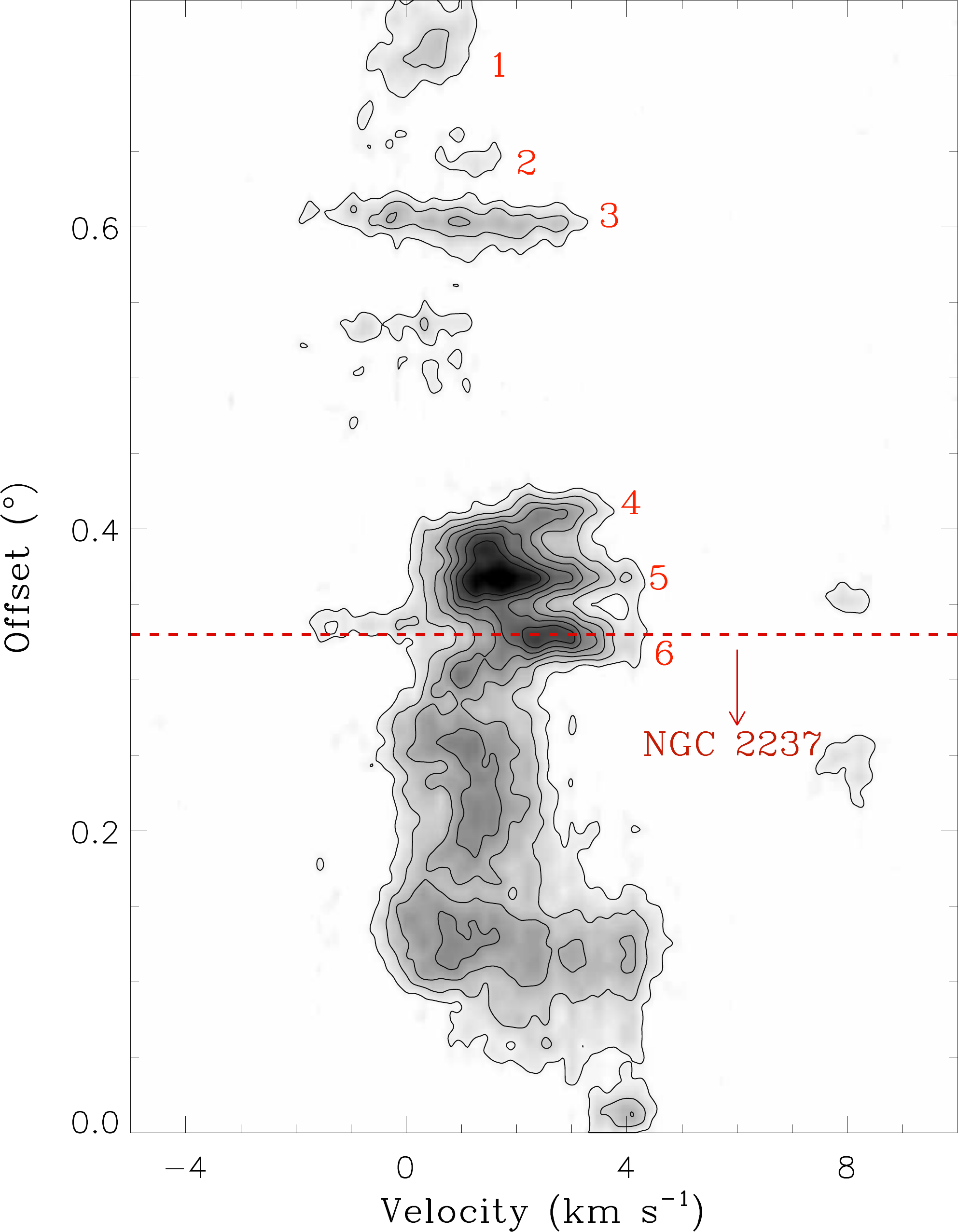}
\includegraphics[scale=.3]{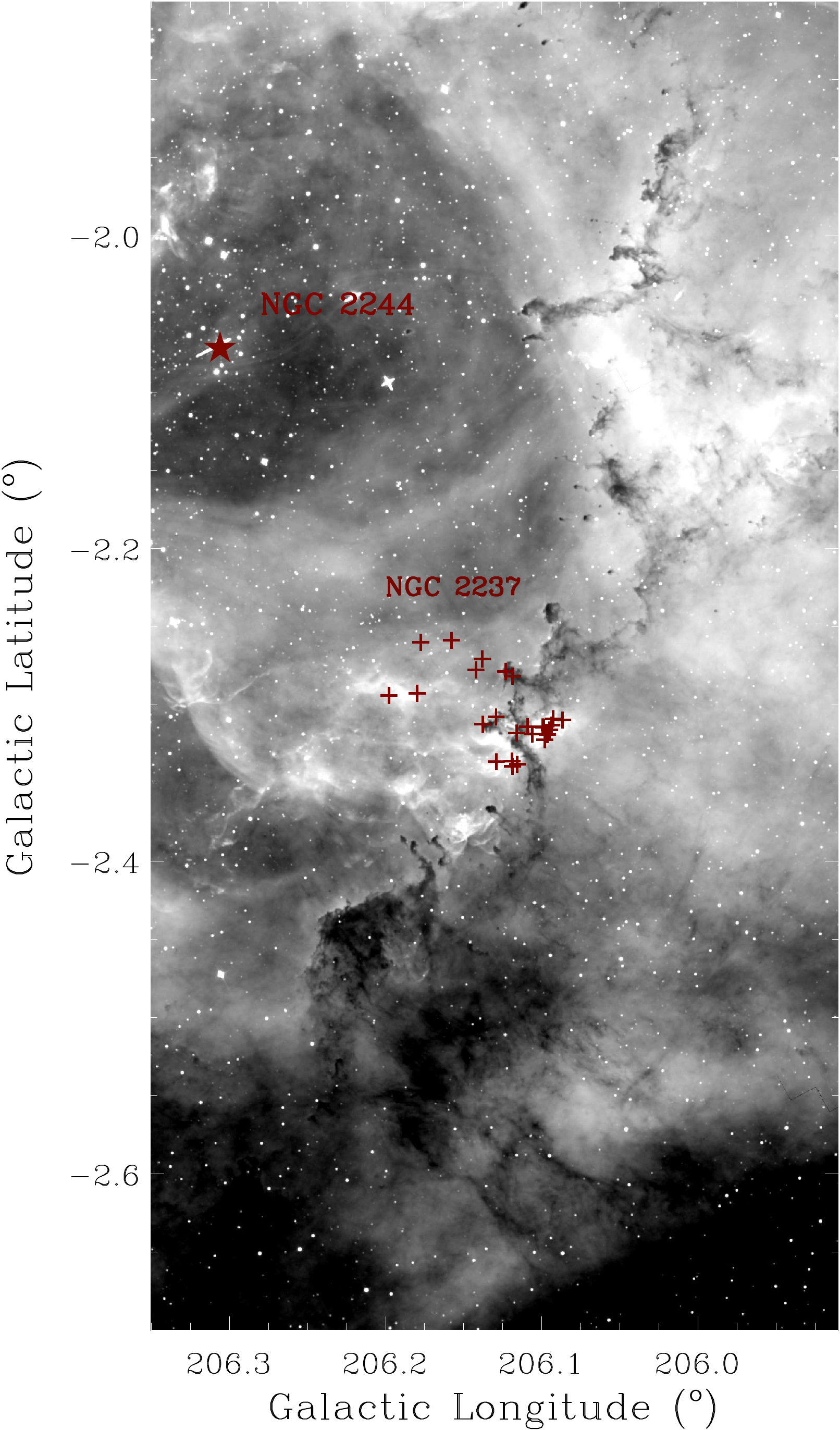}
\caption{Left (a): colour-coded image of $^{12}$CO emission of the C1 cloud, where the blue represents the integrated intensity in the velocity range from -2.8 km s$^{-1}$ to -0.3 km s$^{-1}$, the green from -0.3 km s$^{-1}$ to 2.2 km s$^{-1}$, and the red from 2.2 km s$^{-1}$ to 4.7 km s$^{-1}$. The center of the NGC 2244 cluster is marked with an asterisk and the members of the NGC 2237 OB cluster are indicated with pluses. The cometary structures with remarkable velocity gradients are labeled with numerals 1-6. Middle (b): position-velocity map of C1 in $^{12}$CO emission along the arrow marked in the left panel. The minimal level and the interval of the overlaid contours are both 0.1 $\times$ peak of $^{12}$CO brightness. Right (c): H$_{\alpha}$ image of the same region from the IPHAS survey.}
\label{fig6}
\end{figure}

The molecular cloud C1 (Figure \ref{fig20} in Appendix) possesses the lowest velocity, -2 km s$^{-1}$ to 3 km s$^{-1}$, in the RMC region. It is located within the \ion{H}{2}-region cavity and shows a filamentary distribution, which we refer to as filament a. Figures \ref{fig6}a and \ref{fig6}b present the velocity distribution and position-velocity map of cloud C1, and Figure \ref{fig6}c is the H$_{\alpha}$ image of the same region from the IPHAS survey \citep{2005MNRAS.362..753D}. From Figures \ref{fig6}a and \ref{fig6}b we can see that the C1 cloud exhibits an apparent velocity gradient in the direction from north to south with the south end of the cloud possessing a larger, i.e. relatively red-shifted, velocity, than the north end. Assuming a distance of 1.4 kpc for RMC, this velocity gradient is estimated to be 0.5 km s$^{-1}$ pc$^{-1}$. Six cometary clumps show remarkable velocity gradients in the direction from west to east, with the eastern side, i.e. the side facing to the NGC 2244 and NGC 2237 OB clusters, being red-shifted compared to the western side. This fact indicates that the nearby NGC 2244 and NGC 2237 OB clusters may have strong influence on these clumps and perhaps on the whole C1 cloud. \citet{2009MNRAS.395.1805D} also identified these cometary structures and derived the velocity gradients from the CO J = 3-2 data. They divided the C1 cloud into 13 clumps and derived the physical properties for these clumps. We note that clump 3 marked in Figure \ref{fig6}a corresponds to the Wrench Trunk in \citet{2006A&A...454..201G}. Using the $^{12}$CO data \citet{2006A&A...454..201G} found that the Wrench Trunk possesses a velocity gradient of 0.17 km s$^{-1}$ per steps of 10\arcsec along the trunk axis. At the same time they found that the Wrench Trunk is rotating as a rigid body with a velocity gradient of 0.16 km s$^{-1}$ per steps of 10\arcsec perpendicular to the trunk axis. Our $^{12}$CO data confirms the existence of both velocity gradients. Our measurements of the velocity gradient along and perpendicular to the trunk axis are, respectively, 0.20 km s$^{-1}$ and 0.13 km s$^{-1}$ per steps of 10\arcsec, which are consistent with their results. The clumpy appearance of the whole C1 cloud indicates that this filament is relatively evolved \citep{2010A&A...518L.103M}. We note that clumps 4-6 are well associated with the NGC 2237 OB cluster. The NGC 2237 OB cluster, with its youngest member at a age of only 1.1 $\times$ 10$^4$ yr \citep{2013A&A...557A..29C}, is much younger than the nearby NGC 2244 OB cluster. It has been suggested that the formation of the NGC 2237 OB cluster is triggered by the NGC 2244 OB cluster \citep{2010ApJ...716..474W,2008hsf1.book..928R}. If this scenario proves true, it is possible that the NGC 2237 OB cluster was born within the C1 cloud. From Figure \ref{fig6}c we can see that nearly all components of C1 in Figure \ref{fig6}a have a corresponding extinction feature in the H$_{\alpha}$ emission image, which shows that the C1 cloud lies in front of the Rosette Nebula.

\begin{figure}[h]
\centering
\includegraphics[width=0.4\textwidth]{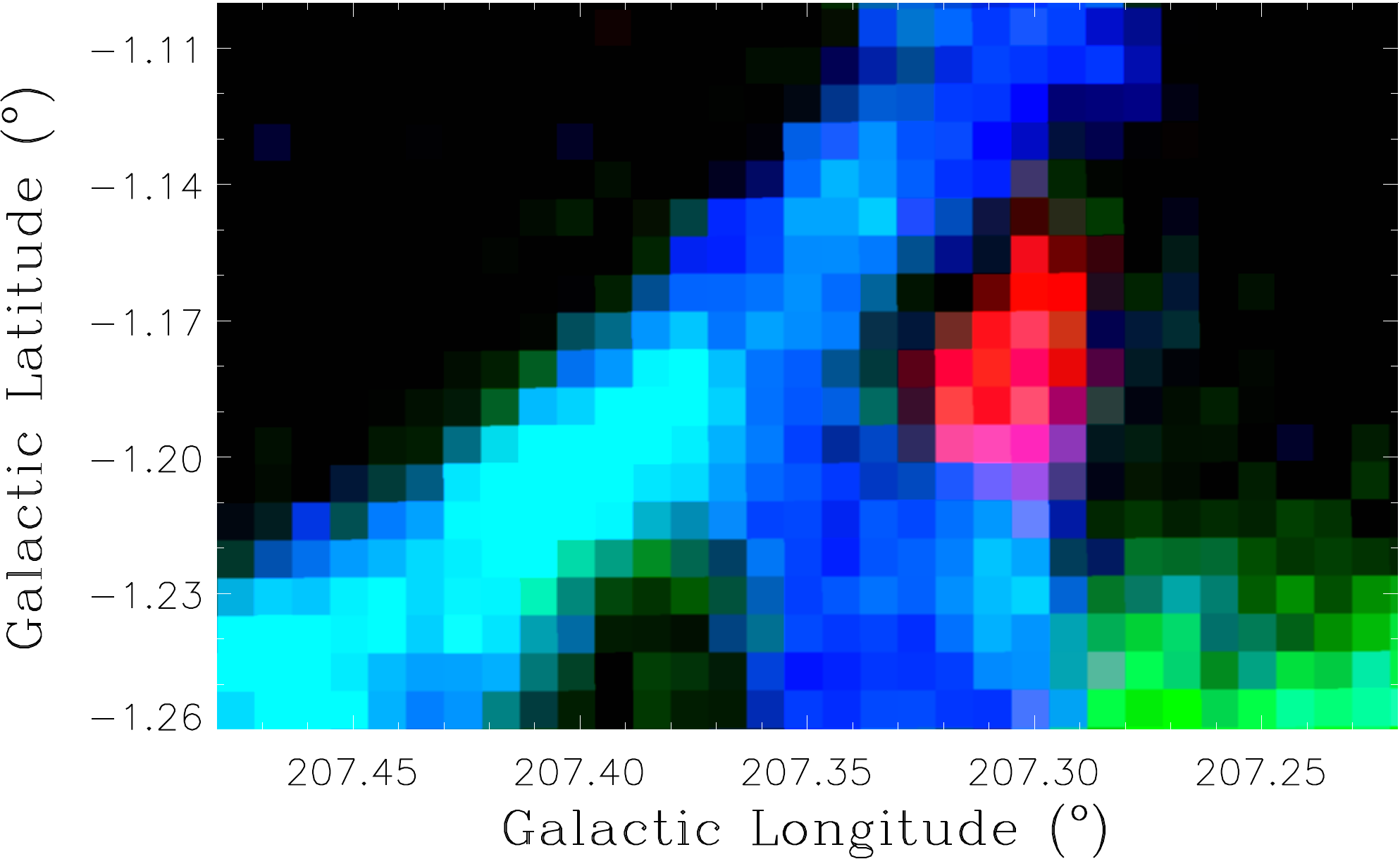}
\includegraphics[width=0.4\textwidth]{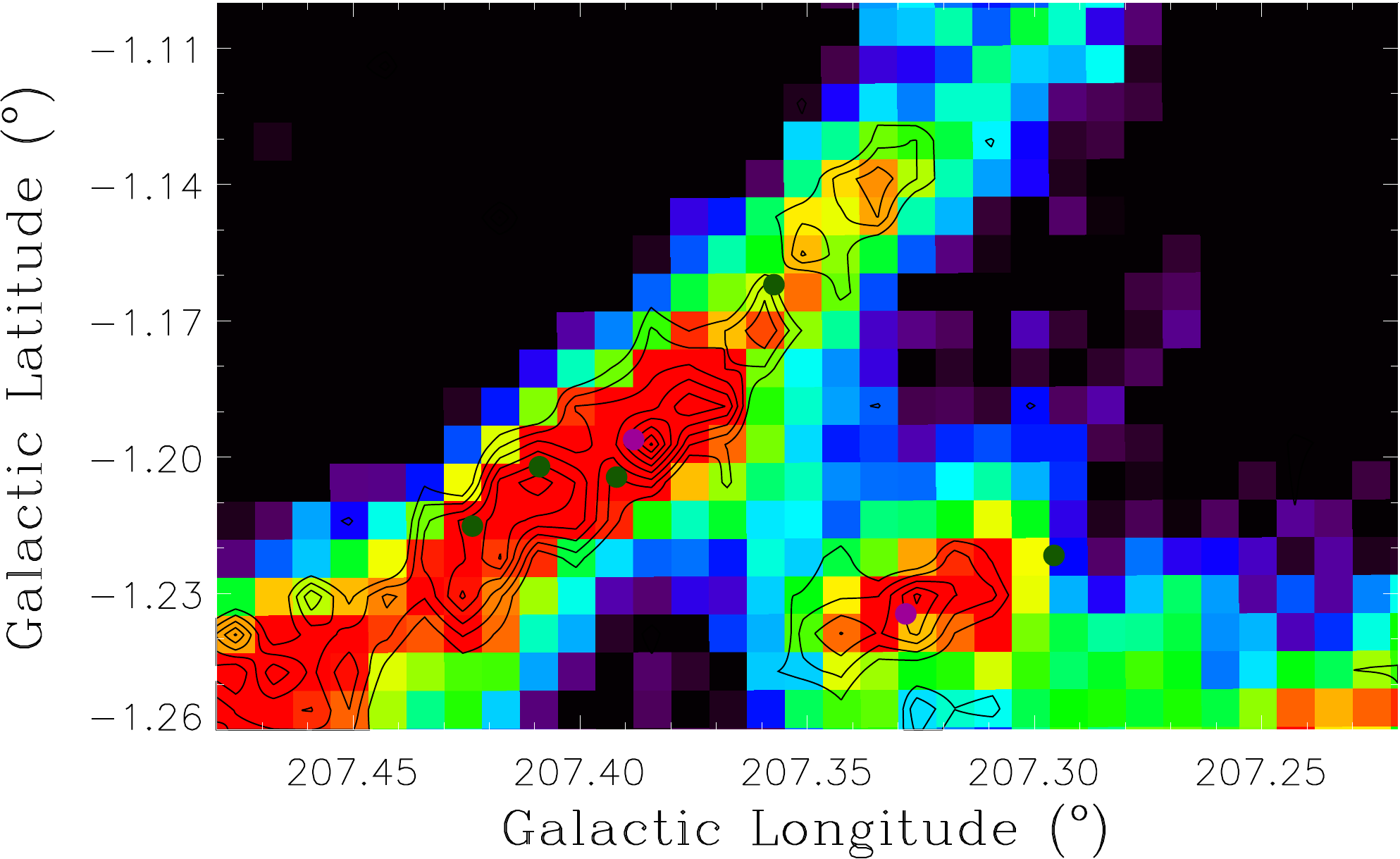}
\includegraphics[width=0.4\textwidth]{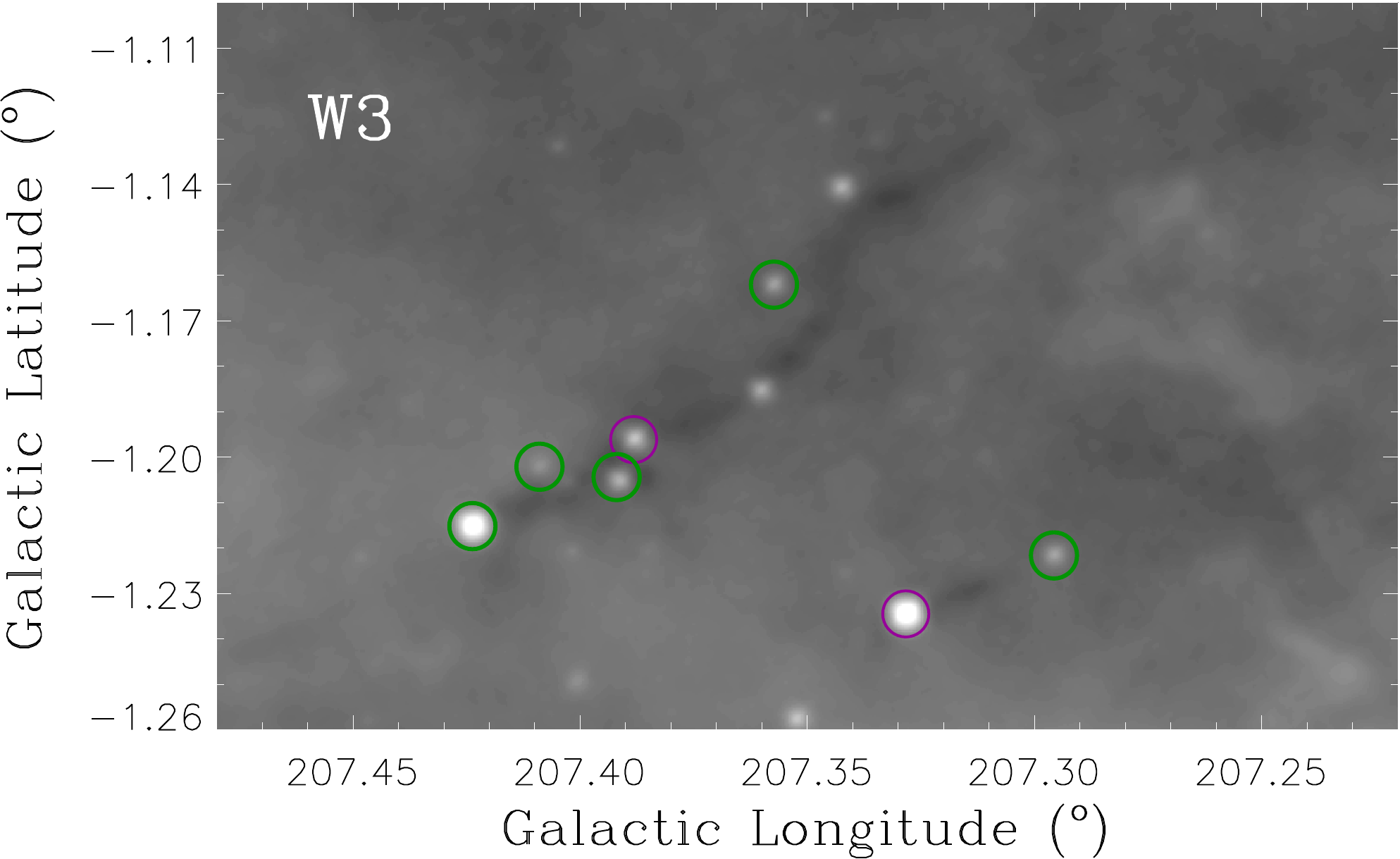}
\includegraphics[width=0.4\textwidth]{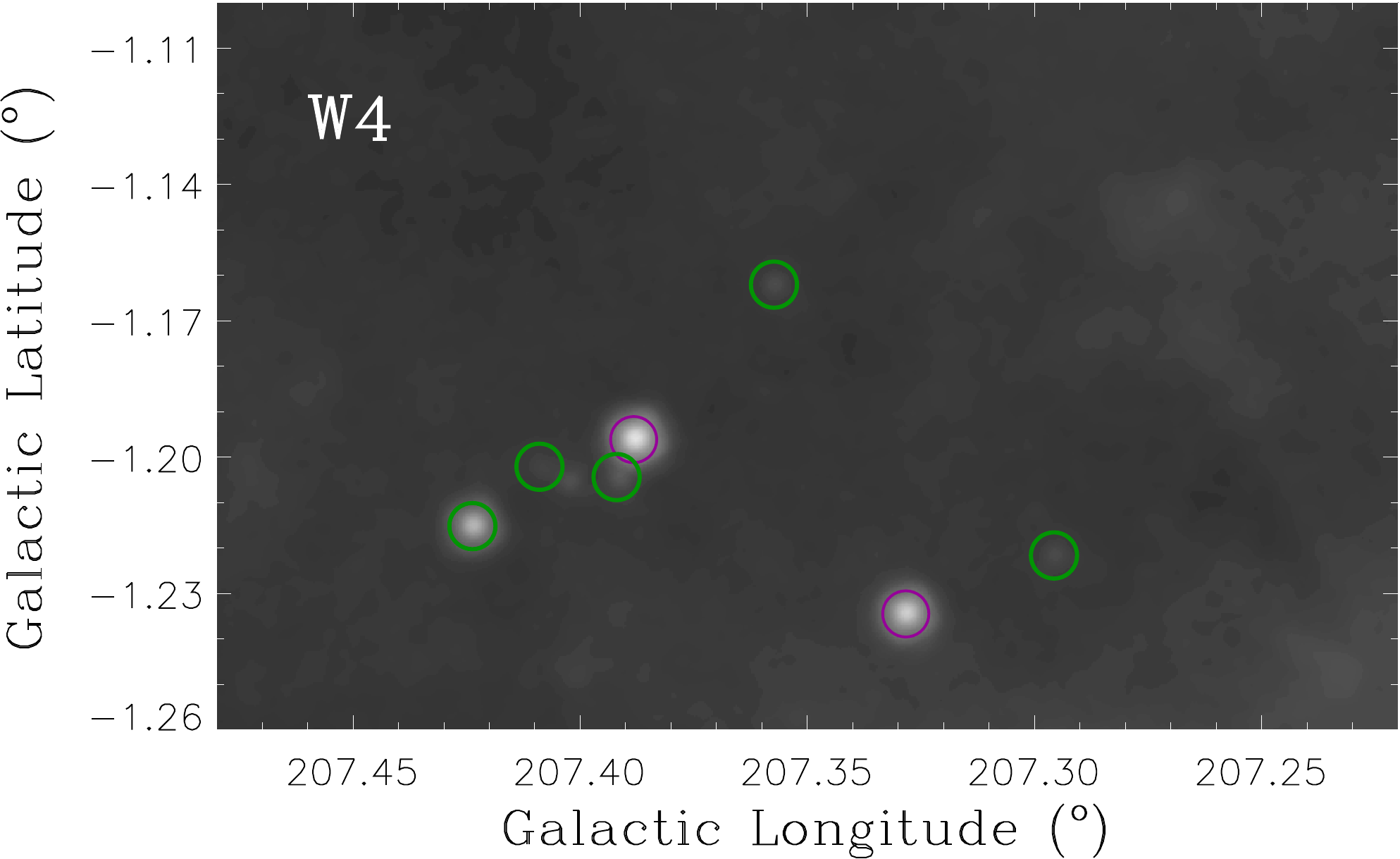}
\caption{Top left (a): colour-coded $^{12}$CO emission image showing the filament in the north part of C2. The different velocity ranges are coded with the blue for 2.7 km s$^{-1}$ to 4.2 km s$^{-1}$, the green for 4.2 km s$^{-1}$ to 5.7 km s$^{-1}$, and the red for 5.7 km s$^{-1}$ to 6.2 km s$^{-1}$. Top right (b): $^{13}$CO integrated intensity map superimposed with C$^{18}$O emission contours. The minimal level for the contours is 0.3 K km s$^{-1}$ and the interval is 0.12 K km s$^{-1}$. Bottom left (c): WISE band 3 image of the same region. The five Class II YSOs and two Class I protostars identified in this work are marked with green and purple circles, respectively. Bottom right (d): The WISE band 4 image of the region.}
\label{fig7}
\end{figure}

The $^{12}$CO and $^{13}$CO emission maps of the C2 cloud are presented in Figure \ref{fig21} in Appendix, where the integrated velocity range for both maps is from 3 km s$^{-1}$ to 7 km s$^{-1}$. The color-coded velocity distribution of C2 is also shown in Figure \ref{fig21}. We can see that cloud C2 consists of two parts, the northern and the southern part. The southern part shows a filamentary shape, which we refer to as filament b. The whole cloud of C2 coincides with the northeastern edge of the radio continuum emission of the Rosette \ion{H}{2} region. In particular, the concave arc-like shape of the southwestern side of the northern part of C2 coincides very well with the protruding arc-like edge of the radio continuum emission. These facts show that there may be interaction between the Rosette \ion{H}{2} region and cloud C2. The strongest $^{12}$CO and $^{13}$CO emission of C2 is located in the northern part and it exhibits a filamentary shape oriented in the northwestern-southeastern direction (see Figures \ref{fig7}a and \ref{fig7}b), which we refer to as filament c. The filament also shows up in the WISE band 3 image as an extinction feature (Figure \ref{fig7}c). Using the archive data of UKIDSS and 2MASS, we identified 5 Class II YSOs in the region, among which 4 YSOs are located in the filament. In this region we also identified two Class I protostars, among which one is located in filament c and the other one is associated with a small dark extinction feature (see Figure \ref{fig7}c).

\begin{figure}[h]
\centering
\includegraphics[width=0.2\textwidth]{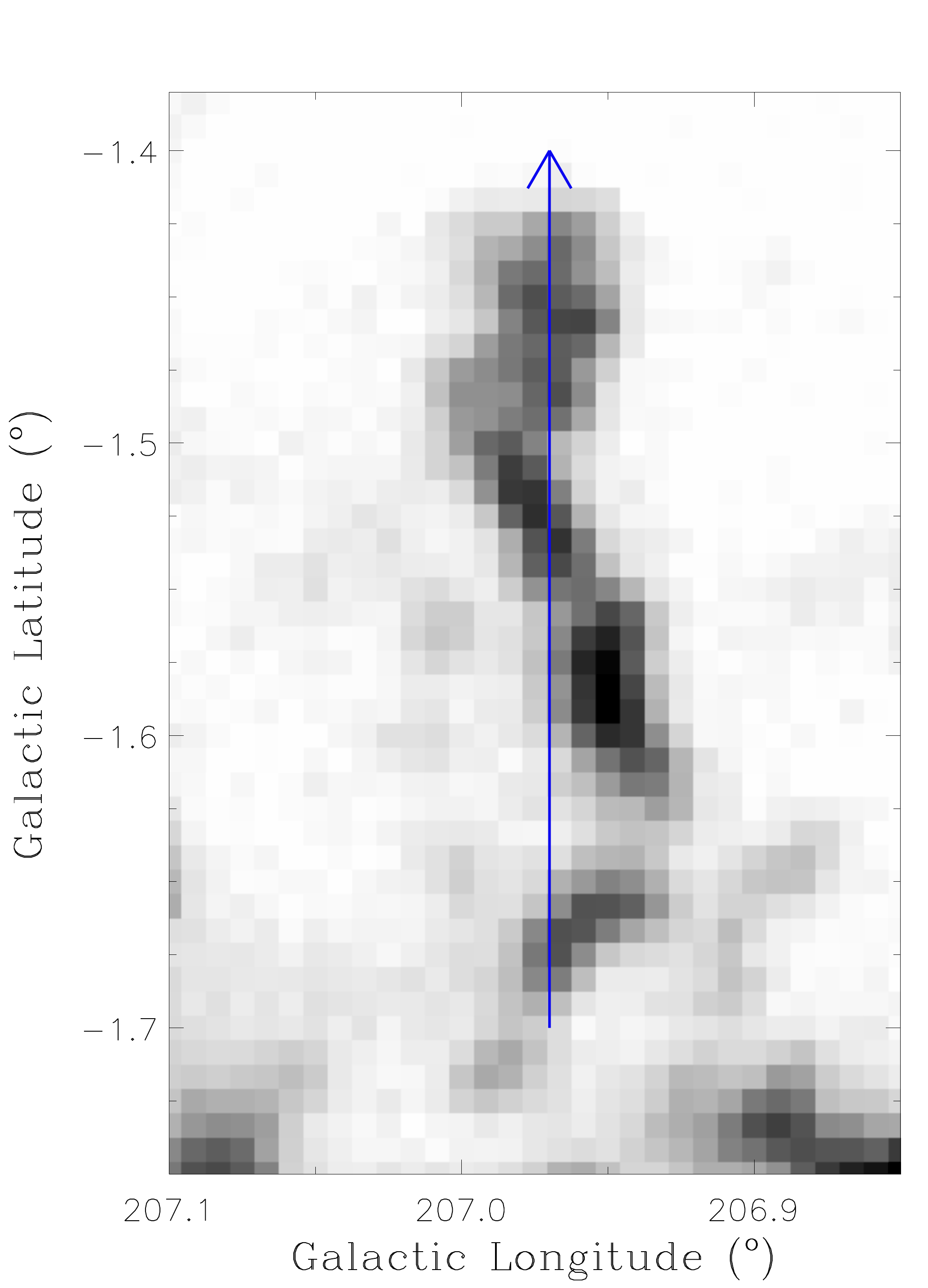}
\includegraphics[width=0.2\textwidth]{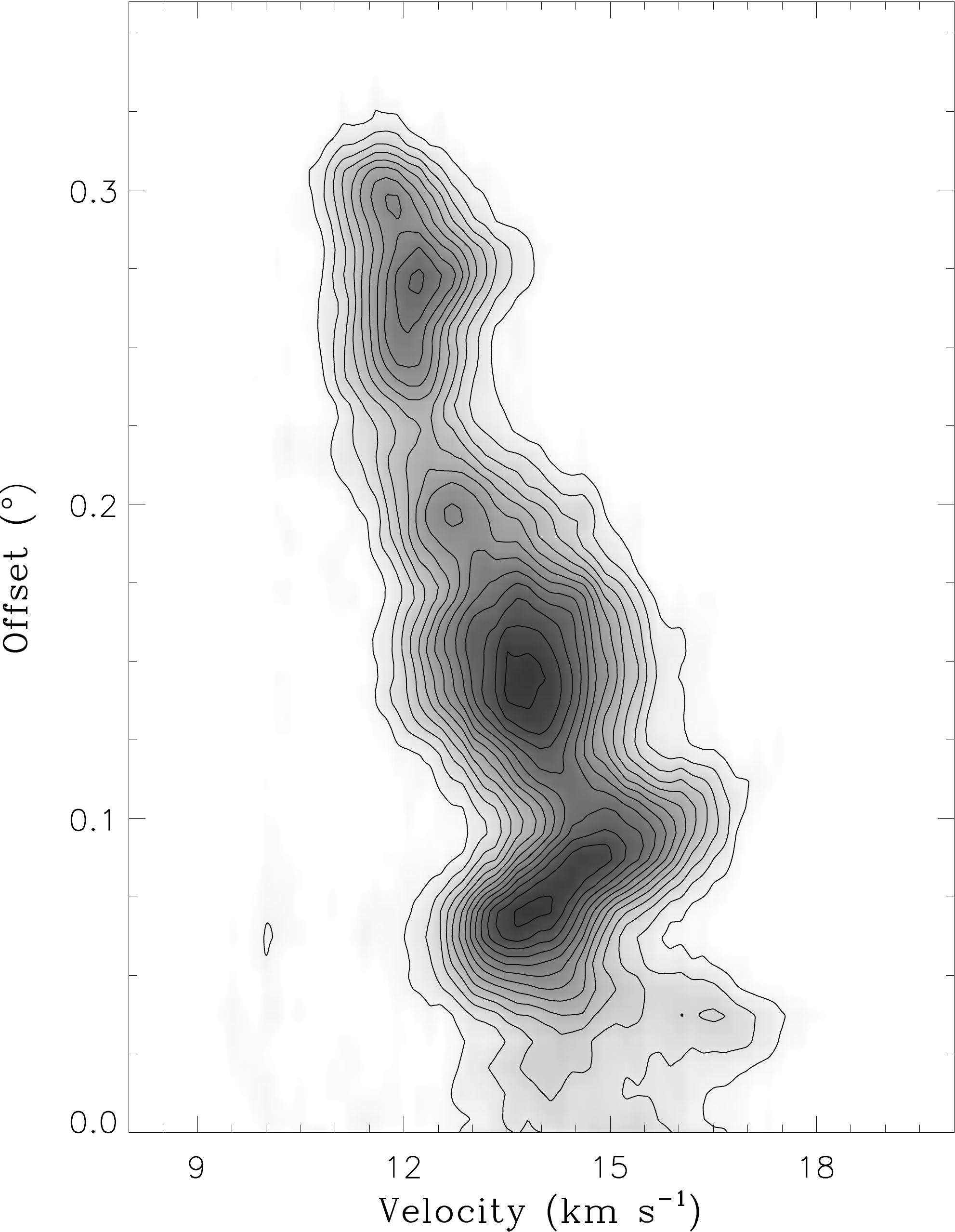}
\caption{Left: $^{12}$CO integrated intensity map of the integral-shaped filament. Right: position-velocity map of $^{12}$CO emission along the arrow marked in the left panel. The overlaid contours are $^{12}$CO emission with the minimal level and the interval both being 0.1 $\times$ peak of $^{12}$CO brightness.}
\label{fig8}
\end{figure}

\begin{figure}[h]
\centering
\includegraphics[scale=.3]{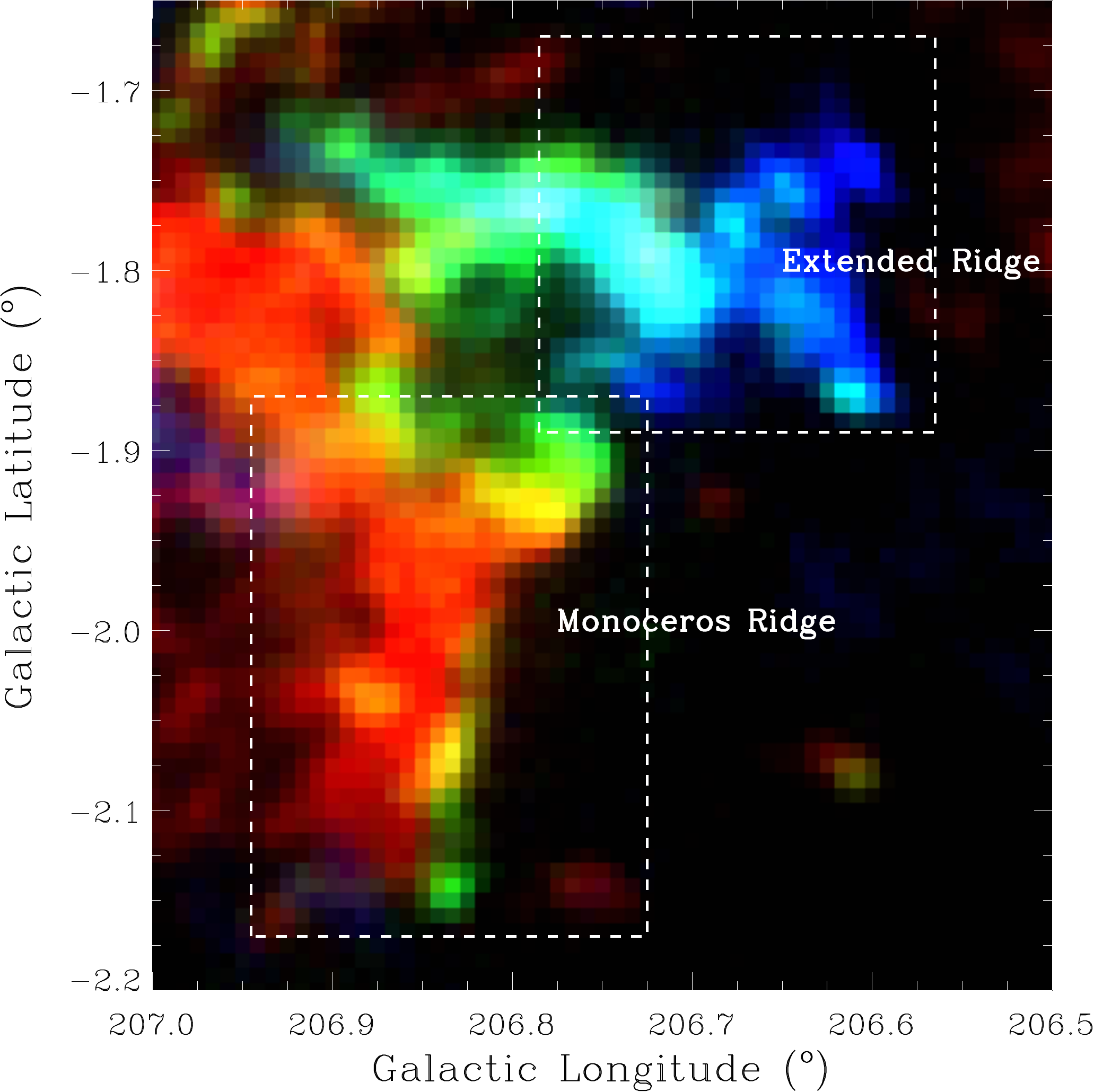}
\includegraphics[scale=.3]{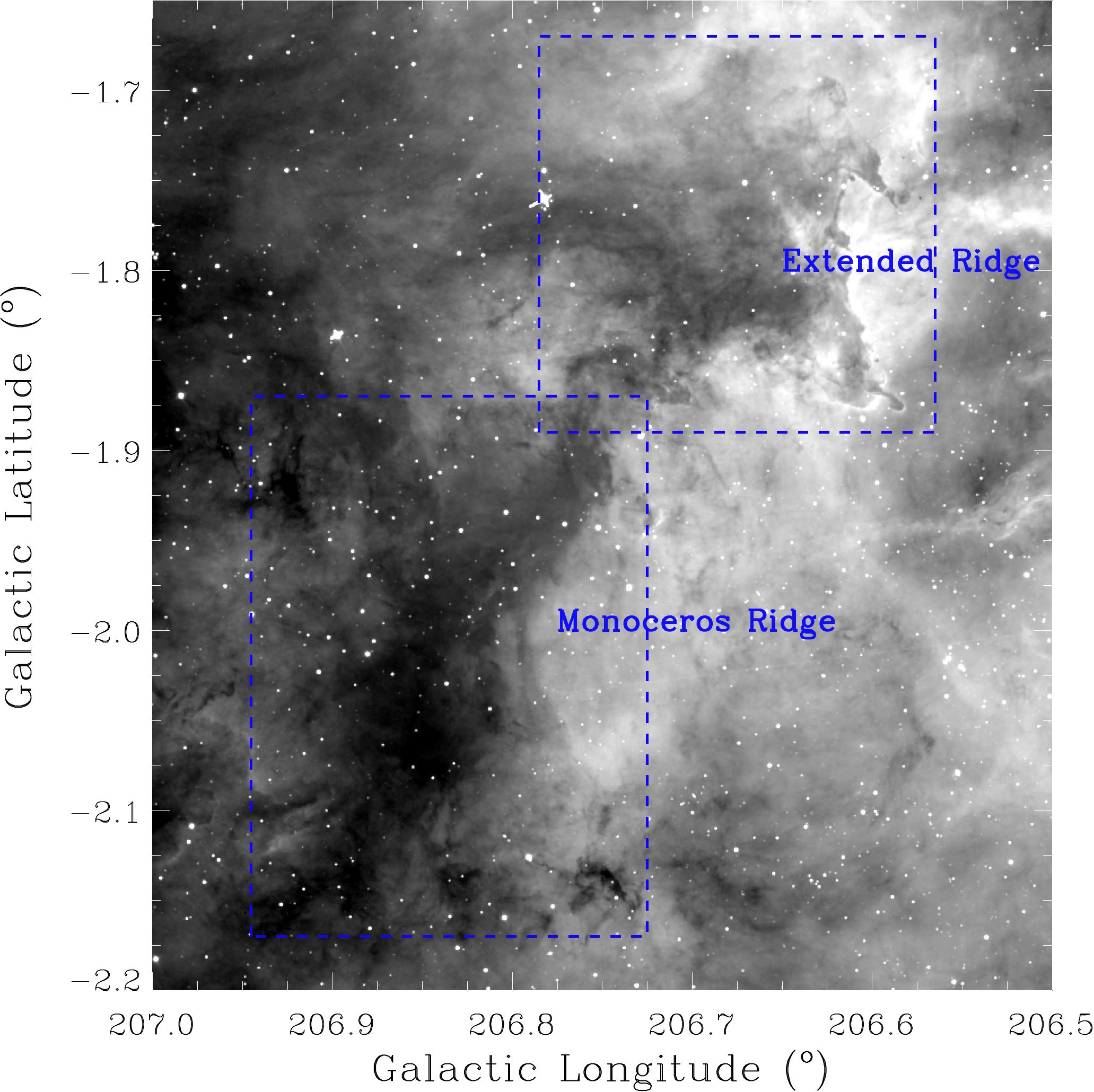}
\caption{Left (a): colour-coded image of the velocity distribution of $^{12}$CO emission for the Extended Ridge and the Monoceros Ridge. The different velocity ranges are represented by different colours, where the blue is from 8.8 km s$^{-1}$ to 11.8 km s$^{-1}$, the green from 11.6 km s$^{-1}$ to 13.4 km s$^{-1}$, and the red from 13.4 km s$^{-1}$ to 17.2 km s$^{-1}$. Right (b): H$_{\alpha}$ image of the same region from the IPHAS survey.}
\label{fig9}
\end{figure}

Figure \ref{fig22} in Appendix shows the distribution of molecular clouds in the velocity range from 7 km s$^{-1}$ to 14 km s$^{-1}$. The $^{13}$CO emission in this velocity range is mainly concentrated to the eastern part of the region where the clouds exhibits further velocity difference with the western part being relatively red-shifted (see the top panel of Figure \ref{fig22}). Therefore we divide the clouds in the velocity range from 7 km s$^{-1}$ to 14 km s$^{-1}$ into two clouds, C3 and C4, as outlined in the middle and bottom panels of Figure \ref{fig22}. The C3 cloud consists of the A, B, and C emission maxima identified by \cite{1980ApJ...241..676B} and the Center region of \citet{2008hsf1.book..928R}. The Center region is the most compact part of the RMC. Much of previous work on RMC has been focused on this region because of its active star formation, for example, it hosts 5 embedded young clusters, PL4, PL5, PL6 \citep{1997ApJ...477..176P}, REFL8 \citep{2008ApJ...672..861R}, and PouE \citep{2008MNRAS.384.1249P}. Cloud C4 consists of the Extended Ridge, the Monoceros Ridge, and the Shell region identified by \citet{2008hsf1.book..928R}. We note that from the top panel of Figure \ref{fig22}, an integral-shaped filament, which we refer to as filament d, can be seen to the northeast of the Extended Ridge, lying in the direction of from north to south. The $^{12}$CO emission and position-velocity maps of this filament are presented in Figure \ref{fig8}. Cloud C4 coincides with the eastern rim of the Rosette Nebula (see Figures \ref{fig2} and \ref{fig3}) and therefore may have been strongly influenced by the expanding Rosette \ion{H}{2} region. The Extended Ridge has a velocity range from 7 km s$^{-1}$ to 13 km s$^{-1}$, which is relatively blue-shifted compared with its nearby molecular clouds (see Figure \ref{fig9}a). Although the Extended Ridge is located very close to the NGC 2244 OB cluster, no embedded young cluster has been found within it, which confirms that the impact of the \ion{H}{2} region does not always promote star formation in the influenced clouds \citep{2013A&A...557A..29C}. Two embedded clusters, PL2 \citep{1997ApJ...477..176P} and PouC \citep{2008MNRAS.384.1249P}, are located within the Monoceros Ridge. As shown in Figure \ref{fig9}b, both the Extended Ridge and the Monoceros Ridge coincide well with extinction features in the IPHAS H$_{\alpha}$ image, showing that they are located in front of the Rosette Nebula.

\begin{figure}[h]
\centering
\includegraphics[width=0.3\textwidth]{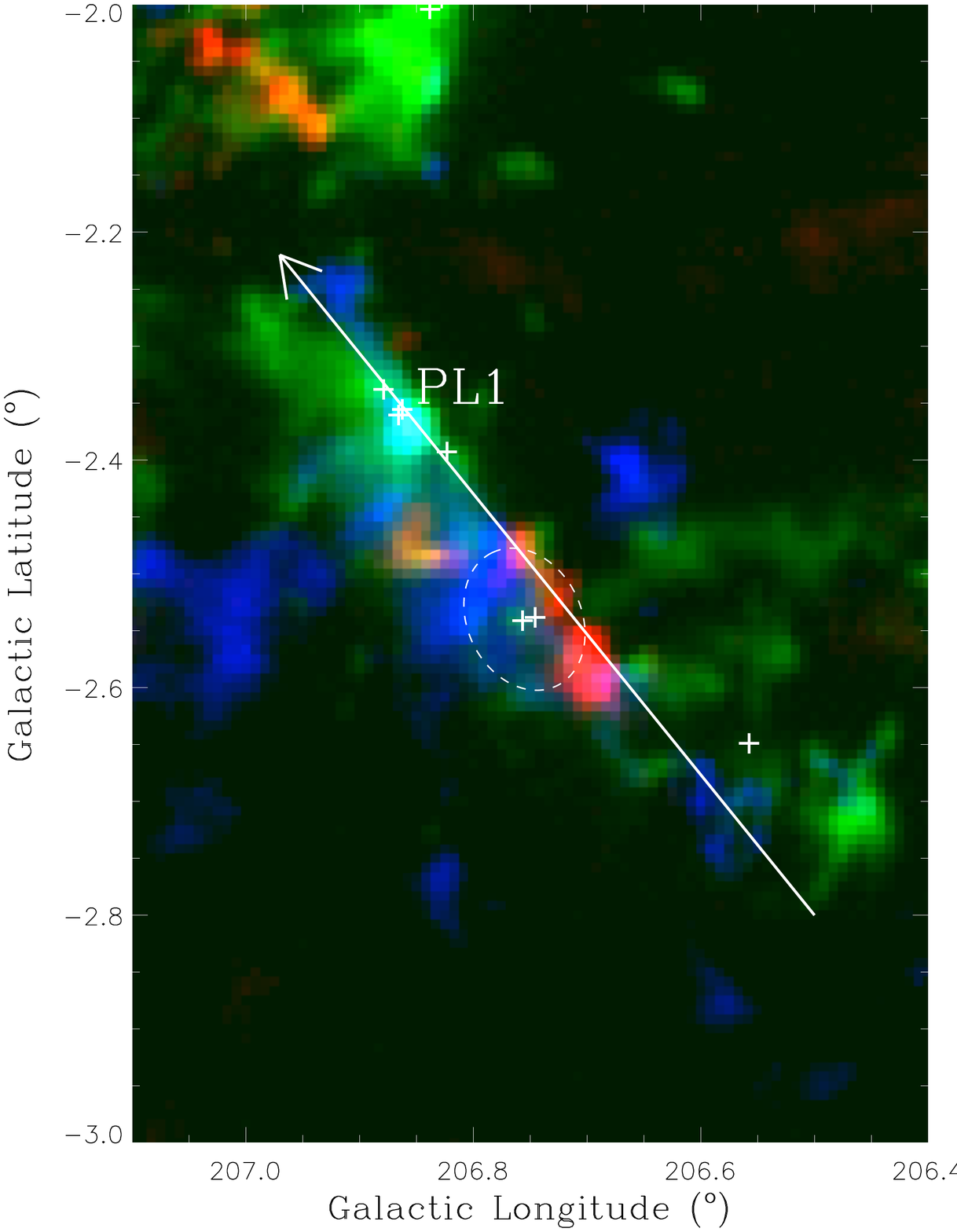}
\includegraphics[width=0.3\textwidth]{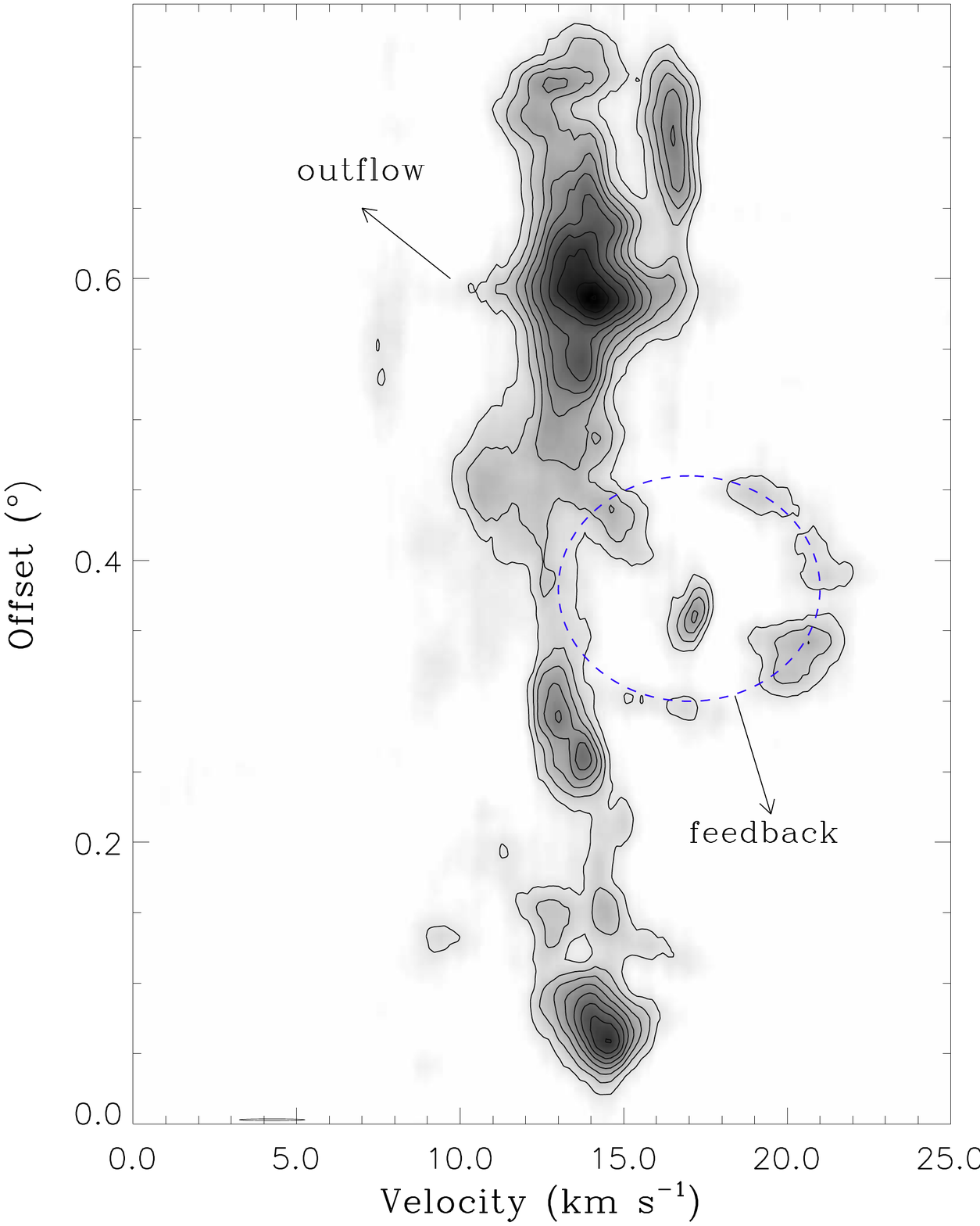}
\caption{Left (a): colour-coded image of the velocity distribution of filament e in $^{12}$CO emission. Different velocity ranges are shown by different colours, where the blue is from 8 km s$^{-1}$ to 13 km s$^{-1}$, the green from 13 km s$^{-1}$ to 18 km s$^{-1}$, and the red from 18 km s$^{-1}$ to 23 km s$^{-1}$. The pluses indicate the positions of Class I protostars in the region.  Right: position-velocity map of $^{12}$CO emission along the arrow marked in the left panel overlaid with $^{12}$CO emission contours. The minimal level and the interval of the contours are both 0.1 $\times$ peak of $^{12}$CO brightness.}
\label{fig10}
\end{figure}

The Shell region exhibits a shape of filament, which we refer to as filament e in the RMC. The young cluster, PL1, is located at the CO emission peak of the filament. From the photometry of 2MASS and WISE, we identify eight Class I YSOs located in this filament. The $^{12}$CO integrated intensity map and the position-velocity map of filament e are presented in Figure \ref{fig10}. An outflow is found to be associated with PL1 (see also \citep{2009MNRAS.395.1805D}). In the middle region of the filament there are two Class I YSOs which are surrounded by a ring-like structure (Figure \ref{fig10}a). The northwestern part of this ring-like structure is red-shifted while the southeastern part is blue-shifted. This expanding ring structure is also demonstrated in the position-velocity map (Figure \ref{fig10}b), showing that the central two Class I protostars have already exerted strong feedbacks on their surrounding molecular clouds.

Molecular clouds of velocities in the range of 14 km s$^{-1}$ to 17.5 km s$^{-1}$ are shown in Figure \ref{fig23} and they can be divided into four regions, C5-C8. Cloud C5 is spatially coincident with the Center region of cloud C3, but its velocity is different from the Center region cloud. The dense region of C5 as traced by the $^{13}$CO emission has a filamentary morphology, which we refer to as filament f. Ten embedded young stellar clusters have been found in the C5 cloud area. However, at present we cannot determine whether these young stellar clusters are physically associated with cloud C3 or cloud C5. Cloud C6 has a velocity range from 13.5 km s$^{-1}$ to 17.5 km s$^{-1}$. It corresponds to the G $^{12}$CO emission maximum in \cite{1980ApJ...241..676B}. The embedded young cluster PL3 is located at the emission peak of C6.

\begin{figure}[h]
\centering
\includegraphics[width=0.2\textwidth]{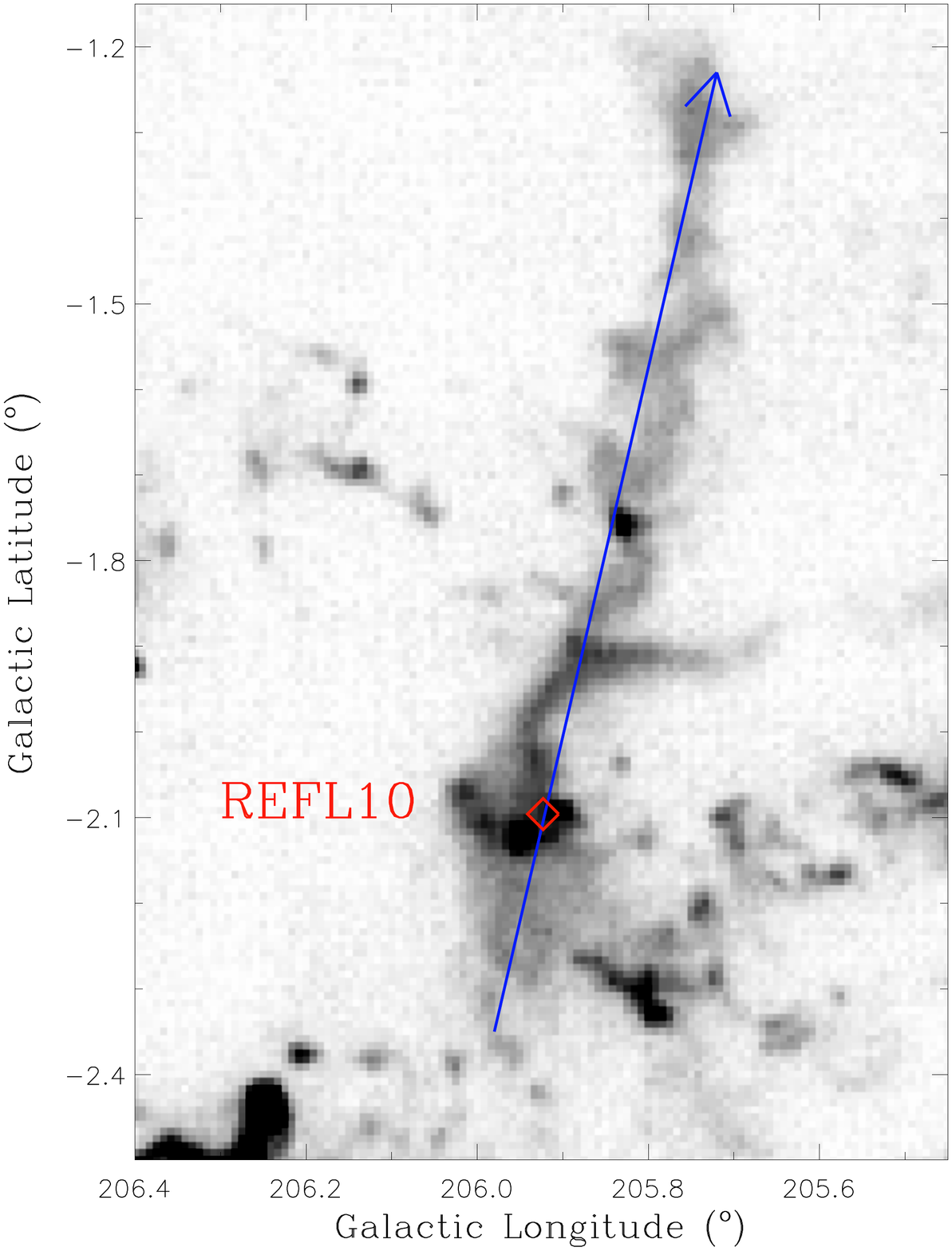}
\includegraphics[width=0.2\textwidth]{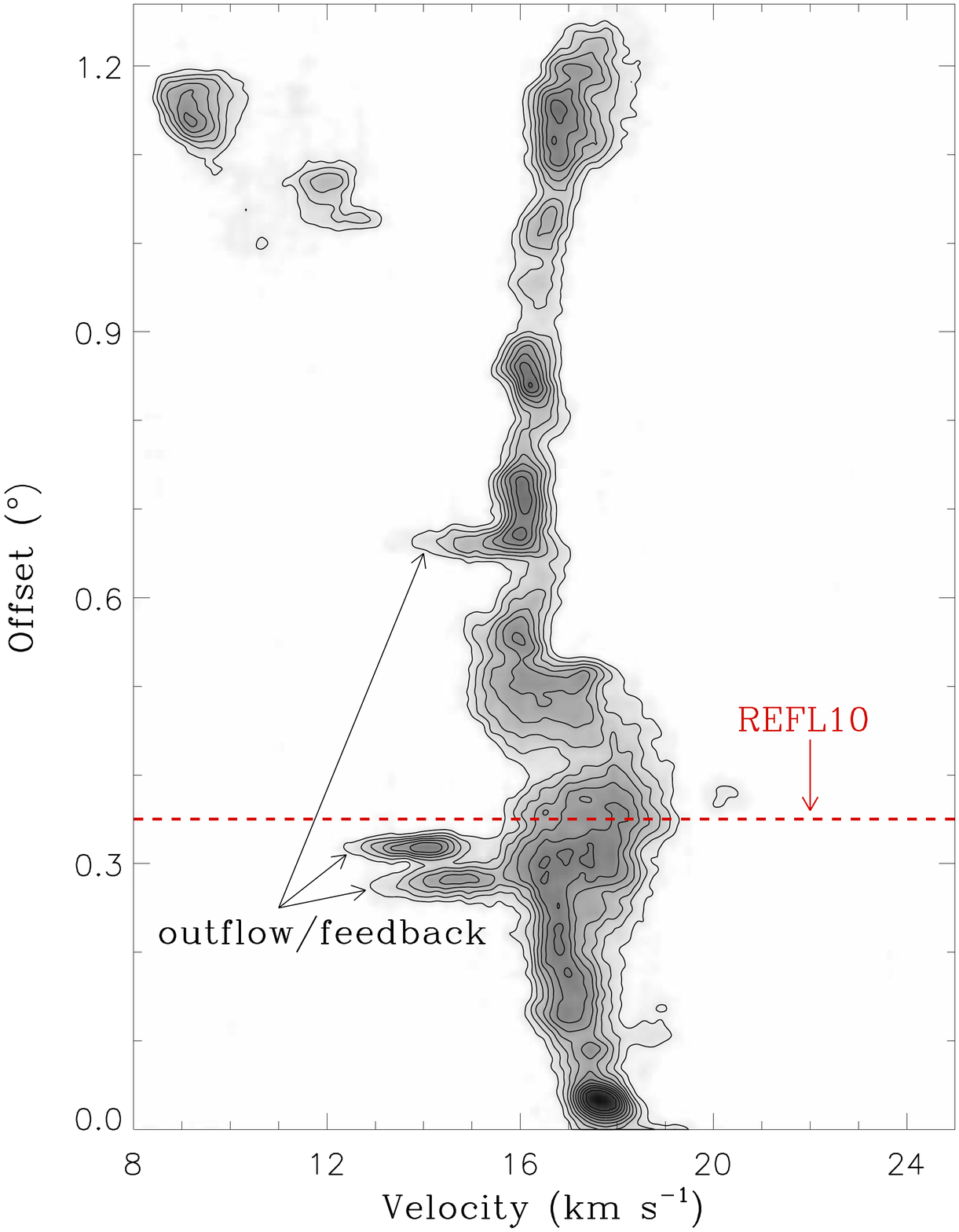}
\caption{Left (a): integrated intensity map of $^{12}$CO of filament g. Right (b): position-velocity map of $^{12}$CO emission along the arrow marked in the left panel. The overlaid contours are $^{12}$CO emission with the minimal level and the interval both being 0.1 $\times$ peak of $^{12}$CO brightness.}
\label{fig11}
\end{figure}

The $^{12}$CO and $^{13}$CO emission of clouds C7 and C8 starts to appear in the velocity range represented by Figure \ref{fig23} and extends to the velocity range from 17.5 km s$^{-1}$ to 20.5 km s$^{-1}$ (see Figure \ref{fig24}). They are coincident with the western rim of the Rosette Nebula. In the velocity range from 14 km s$^{-1}$ to 17.5 km s$^{-1}$, cloud C7 exhibits a long filamentary shape with a length of about 30 pc at the distance of 1.4 pc, which we refer to as filament g. The southern end of this filament corresponds to the $^{12}$CO emission maximum H in \cite{1980ApJ...241..676B} (see Figure \ref{fig23}) where the embedded clusters REFL10 is located. The $^{12}$CO integrated intensity map and position-velocity map of filament g are presented in Figure \ref{fig11}. Three feedback or outflow features can be found. We note that the direction of the velocity gradient along filament g is reverted at the location of the REFL10 young stellar cluster.

In the velocity range from 14 km s$^{-1}$ to 17.5 km s$^{-1}$ cloud C8 has a clumpy appearance (Figure \ref{fig23}). From Figure \ref{fig24} we can see that clouds C7 and C8 appear diffuse in the velocity range from 17.5 km s$^{-1}$ to 20.5 km s$^{-1}$. Cloud C8 corresponds to the $^{12}$CO emission maxima E and I in \cite{1980ApJ...241..676B}. With velocities ranging from 12 km s$^{-1}$ to 20.5 km s$^{-1}$, cloud C8 shows the largest velocity dispersion among the RMC. One possible reason for the large velocity dispersion is that cloud C8 may have been strongly influenced by the nearby NGC 2237 OB cluster, in addition to the impacts from the NGC 2244 OB cluster. Moreover, one molecular outflow has been found in cloud C8 \citep{2009MNRAS.395.1805D} which may has further influenced cloud C8. Cloud C9 is located to the west of C8. It is newly discovered in our survey of the RMC region benefiting from our large sky coverage. The velocity range of cloud C9 is from 15.5 km s$^{-1}$ to 20.5 km s$^{-1}$. The embedded cluster, CMFT10 \citep{2013A&A...557A..29C}, is located within C9.

\begin{deluxetable}{rcccrrcccrrc}
\decimals
\rotate
\tabletypesize{\footnotesize}
\tablewidth{0pt}
\tablenum{1}
\tablecaption{The independent clouds of RMC\label{tab1}}
\tablehead{
\colhead{ID} & \colhead{Cloud Name} & \colhead{l} & \colhead{b} & \colhead{$\rm{v_{lsr}}$} & \colhead{$\rm{T_{ex}}$} & \colhead{$\rm{T_{peak}(^{13}CO)}$} & \colhead{$\rm{T_{peak}(C^{18}O)}$} & \colhead{$\rm{\Delta v(^{13}CO)}$} & \colhead{$\rm{R_{13,18}}$} & \colhead{Mass}  & \colhead{Notes}\\
\colhead{ }  & \colhead{ }   & \colhead{($\arcdeg$)}  & \colhead{($\arcdeg$)}  & \colhead{$\rm{(km s^{-1})}$}  & \colhead{(K)}  & \colhead{(K)}   & \colhead{(K)}  & \colhead{($\rm{km s^{-1}}$)}  & \colhead{ }  & \colhead{($\rm{M_{\odot}}$)} & \colhead{ }}
\startdata
C1 & MWISP G206.192-2.392    &    206.192 & -2.392 & 3.96  & 15.3 & 4.0 & \nodata  & 2.47 & \nodata &  8.1$\rm{\times10^2}$  &  \nodata \\
C2 & MWISP G207.385-1.252    &    207.385 & -1.252 & 4.61  & 9.5  & 3.9 & 0.9 & 1.34 & 7.3     & 2.6$\rm{\times10^3}$ &    C\tablenotemark{a} \\
C3 & MWISP G207.568-1.735    &    207.568 & -1.735 & 12.34 & 11.9 & 7.9 & 1.8 & 2.23 & 12.2    & 4.4$\rm{\times10^4}$  &  A, B and C\tablenotemark{a} \\
C4 & MWISP G206.853-1.991    &    206.853 & -1.991 & 14.91 & 20.8 & 8.3 & 1.1 & 2.00 & 16.6 &  8.6$\rm{\times10^3}$ &    Monoceros Ridge, Extended Ridge and Shell\tablenotemark{b}\\
C5 & MWISP G207.774-1.877    &    207.774 & -1.877 & 11.20 & 10.8 & 7.6 & 1.8 & 1.87 & 11.7     & 1.7$\rm{\times10^4}$ &    New giant filament\\
C6 & MWISP G207.314-2.352    &    207.314 & -2.352 & 15.50 & 16.7 & 8.8 & 0.6 & 1.26 & 15.1     & 4.5$\rm{\times10^3}$ &    G\tablenotemark{a}\\
C7 & MWISP G205.858-2.199    &    205.858 & -2.199 & 17.69 & 9.8  & 2.9 & \nodata  & 1.05 & \nodata & 2.9$\rm{\times10^3}$ &    New giant filament\\
C8 & MWISP G206.031-2.606    &    206.031 & -2.606 & 16.38 & 17.3 & 6.0 & 0.7 & 3.43 & 24.2     & 6.8$\rm{\times10^3}$ &   E and I\tablenotemark{a}\\
C9 & MWISP G205.342-2.773    &    205.342 & -2.773 & 18.24 & 9.9  & 5.5 & 1.0 & 1.62 & 10.5 & 2.1$\rm{\times10^3}$ &    New molecular cloud\\
\enddata
\tablenotetext{a}{$^{12}$CO emission maxima in \citet{1980ApJ...241..676B}.}
\tablenotetext{b}{Structures noted in \citet{2008hsf1.book..928R}.}
\tablecomments{Columns 3-5 give the position centroids of the clumps in the PPV space. The temperature information of the three isotopologues are included in columns 6-8. Columns 9-11 give the line width, the abundance ratio of $^{13}$CO to C$^{18}$O, and the total mass derived from the X$\rm{_{CO}}$ factor, respectively. The position and velocity centroids of each cloud are derived from the $^{13}$CO emission. The last column gives other nomenclatures of the clouds in previous work.}
\end{deluxetable}

The cloud name within the MWISP survey, the galactic coordinates of the $^{12}$CO emission peak, the LSR velocity, the excitation temperature, $^{13}$CO and C$^{18}$O peak intensity, and the $^{13}$CO velocity dispersion for clouds C1-C9 are list in Table \ref{tab1}. Along with these information, the abundance ratio of $^{13}$CO to C$^{18}$O and the mass for cloud C1-C9, which are derived in Section \ref{sec3-4}, are also presented in Table \ref{tab1}.

\subsubsection{RMC background molecular clouds}

As shown in Figure \ref{fig1}, the $^{12}$CO emission in our surveyed area is concentrated to the velocity range from -2 km s$^{-1}$ to 20.5 km s$^{-1}$, which we ascribe to molecular clouds in the RMC complex at the distance of 1.4 kpc. However we can see that there are molecular clouds with velocities in the range from 20.5 km s$^{-1}$ to 58 km s$^{-1}$. According to the rotation model of \citet{2014ApJ...783..130R}, these molecular clouds are located at distances from 2.4 kpc to 11 kpc, which are behind the RMC complex and therefore we infer to them as the RMC background molecular clouds.

To identify and characterize the physical properties of clumps in the RMC background molecular clouds, we used the Fellwalker algorithm from Starlink \citep{2015A&C....10...22B}. The Fellwalker algorithm is a gradient-tracing scheme using all available data values. Results produced by the Fellwalker algorithm are less dependent on specific parameter settings than other clump finding algorithms such as CLUMPFIND. Using this algorithm, we have identified 73 molecular clumps. The integrated $^{12}$CO emission maps of these clumps are presented in Figure \ref{fig25} and the parameters fit for these clumps are given in Table \ref{tab2}. In Table \ref{tab2} the distances are derived using the kinematics distance model of \citet{2014ApJ...783..130R} and the mass for each clump is calculated in the way discussed in Section \ref{sec3-4}. The largest distance of these clumps from the Sun is 11 kpc which is located in the outer arm \citep{2008AJ....135.1301V,2009ApJ...700..137R,2014ApJ...783..130R}.

Some of these background clouds may be globules accelerated by the expansion of the Rosette Nebula \ion{H}{2} region. With near-infrared imaging and mid-infrared and far-infrared data from Spitzer IRAC and Herschel PACS, \citet{2014A&A...567A.108M,2017A&A...605A..82M} investigated the physical properties and star formation activity of the globulettes and globules that lie in front of the Rosette Nebula and are associated with the C1 cloud. They found that the masses of globulettes are subsolar and the masses of globules are 9.5-11.6 M$_{\odot}$ and that the globules RN A and RN E have velocities around 3 km s$^{-1}$ which is similar to the velocity of cloud C1. The average velocity of the Rosette Nebula is 16.7 km s$^{-1}$ \citep{2009MNRAS.395.1805D}, therefore, it can be assumed that globules RN A and RN E have been accelerated by the Rosette Nebula \ion{H}{2} region about 13.7 km s$^{-1}$ toward the earth. Taking the results of \citet{2014A&A...567A.108M,2017A&A...605A..82M} as a reference on the masses of globules and the acceleration by the Rosette Nebula \ion{H}{2} region, we consider the background molecular clouds that have masses less than 15 M$_{\odot}$ and velocities less than 30.4 km s$^{-1}$ as the possible globules blown out of the \ion{H}{2} region on the far side. We recalculated the masses of the background molecular clouds with a fixed distance of 1.4 kpc rather than the kinematic distances derived from the Galactic rotation curve and it is found that 17 background molecular clouds satisfy the mass and velocity criterion. These 17 clouds are possible globules that have been blown out of the \ion{H}{2} region on the far side of the Rosette Nebula. Table \ref{tab3} lists the names of these clouds along with their masses at the distance of 1.4 kpc.

\subsection{Cloud-cloud collision in the RMC}

\begin{figure}[h]
\centering
\includegraphics[width=0.6\textwidth]{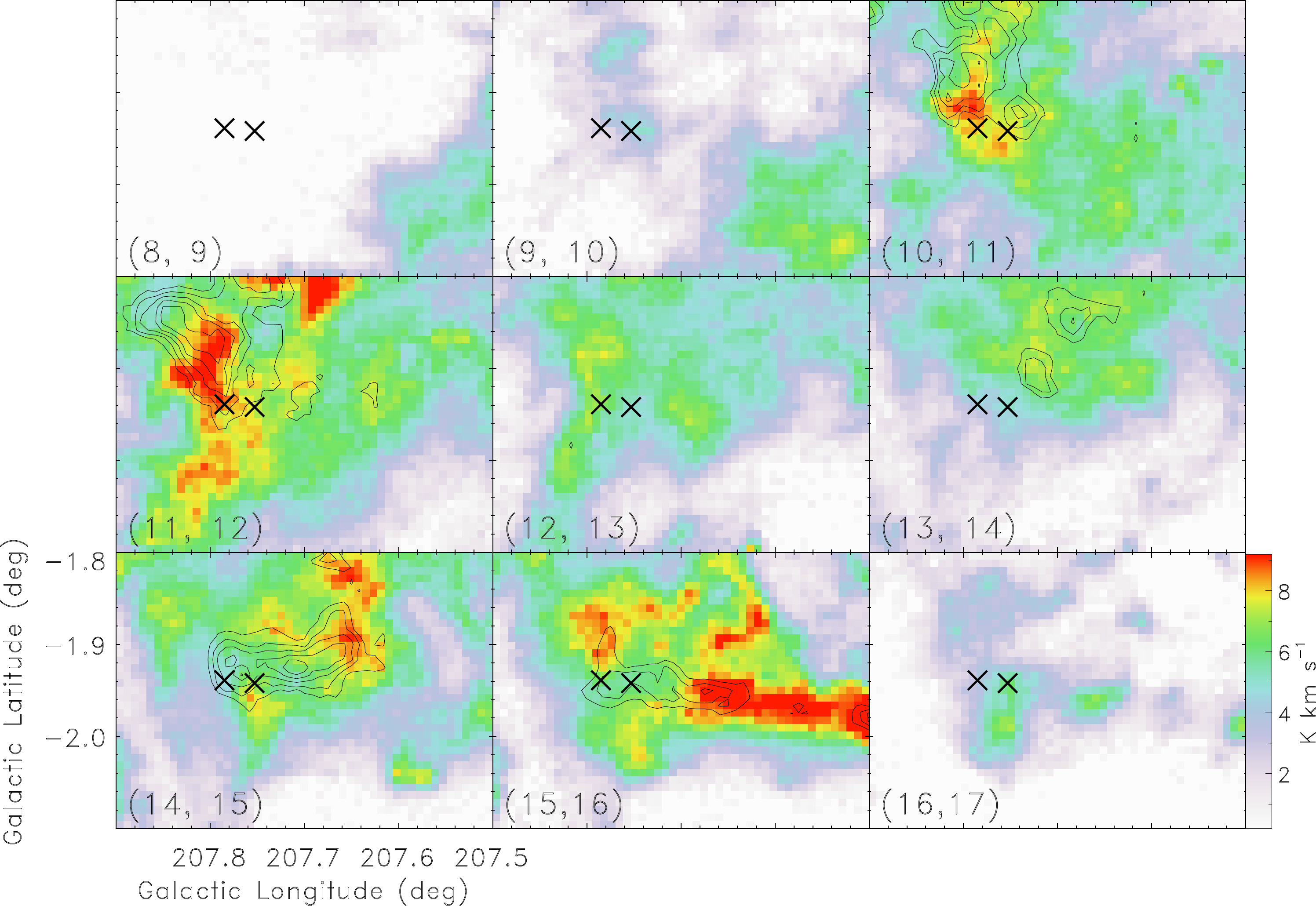}
\caption{Velocity channel map of $^{12}$CO emission (color) of the REFL9 and PouF region. The crosses indicate the positions of the young clusters REFL9 and PouF. The contours represent the emission of $^{13}$CO. The minimal level and the interval are 0.5 and 0.1 times the peak of $^{13}$CO brightness, respectively.}
\label{fig12}
\end{figure}

\begin{figure}[h]
\centering
\includegraphics[width=0.3\textwidth]{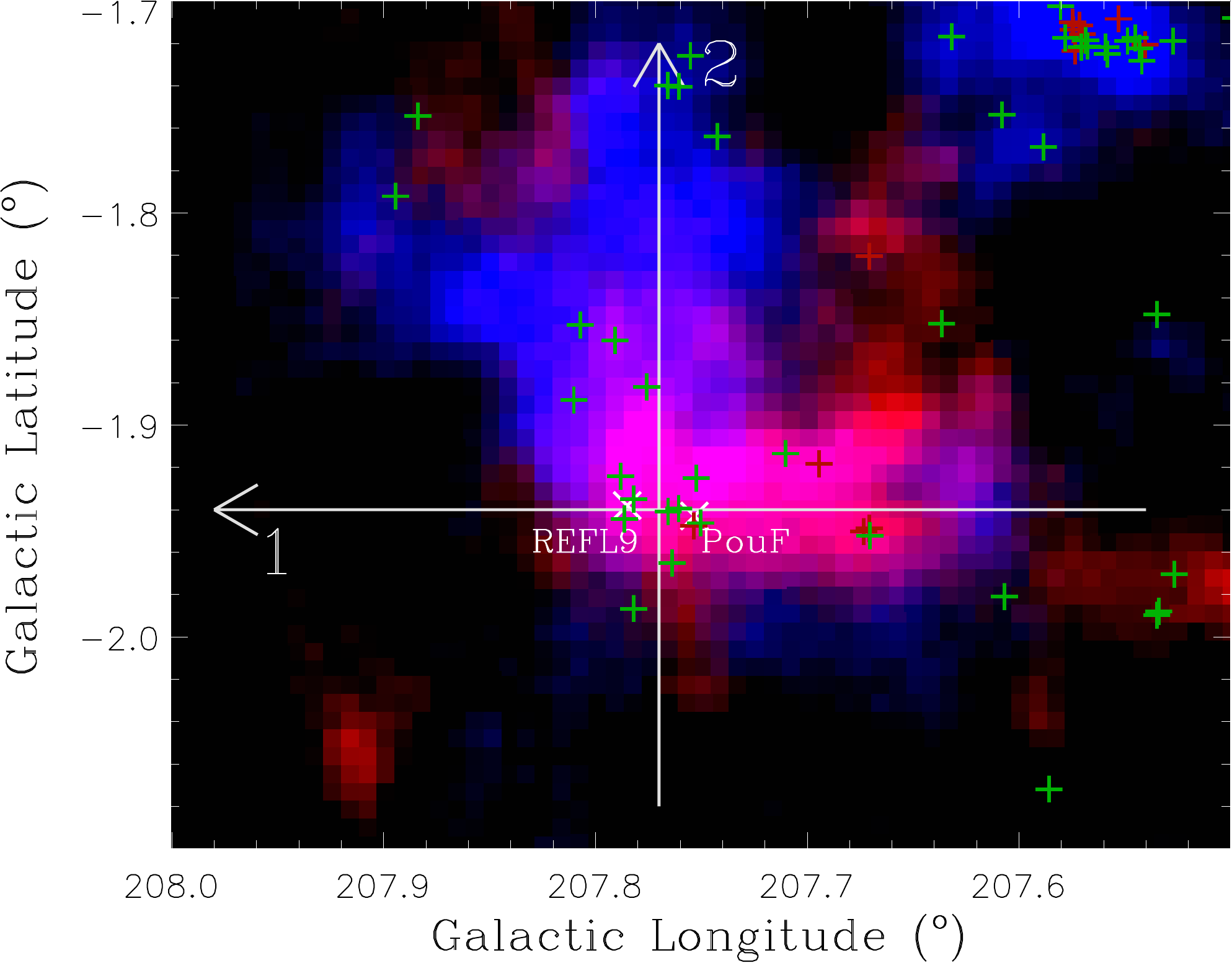}
\includegraphics[width=0.3\textwidth]{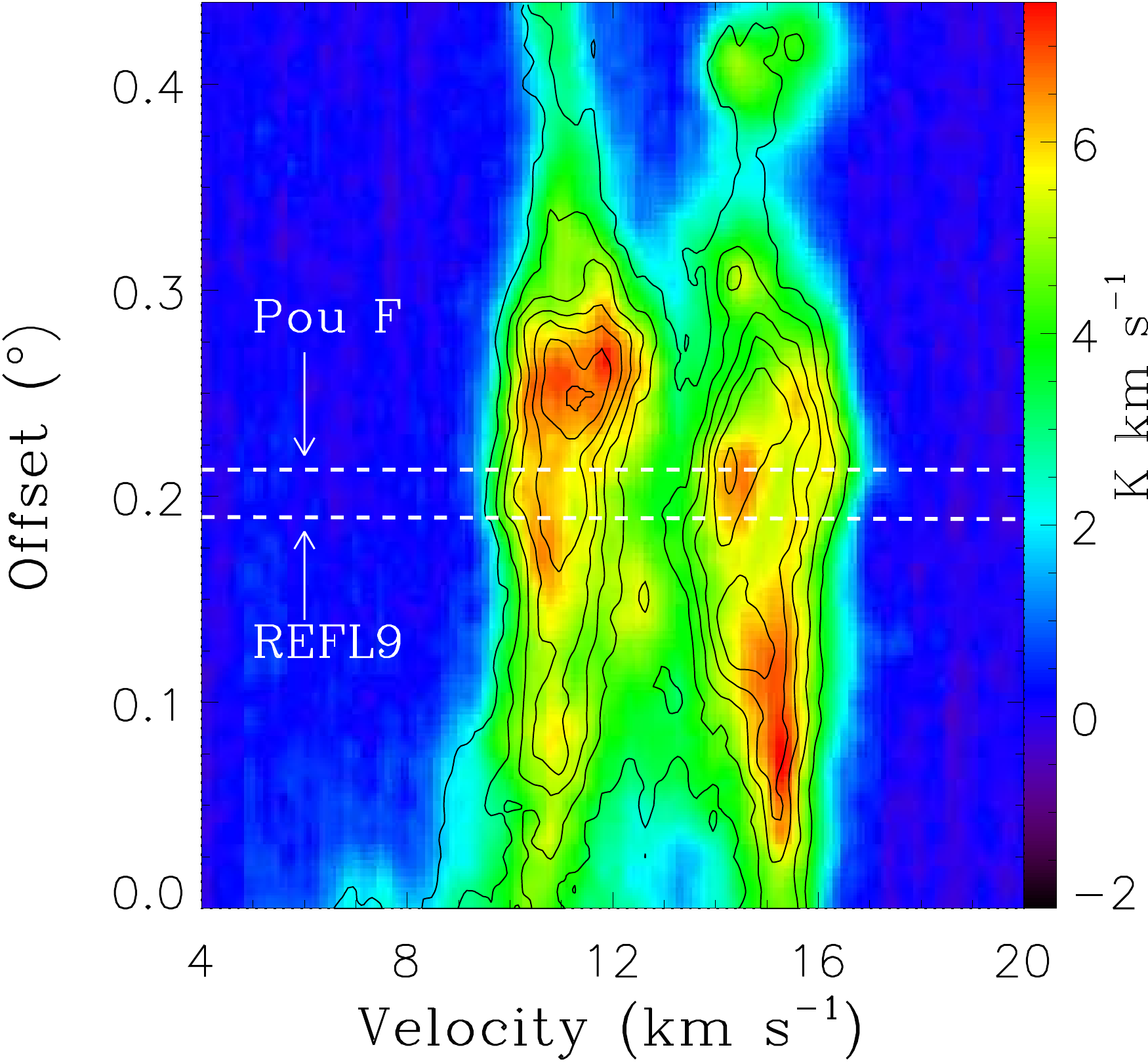}
\includegraphics[width=0.3\textwidth]{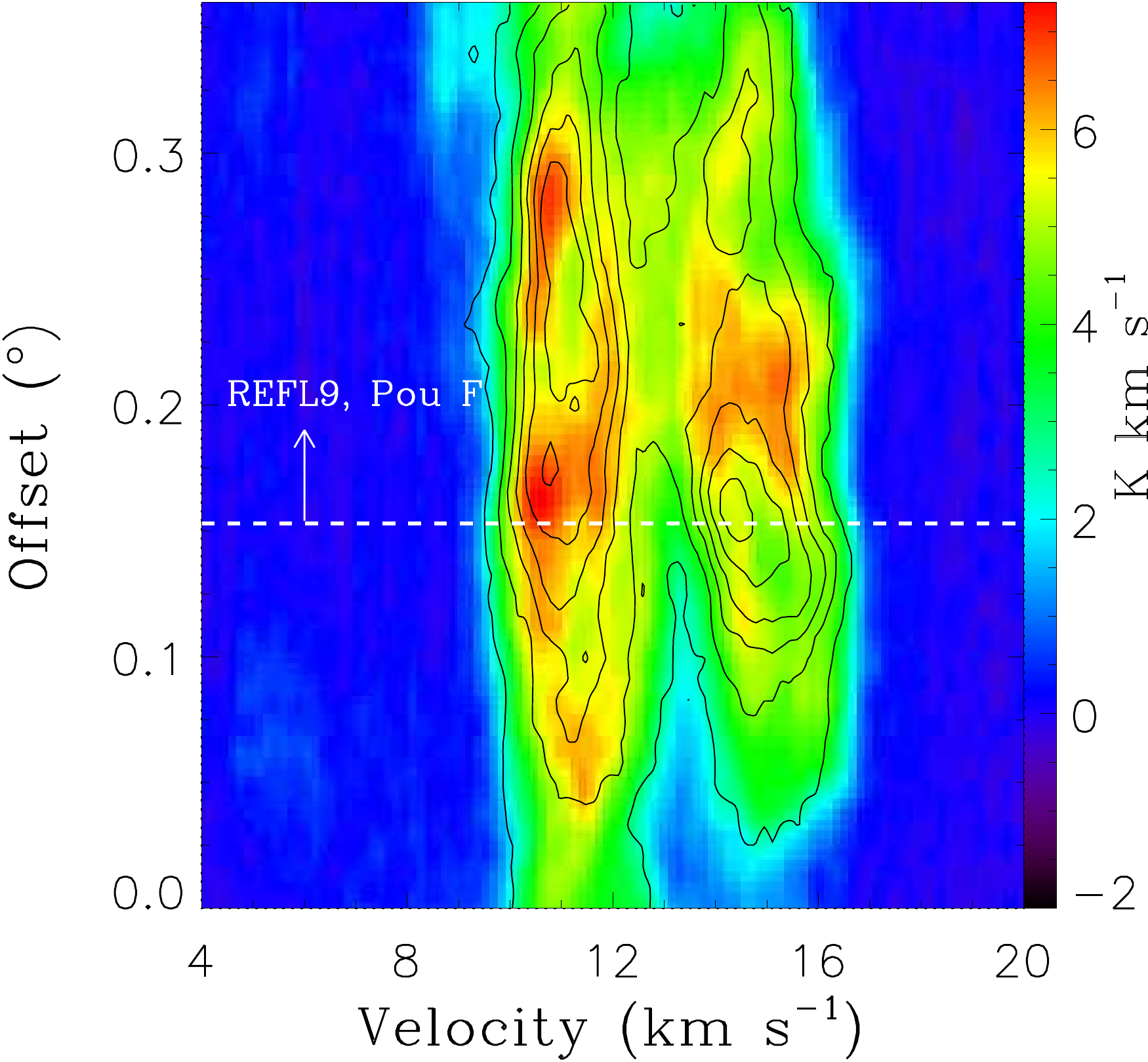}
\caption{Left: velocity distribution map of $^{12}$CO emission of the REFL9 and PouF (see Figure \ref{fig22}) region where the blue represents integrated intensity from 8 to 12 km s$^{-1}$ and the red from 13 to 17 km s$^{-1}$. The two white crosses at the center indicate the positions of the REFL9 and PouF young clusters. The red and the green pluses indicate the Class I and Class II YSOs in the region, respectively. Middle: position-velocity map of $^{12}$CO emission (color) along arrow 1 in the left image with $^{13}$CO emission contours overlaid. Right: position-velocity map along arrow 2 in the left. The dash lines in the middle and right panels indicate the centers of the REFL9 and PouF young clusters.}
\label{fig13}
\end{figure}

The collision between molecular clouds is an important dynamic process for the evolution of molecular clouds and is believed to be a possible approach to triggered star formation \citep{2014ApJ...780...36F, 2015ApJ...807L...4F}.  Figure \ref{fig12} shows the velocity channel map for the molecular clouds in the region of the young clusters REFL9 and PouF. It can be seen that there are two molecular clouds in this region. One cloud is orientated in the north-south direction with velocities in the range from  10 to 12 km s$^{-1}$. The other one is orientated in the east-west direction with velocities in the range from  14 to 16 km s$^{-1}$. The two molecular clouds intersect at the position of the REFL9 and PouF young clusters. The spatial distribution of the members of these two young clusters is presented in the left panel of Figure \ref{fig13}, and the position-velocity maps along the east-west direction (arrow 1) and the north-south direction (arrow 2) through the center of the REFL9 and PouF clusters are shown in the middle and right panels of Figure \ref{fig13}, respectively. From the position-velocity maps we can see velocity bridging between the two clouds for a wide spatial extent. Wide velocity bridging feature is believed to be the signature of cloud-cloud collision \citep{2015MNRAS.450...10H, 2016ApJ...820...26F, 2017ApJ...835L..14G}. Therefore, Figure \ref{fig13} provides evidence for a cloud-cloud collision taking place in the REFL9 and PouF region. The REFL9 and PouF young clusters are probably the result of this cloud-cloud collision.

\subsection{Physical properties of clouds in the RMC complex}\label{sec3-4}

Assuming the molecular clouds are under the local thermodynamic equilibrium (LTE) conditions and that the $^{12}$CO J = 1$-$0 emission is optically thick, the mass of the molecular cloud can be calculated with the measured brightness of $^{12}$CO and $^{13}$CO J = 1$-$0 emission. Taking the cosmic microwave background (CMB) radiation to be 2.7 K and the filling factor of $^{12}$CO J = 1-0 emission to be unity, the excitation temperature can be calculated according to the following formula \citep{1998AJ....116..336N, 2000Obs...120..289R}
\begin{equation}
T_{ex} = \frac{5.53}{ln (1+\frac{5.53}{T_{mb} (^{12}CO)+0.819})},
\end{equation}
where T$_{mb}$ is the main-beam brightness temperature of $^{12}$CO. Assuming the molecular cloud has the same excitation temperature along the line of sight, the optical depth of $^{13}$CO and C$^{18}$O emission can be expressed as following \citep{2010ApJ...721..686P}

\begin{equation}
\label{eqa2}
\tau (^{13}CO)=-ln [1-\frac{T_{mb} (^{13}CO)}{5.29} ([e^{5.29/T_{ex}}-1]^{-1}-0.164)^{-1}].
\end{equation}

\begin{equation}
\label{eqa3}
\tau (C^{18}O)=-ln [1-\frac{T_{mb}(C^{18}O)}{5.27} ([e^{5.27/T_{ex}}-1]^{-1}-0.166)^{-1}]
\end{equation}

The column density of linear rigid rotor molecule is related to its optical depth of transition from the J to the J$+$1 energy level by the formula \citep{1991ApJ...374..540G,1997ApJ...476..781B}

\begin{equation}
\label{eqa4}
N = \frac{3h}{8\pi^3\mu^2}\frac{k(T_{ex}+hB/3k)}{(J+1)hB}\frac{e^{E_J/kT_{ex}}}{1-e^{(-hv/kT_{ex})}}\int \tau dv
\end{equation}

where h and k are the Planck and Boltzmann constants respectively and B is the molecular rotation constant.

For $^{13}$CO and C$^{18}$O J = 1$-$0 emission, formula \ref{eqa4} can be converted to \citep{1991ApJ...374..540G,1997ApJ...476..781B}

\begin{equation}
\label{eqa5}
N(^{13}CO) = 2.42 \times 10^{14} \frac{T_{ex}+0.88}{1-e^{-5.29/T_{ex}}} \int \tau (^{13}CO)dv.
\end{equation}

\begin{equation}
\label{eqa6}
N(C^{18}O) = 2.54 \times 10^{14} \frac{T_{ex}+0.88}{1-e^{5.27/T_{ex}}} \int \tau (C^{18}O)dv
\end{equation}

When the emission is optically thin, T$_{ex} \int \tau dv \approx \frac{\tau }{1-e^{-\tau }} \int T_{mb} dv$ \citep{2009tra..book.....W}. Assuming $[^{12}C/^{13}C = 77]$, $[^{16}O/^{18}O = 560]$ \citep{1994ARA&A..32..191W} and H$_2/^{12}CO = 1.1 \times 10^4$ \citep{1982ApJ...262..590F}, the hydrogen molecule column density can be obtained from the integrated intensity of $^{13}$CO emission

\begin{equation}
N(H_2)^{^{13}CO} =2.05\times 10^{20}\frac{\tau (^{13}CO)}{1-e^{-\tau (^{13}CO)}}\frac{1+0.88/T_{ex}}{1-e^{5.29/T_{ex}}} \int T_{mb} (^{13}CO)dv.
\end{equation}
\begin{equation}
N(H_2)^{C^{18}O} =1.56\times 10^{21} \times \frac{\tau (C^{18}O)}{1-e^{-\tau (C^{18}O)}} \times \frac{1+0.88/T_{ex}}{1-e^{5.27/T_{ex}}} \int T_{mb} (C^{18}O)dv
\end{equation}

Another method to estimate the mass of a cloud is to use the $^{12}$CO-to-H$_{2}$ conversion factor, the so called X factor. Taking the X factor to be 1.8 $\times$ 10$^{20}$ H$_{2}$ cm$^{-2}$ (K km s$^{-1}$)$^{-1}$ \citep{2001ApJ...547..792D}, we can obtain the column density of molecular cloud just with the integrated intensity of $^{12}$CO emission

\begin{equation}
N(H_2) = 1.8\times 10^{20} \int T_{mb} (^{12}CO)dv.
\end{equation}

The masses calculated with the above LTE method for clouds C1-C9 and with the X factor method for the clumps in the RMC background clouds are list in Tables \ref{tab1} and \ref{tab2}.

The mass of molecular clouds in the whole region of our survey is calculated to be 2.0 $\times$ 10$^5$ M$_{\odot}$ with the X factor method, 5.8 $\times$ 10$^4$ M$_{\odot}$ from $^{13}$CO and 4.2 $\times$ 10$^3$ M$_{\odot}$ from C$^{18}$O with the LTE approach. \citet{2006ApJ...643..956H} calculated the mass of the RMC to be 1.6 $\times$ 10$^{5}$ M$_{\odot}$ with an X factor of 1.9 $\times$ 10$^{20}$ H$_{2}$ cm$^{-2}$ (K km s$^{-1}$)$^{-1}$. Assuming the H$_{2}$ to $^{13}$CO abundance ratio to be 8 $\times$ 10$^{5}$, they calculated the mass to be 1.16 $\times$ 10$^{5}$ M$_{\odot}$ with the LTE method. From the dust column density map derived from Herschel observations, \citet{2010A&A...518L..83S} calculated the mass of the eastern region of RMC to be about 1 $\times$ $10^5$ M$_{\odot}$.

\begin{figure}[h]
\centering
\includegraphics[width=0.5\textwidth]{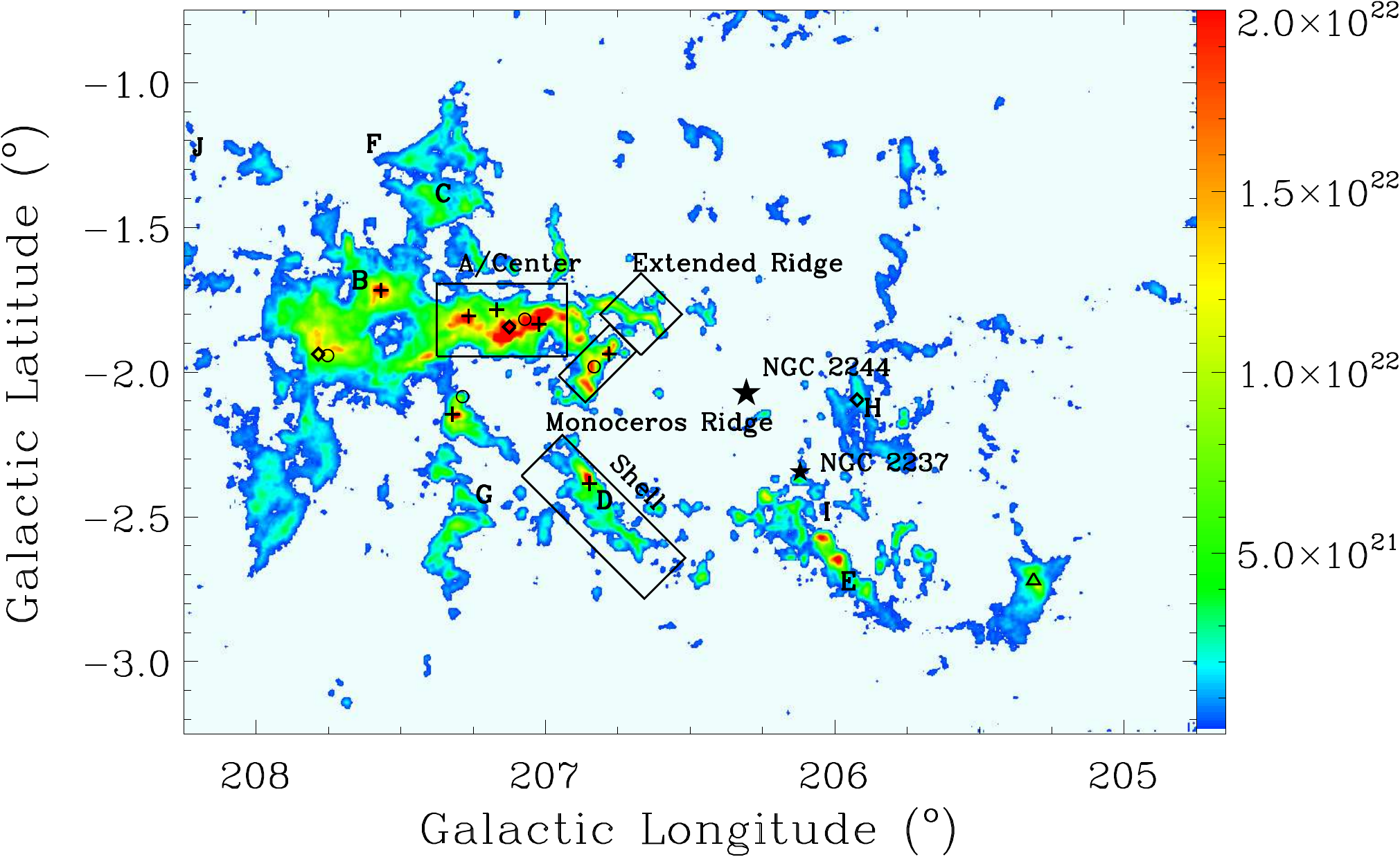}
\caption{Map of H$_2$ column density of the RMC complex. The Center, Extended Ridge, Monoceros Ridge, and Shell regions from \citet{2008hsf1.book..928R} are indicated with rectangles. All the others are the same as in Figure \ref{fig3}.}
\label{fig14}
\end{figure}

The distribution of the column density of hydrogen molecules in the region is presented in Figure \ref{fig14}. From Figure \ref{fig14} we can see that young stellar clusters are preferentially located at high column density regions, which is consistent with the result of \citet{2013ApJ...769..140Y} who found a similar relation in RMC with infrared extinction data.

\clearpage
\startlongtable
\begin{deluxetable}{lcccccccr}
\tabletypesize{\footnotesize}
\tablewidth{10pt}
\tablenum{2}
\tablecaption{Clumps in RMC background molecular clouds}
\label{tab2}
\tablehead{
\colhead{Clumps} & \colhead{l} & \colhead{b} & \colhead{v$_{\rm lsr}$} & \colhead{d$_{\rm k}$} & \colhead{T$\rm{_{peak}}$} & \colhead{$\Delta$v} & \colhead{$\rm{R_{eff}}$} & \colhead{M}\\
\colhead{ } & \colhead{$(\arcdeg)$} & \colhead{$(\arcdeg)$} & \colhead{(km s$^{-1}$)} & \colhead{(kpc)} & \colhead{(K)} & \colhead{(km s$^{-1}$)} & \colhead{(pc)} & \colhead{(M$_{\odot}$)}}
\startdata
MWSIP G204.754-2.637&204.754&-2.637&31.4 & 4.3 &  2.7 & 0.4 &  1.2 &  12.8\\
MWSIP G204.763-0.899&204.763&-0.899&32.2 & 4.4 &  6.8 & 1.2 &  1.8 &  208.9\\
MWSIP G204.766-1.305&204.766&-1.305&30.7 & 4.2 &  3.6 & 0.9 &  1.4 &  58.3\\
MWSIP G204.809-0.951&204.809&-0.951&31.4 & 4.3 &  4.9 & 1.8 &  2.0 &  573.4\\
MWSIP G204.827-1.777&204.827&-1.777&23.6 & 2.9 &  3.5 & 0.6 &  0.9 &  14.0\\
MWSIP G204.859-1.097&204.859&-1.097&32.2 & 4.4 &  5.3 & 1.7 &  3.0 &  1089.1\\
MWSIP G204.865-1.068&204.865&-1.068&26.0 & 3.3 &  3.9 & 2.1 &  2.0 &  260.6\\
MWSIP G204.873-2.093&204.873&-2.093&41.9 & 6.6 &  3.3 & 0.4 &  2.0 &  18.2\\
MWSIP G204.929-2.021&204.929&-2.021&51.0 & 9.4 &  4.6 & 0.3 &  2.8 &  62.1\\
MWSIP G204.941-3.222&204.941&-3.222&40.3 & 6.2 &  3.5 & 1.7 &  3.1 &  626.7\\
MWSIP G205.237-1.916&205.237&-1.916&52.2 & 9.6 &  4.9 & 0.7 &  3.5 &  520.6\\
MWSIP G205.418-2.791&205.418&-2.791&42.0 & 6.4 &  3.4 & 0.5 &  2.0 &  48.8\\
MWSIP G205.423-2.820&205.423&-2.820&39.0 & 5.7 &  3.6 & 1.5 &  1.7 &  462.5\\
MWSIP G205.445-0.973&205.445&-0.973&35.1 & 4.9 &  3.0 & 0.8 &  1.4 &  93.4\\
MWSIP G205.465-3.116&205.465&-3.116&36.6 & 5.1 &  3.5 & 0.4 &  1.7 &  24.1\\
MWSIP G205.485-3.213&205.485&-3.213&41.4 & 6.2 &  2.9 & 0.7 &  2.2 &  70.0\\
MWSIP G205.490-3.064&205.490&-3.064&37.0 & 5.2 &  4.4 & 1.9 &  4.4 &  1285.5\\
MWSIP G205.492-0.886&205.492&-0.886&35.8 & 5.0 &  4.8 & 1.7 &  1.5 &  685.4\\
MWSIP G205.507-3.191&205.507&-3.191&35.6 & 4.9 &  4.2 & 1.3 &  3.1 &  860.9\\
MWSIP G205.561-2.935&205.561&-2.935&35.7 & 4.9 &  4.8 & 1.1 &  1.9 &  204.9\\
MWSIP G205.566-3.009&205.566&-3.009&36.6 & 5.1 &  4.0 & 1.4 &  2.2 &  332.6\\
MWSIP G205.596-0.790&205.596&-0.790&39.3 & 5.7 &  4.7 & 1.0 &  2.3 &  245.7\\
MWSIP G205.647-2.867&205.647&-2.867&35.8 & 4.9 &  4.9 & 2.9 &  2.5 &  1240.9\\
MWSIP G205.738-2.906&205.738&-2.906&37.2 & 5.2 &  6.0 & 1.9 &  4.9 &  3174.9\\
MWSIP G205.817-1.703&205.817&-1.703&45.3 & 7.1 &  4.7 & 1.1 &  3.9 &  1249.7\\
MWSIP G205.888-2.686&205.888&-2.686&38.2 & 5.4 &  3.6 & 0.9 &  2.5 &  117.1\\
MWSIP G205.969-2.649&205.969&-2.649&37.6 & 5.2 &  3.4 & 0.7 &  2.0 &  128.5\\
MWSIP G205.975-2.547&205.975&-2.547&31.7 & 4.1 &  3.5 & 0.3 &  1.2 &  8.3\\
MWSIP G205.988-1.196&205.988&-1.196&46.8 & 7.5 &  4.1 & 0.5 &  2.5 &  97.8\\
MWSIP G206.041-1.116&206.041&-1.116&30.0 & 3.8 &  3.8 & 0.4 &  1.2 &  20.2\\
MWSIP G206.072-2.772&206.072&-2.772&37.6 & 5.2 &  3.5 & 0.9 &  2.1 &  111.4\\
MWSIP G206.138-0.895&206.138&-0.895&46.6 & 7.3 &  3.1 & 0.4 &  2.4 &  60.6\\
MWSIP G206.200-2.820&206.200&-2.820&47.2 & 7.5 &  2.3 & 0.3 &  2.3 &  18.6\\
MWSIP G206.229-1.207&206.229&-1.207&41.5 & 6.0 &  3.2 & 0.6 &  1.9 &  38.7\\
MWSIP G206.231-0.841&206.231&-0.841&46.7 & 7.3 &  4.4 & 0.4 &  2.2 &  44.9\\
MWSIP G206.325-2.188&206.325&-2.188&21.6 & 2.5 &  4.4 & 0.9 &  1.2 &  78.6\\
MWSIP G206.357-1.613&206.357&-1.613&39.1 & 5.4 &  2.9 & 0.6 &  1.9 &  51.0\\
MWSIP G206.408-2.415&206.408&-2.415&41.0 & 5.8 &  5.4 & 0.8 &  2.3 &  284.0\\
MWSIP G206.412-2.176&206.412&-2.176&21.8 & 2.5 &  4.2 & 0.8 &  1.2 &  67.1\\
MWSIP G206.479-2.354&206.479&-2.354&41.4 & 5.9 &  5.3 & 0.7 &  3.1 &  408.5\\
MWSIP G206.501-2.207&206.501&-2.207&21.7 & 2.5 &  4.1 & 0.5 &  1.2 &  39.8\\
MWSIP G206.547-1.592&206.547&-1.592&21.6 & 2.4 &  2.9 & 1.2 &  0.7 &  5.9\\
MWSIP G206.562-1.923&206.562&-1.923&58.1 & 10.9 & 3.4 & 0.3 &  3.3 &  53.3\\
MWSIP G206.565-0.792&206.565&-0.792&22.0 & 2.5 &  3.6 & 0.3 &  0.8 &  7.3\\
MWSIP G206.603-2.122&206.603&-2.122&43.3 & 6.3 &  5.7 & 0.8 &  3.2 &  786.8\\
MWSIP G206.635-1.426&206.635&-1.426&21.7 & 2.4 &  4.0 & 1.5 &  0.7 &  20.4\\
MWSIP G206.648-1.589&206.648&-1.589&54.3 & 9.4 &  5.1 & 0.6 &  2.9 &  126.3\\
MWSIP G206.699-0.902&206.699&-0.902&21.7 & 2.4 &  2.8 & 0.6 &  0.8 &  11.7\\
MWSIP G206.709-1.493&206.709&-1.493&21.5 & 2.4 &  4.8 & 4.5 &  0.9 &  184.7\\
MWSIP G206.774-1.868&206.774&-1.868&39.3 & 5.4 &  8.6 & 1.2 &  2.4 &  597.8\\
MWSIP G206.777-1.897&206.777&-1.897&21.9 & 2.5 &  11.7 & 2.1 &  1.1 &  285.3\\
MWSIP G206.788-0.898&206.788&-0.898&21.7 & 2.4 &  5.4 & 0.7 &  1.3 &  77.5\\
MWSIP G206.861-2.191&206.861&-2.191&42.4 & 6.0 &  2.5 & 0.4 &  1.8 &  19.5\\
MWSIP G206.885-0.783&206.885&-0.783&39.2 & 5.3 &  3.1 & 0.3 &  1.6 &  17.0\\
MWSIP G206.897-1.932&206.897&-1.932&22.6 & 2.5 &  6.2 & 1.0 &  1.1 &  47.5\\
MWSIP G206.933-1.456&206.933&-1.456&28.6 & 3.4 &  4.1 & 0.7 &  1.3 &  57.3\\
MWSIP G206.946-0.843&206.946&-0.843&39.3 & 5.3 &  3.8 & 0.3 &  1.7 &  23.3\\
MWSIP G206.962-2.156&206.962&-2.156&28.3 & 3.4 &  3.5 & 0.7 &  1.6 &  95.9\\
MWSIP G207.022-1.053&207.022&-1.053&39.8 & 5.4 &  5.8 & 0.9 &  1.8 &  179.2\\
MWSIP G207.071-1.906&207.071&-1.906&41.8 & 5.8 &  4.5 & 0.5 &  2.0 &  127.0\\
MWSIP G207.103-2.368&207.103&-2.368&21.5 & 2.4 &  2.9 & 0.4 &  1.0 &  17.7\\
MWSIP G207.198-2.567&207.198&-2.567&26.1 & 3.0 &  2.6 & 0.4 &  1.0 &  9.1\\
MWSIP G207.217-2.133&207.217&-2.133&21.8 & 2.4 &  4.2 & 1.4 &  1.2 &  89.1\\
MWSIP G207.387-1.149&207.387&-1.149&28.4 & 3.3 &  2.5 & 0.4 &  1.0 &  27.9\\
MWSIP G207.408-2.333&207.408&-2.333&21.0 & 2.3 &  2.2 & 0.5 &  0.6 &  1.1\\
MWSIP G207.508-1.589&207.508&-1.589&22.5 & 2.5 &  4.2 & 0.4 &  0.9 &  18.5\\
MWSIP G207.648-1.107&207.648&-1.107&23.9 & 2.6 &  5.0 & 1.0 &  1.1 &  114.1\\
MWSIP G207.777-2.395&207.777&-2.395&52.4 & 8.2 &  6.0 & 0.9 &  3.2 &  641.9\\
MWSIP G207.787-2.288&207.787&-2.288&52.2 & 8.1 &  4.4 & 0.9 &  4.1 &  925.9\\
MWSIP G207.832-0.933&207.832&-0.933&23.8 & 2.6 &  2.9 & 0.5 &  1.5 &  32.3\\
MWSIP G208.007-0.985&208.007&-0.985&24.2 & 2.6 &  2.8 & 0.3 &  0.9 &  7.6\\
MWSIP G208.033-0.778&208.033&-0.778&25.7 & 2.8 &  4.8 & 1.1 &  1.1 &  71.7\\
MWSIP G208.053-1.198&208.053&-1.198&29.4 & 3.4 &  2.7 & 0.3 &  1.0 &  7.5\\
\enddata
\tablecomments{Columns 2-4 give the position centroids of the clumps in the PPV space, column 5 gives the kinematic distance derived from model A5 in \cite{2014ApJ...783..130R}. Columns 5-8 give physical parameters of the clumps, i.e., peak brightness temperature, line width, effective radius, and the total mass derived from the X$\rm{_{CO}}$ factor, respectively.}
\end{deluxetable}

\clearpage
\startlongtable
\begin{deluxetable}{lccccr}
\tabletypesize{\footnotesize}
\tablewidth{10pt}
\tablenum{3}
\tablecaption{Possible accelerated globules}
\label{tab3}
\tablehead{
\colhead{Clumps} & \colhead{l} & \colhead{b} & \colhead{v$_{\rm lsr}$}  & \colhead{$\rm{R_{eff}}$} & \colhead{M}\\
\colhead{ } & \colhead{$(\arcdeg)$} & \colhead{$(\arcdeg)$} & \colhead{(km s$^{-1}$)}   & \colhead{(pc)} & \colhead{(M$_{\odot}$)}}
\startdata
MWSIP G204.827-1.777&204.827&-1.777&23.6  &  0.5 &  3.1\\
MWSIP G206.041-1.116&206.041&-1.116&30.0  &   0.5 &  2.7\\
MWSIP G206.501-2.207&206.501&-2.207&21.7  &   0.7 &  12.5\\
MWSIP G206.547-1.592&206.547&-1.592&21.6  &  0.4 &  1.9\\
MWSIP G206.565-0.792&206.565&-0.792&22.0  &  0.5 &  2.2\\
MWSIP G206.635-1.426&206.635&-1.426&21.7  & 0.4  &  6.5\\
MWSIP G206.699-0.902&206.699&-0.902&21.7  &  0.5 &  3.7\\
MWSIP G206.897-1.932&206.897&-1.932&22.6  &  0.6 &  14.0\\
MWSIP G206.933-1.456&206.933&-1.456&28.6  &  0.6 &  9.4\\
MWSIP G207.103-2.368&207.103&-2.368&21.5  &   0.6 &  5.9\\
MWSIP G207.198-2.567&207.198&-2.567&26.1  &  0.5 &  1.9\\
MWSIP G207.387-1.149&207.387&-1.149&28.4  &  0.4 &  4.9\\
MWSIP G207.408-2.333&207.408&-2.333&21.0  &  0.4 &  0.4\\
MWSIP G207.508-1.589&207.508&-1.589&22.5  &  0.5 &  5.8\\
MWSIP G207.832-0.933&207.832&-0.933&23.8  &  0.8 &  9.0\\
MWSIP G208.007-0.985&208.007&-0.985&24.2  &  0.5 &  2.1\\
MWSIP G208.053-1.198&208.053&-1.198&29.4  &  0.4 &  1.3\\
\enddata
\tablecomments{Columns 2-4 give the position centroids of the globules in the PPV space. Columns 5-6 give the effective radius and the mass derived from the X$\rm{_{CO}}$ factor of the possible accelerated globules.}
\end{deluxetable}

\section{Discussion}

\subsection{Distribution of velocity, excitation temperature, and velocity dispersion of RMC}\label{discussion:1}

\begin{figure}[h]
\centering
\includegraphics[width=0.4\textwidth]{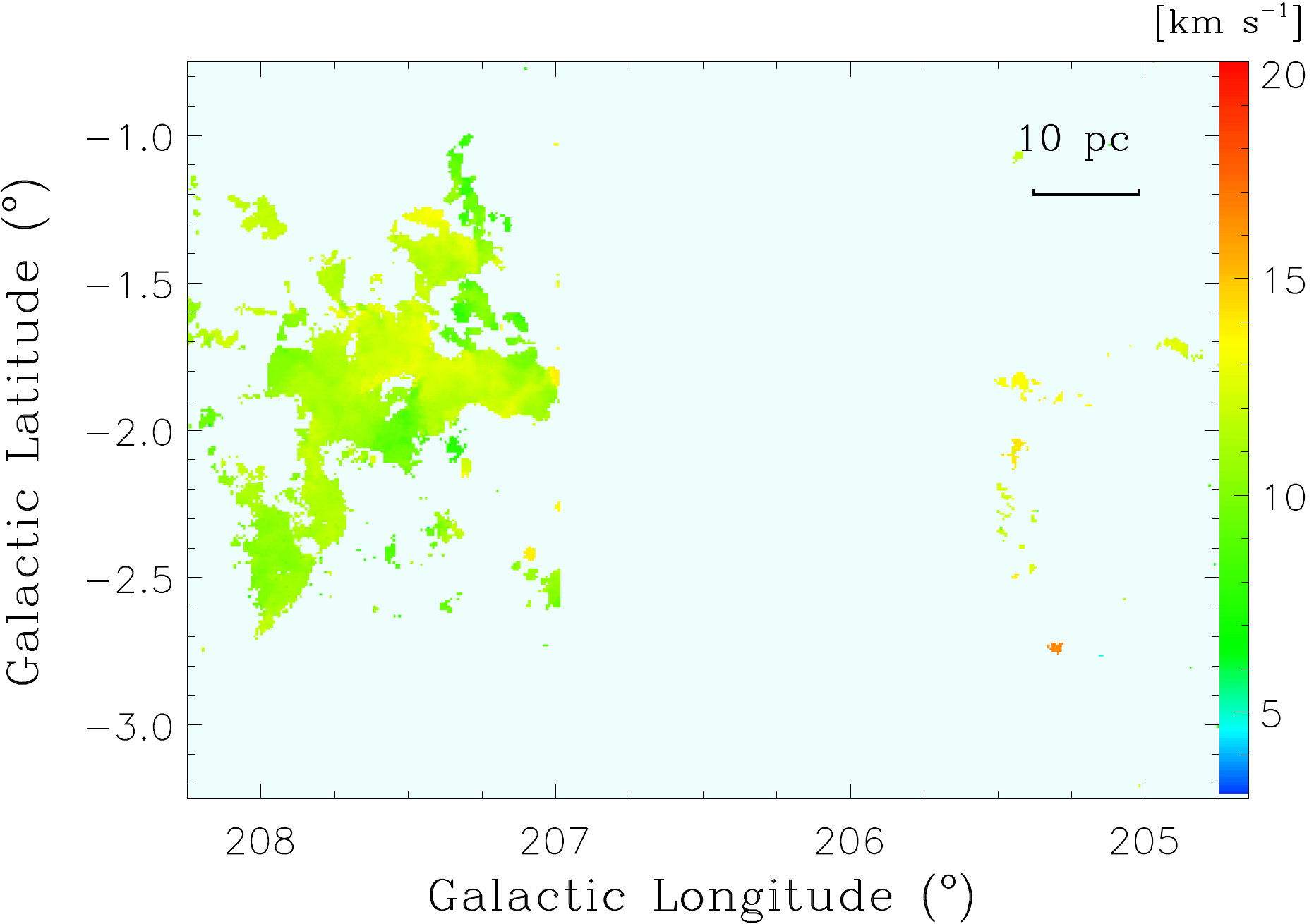}
\includegraphics[width=0.4\textwidth]{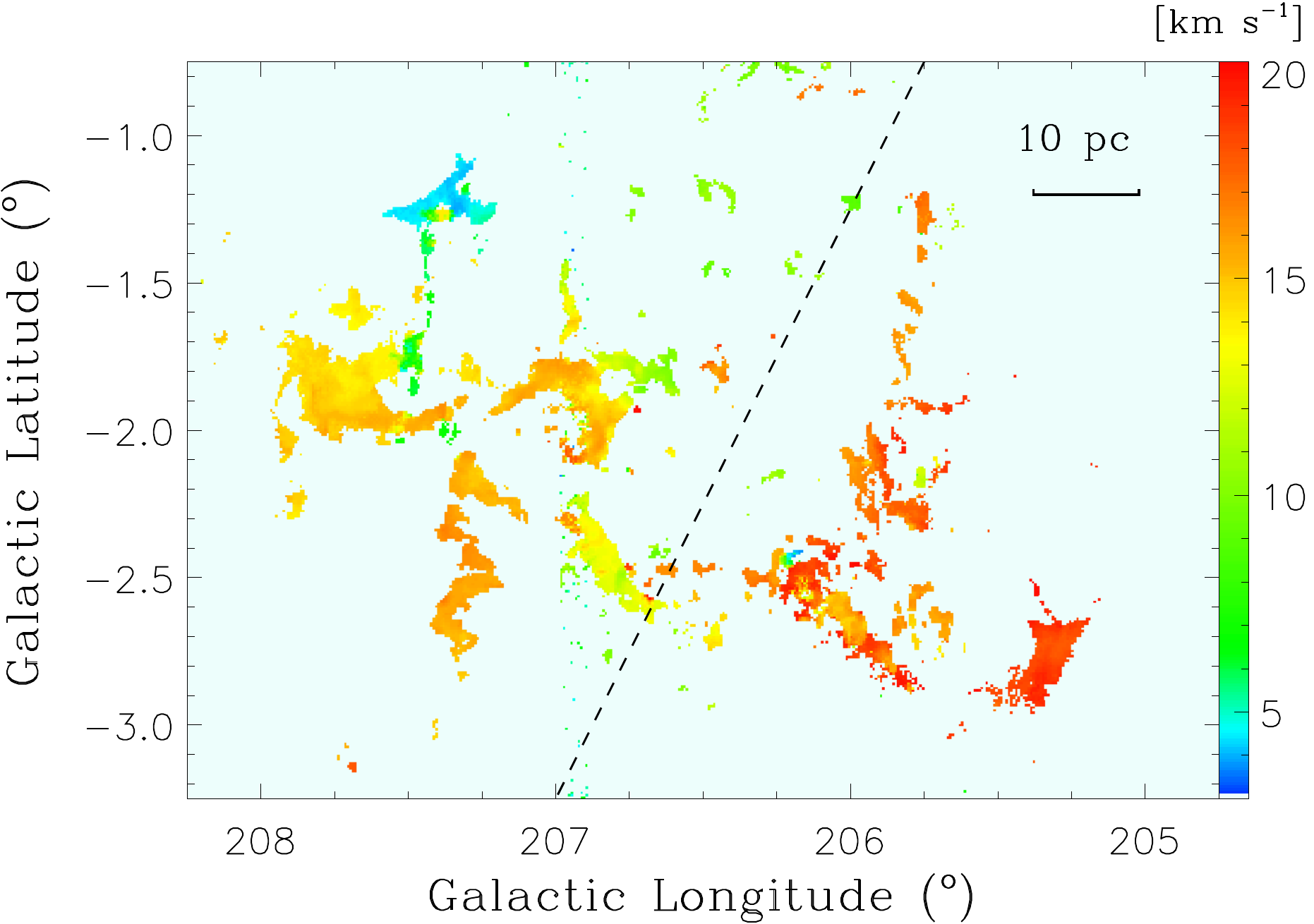}
\caption{Left: velocity distribution of cloud C3 from $^{13}$CO emission. Right: velocity distribution from $^{13}$CO emission for other clouds in the region, i.e., C1-C2 and C4-C9. The dash line shows the rotation axis from \cite{1995ApJ...451..252W}.}
\label{fig15}
\end{figure}

The velocity distribution of molecular clouds in the whole RMC complex is presented in Figure \ref{fig15}. As the C3 cloud spatially overlaps with clouds C2 and C5, its velocity distribution is presented separately in the left panel of Figure \ref{fig15} while the velocity distribution for other clouds in the region, i.e. C1-C2 and C4-C9, is presented in the bottom panel. From Figure \ref{fig15} we can see that clouds of relatively red-shifted velocities are distributed to the southwest of the NGC 2244 cluster while the relatively blue-shifted velocity clouds, in particular clouds C2 and C3, are located to the northeast of the NGC 2244 cluster. This velocity distribution has lead \cite{1995ApJ...451..252W} to conclude that the RMC as a whole possesses a large-scale rotation. However we note that the velocity distribution as shown in Figure \ref{fig15} does not fit well with the rotation scenario, for example, clouds C5 and C6 are located to the left of the dashed line and they have similar velocities to clouds C7-C9 on the right side. Therefore, we propose that no large-scale rotation exists for RMC complex.

\begin{figure}[h]
\centering
\includegraphics[width=0.4\textwidth]{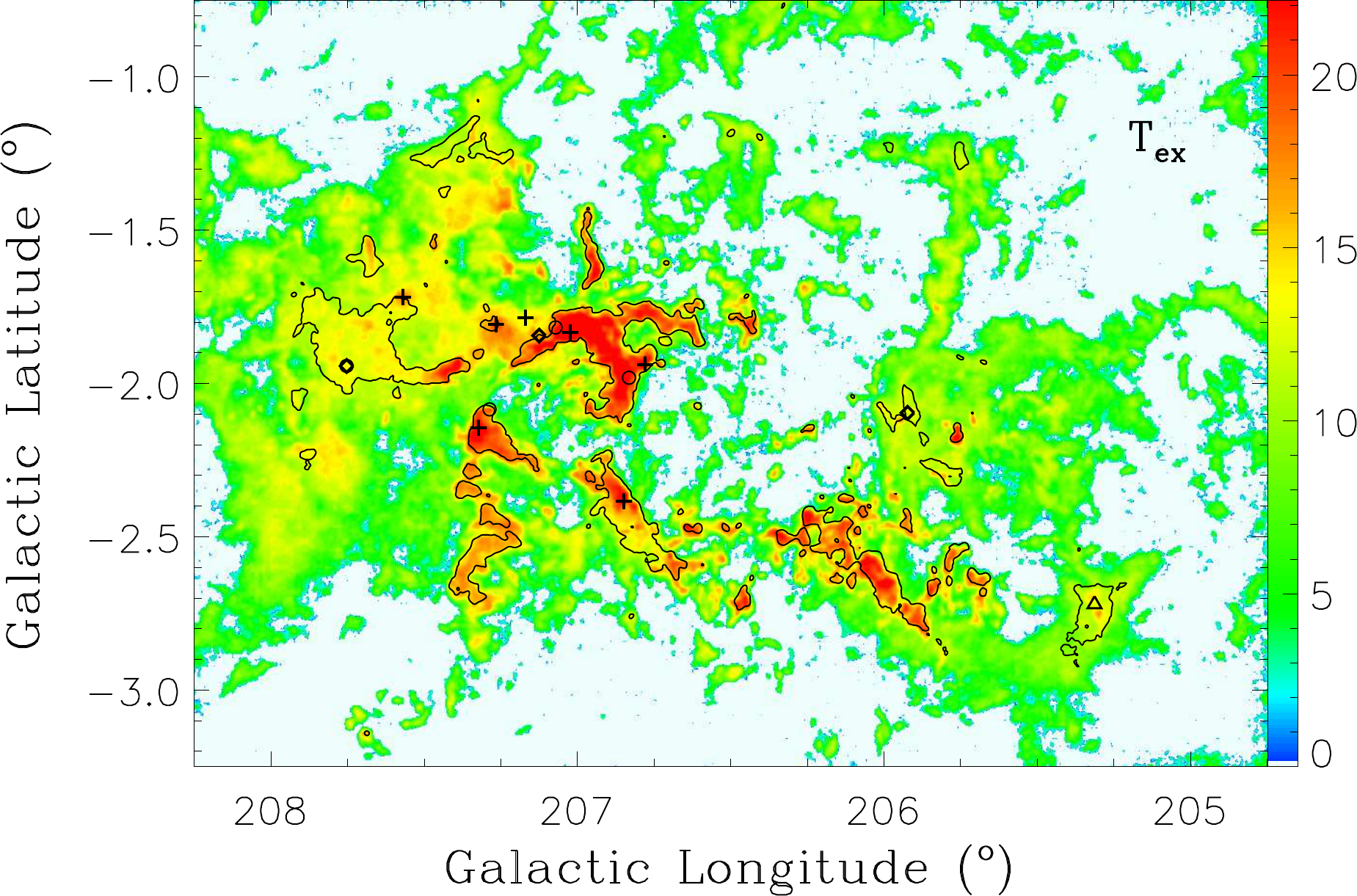}
\caption{Distribution of excitation temperature of the RMC complex. The contours outline $^{13}$CO emission region of clouds C1-C2 and C4-C9.}
\label{fig16}
\end{figure}

Figure \ref{fig16} shows the excitation temperature distribution of the RMC complex. High temperature ($\sim$ 25 K) occurs at the interface between the Rosette Nebula and the surrounding molecular clouds. The highest temperature (37 K) occurs at the western part of cloud C5. Away from the interaction interface the excitation temperature gradually decreases to a value of around 10-15 K. This temperature distribution is consistent with the Hershel result on dust temperature in the region \citep{2010A&A...518L..83S}. This temperature gradient was first found by \citet{1990A&A...230..181C} from analysis of IRAS images. They showed that the outer part of RMC has stronger IRAS 60 $\mu$m and 100 $\mu$m emission than the interaction interface while the IRAS 12 $\mu$m emission is strong around and inside the \ion{H}{2} region. We note that although cloud C3 is a major cloud in the RMC region, no excitation temperature higher than 20 K is found in this cloud. For a comparison, clouds C5 and C6 are located at a similar distance from the NGC 2244 cluster, but both of them possess temperatures higher than cloud C3. It appears that cloud 3 has not been influenced by the NGC 2244 cluster in view of its low excitation temperature.

\begin{figure}[h]
\centering
\includegraphics[width=0.4\textwidth]{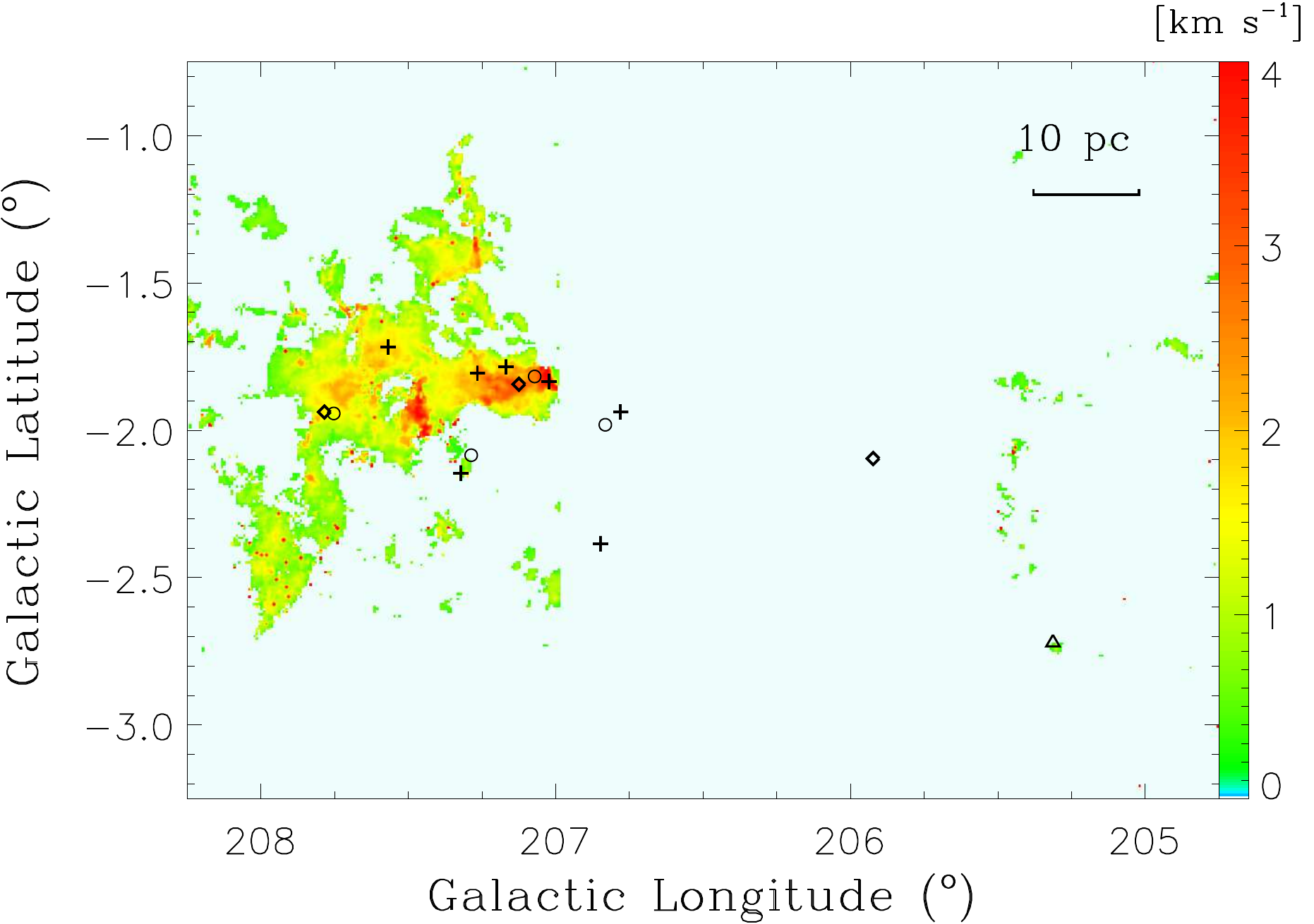}
\includegraphics[width=0.4\textwidth]{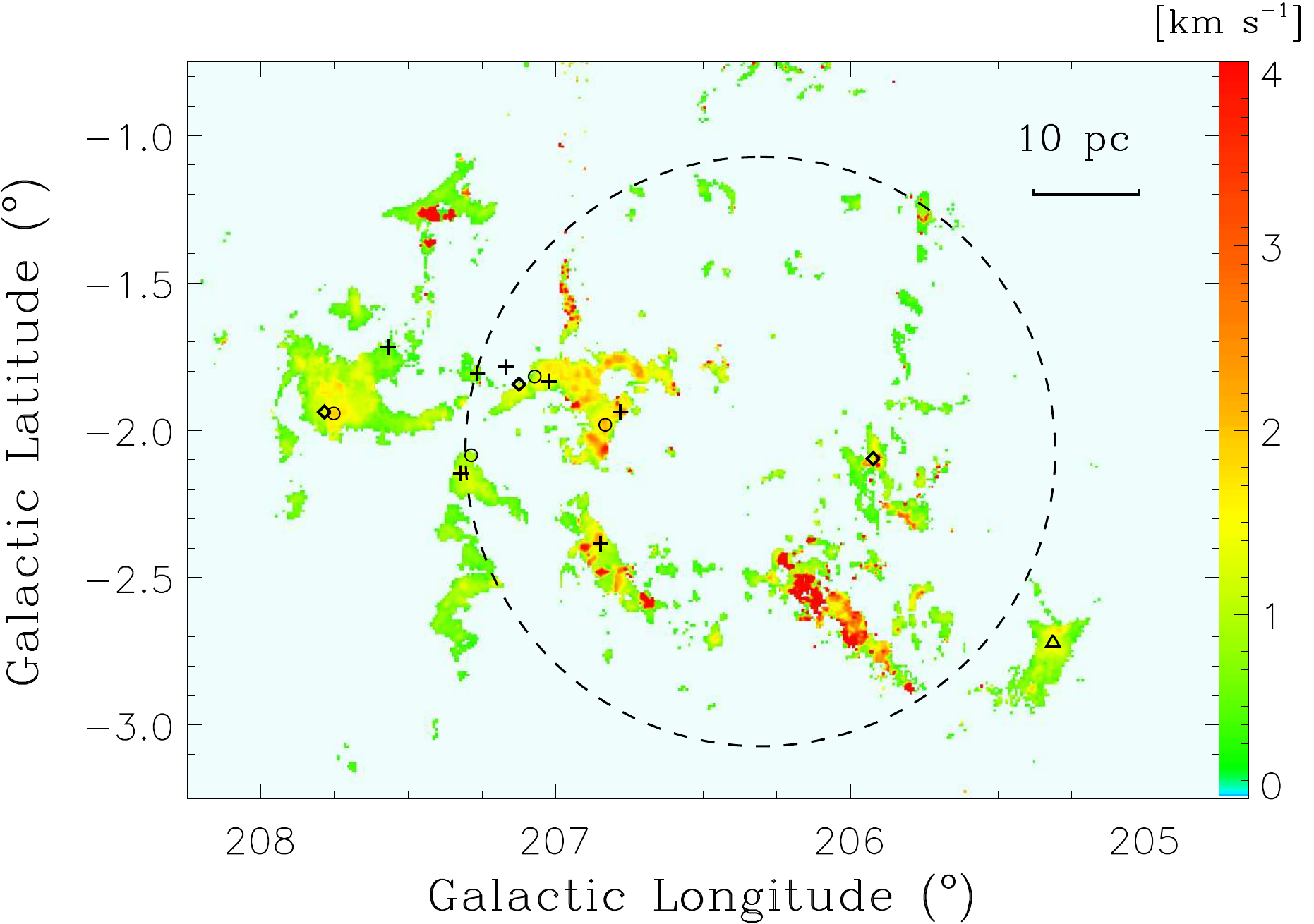}
\caption{Left: velocity dispersion of cloud C3 in $^{13}$CO emission. Right: velocity dispersion in $^{13}$CO for clouds C1-C2 and C4-C9. The dashed circle indicates the range of influence of the NGC 2244 OB cluster. The circles, squares, and triangles indicate the locations of embedded young stellar clusters as in Figure \ref{fig3}.}
\label{fig17}
\end{figure}

The velocity dispersion of the RMC complex is shown in Figure \ref{fig17}. As in Figure \ref{fig3}, the circles, squares, and triangles indicate the locations of embedded young stellar clusters in the region. From Figure \ref{fig17} we can see that most of the regions of high velocity dispersion in cloud C3 are associated with embedded young stellar clusters, showing that feedbacks from young stellar clusters are the main cause for the high velocity dispersion. On the other hand, we find that for clouds C1-C2 and C4-C9 the regions of high velocity dispersion are mainly distributed along the \ion{H}{2} region interface. In particular, as we discussed previously in Section 3, the velocity dispersion in cloud C8 is unusually strong and may be caused by the combined influences from the NGC 2244 and NGC 2237 OB clusters.

\begin{figure}[h]
\centering
\includegraphics[width=0.4\textwidth]{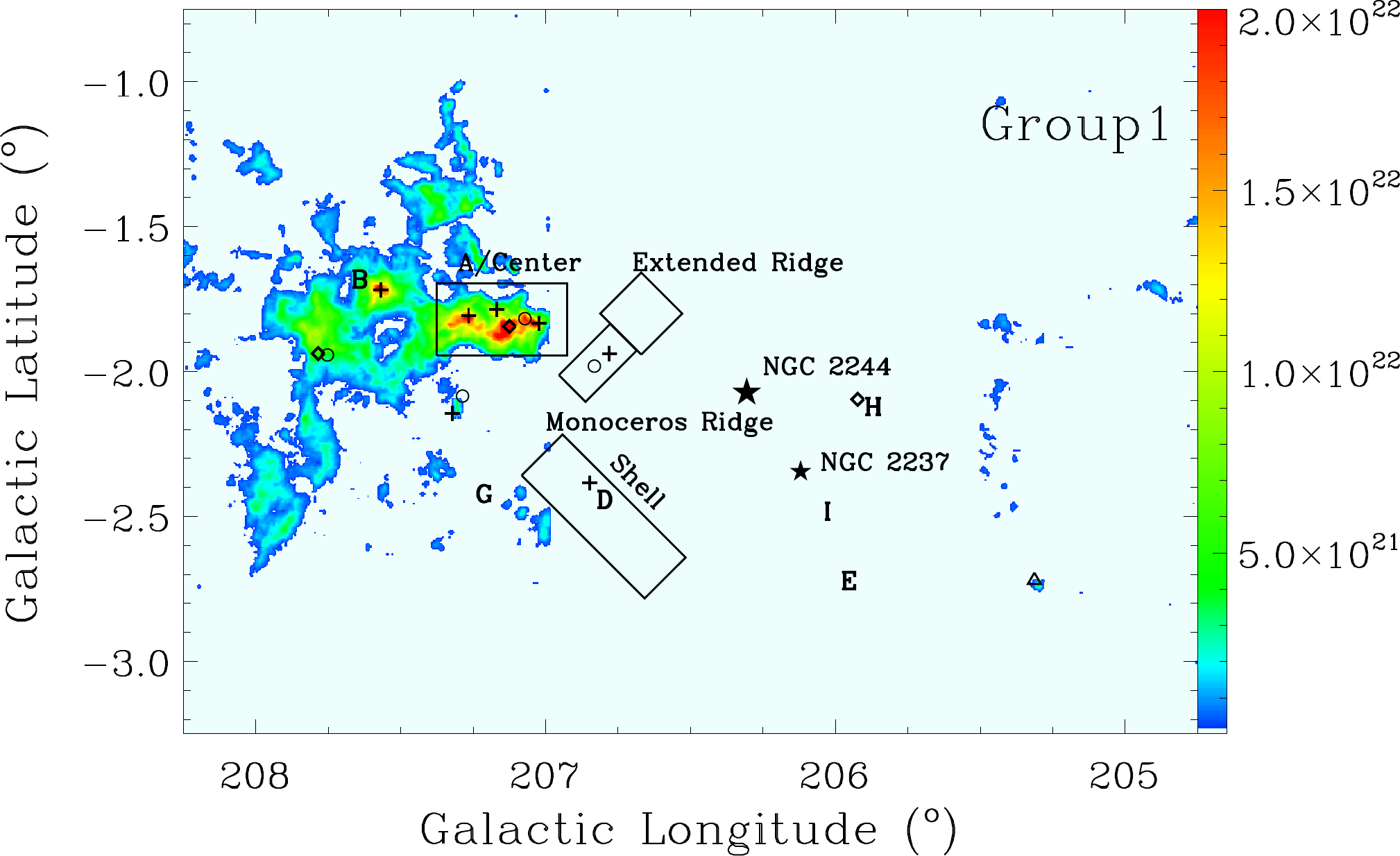}
\includegraphics[width=0.4\textwidth]{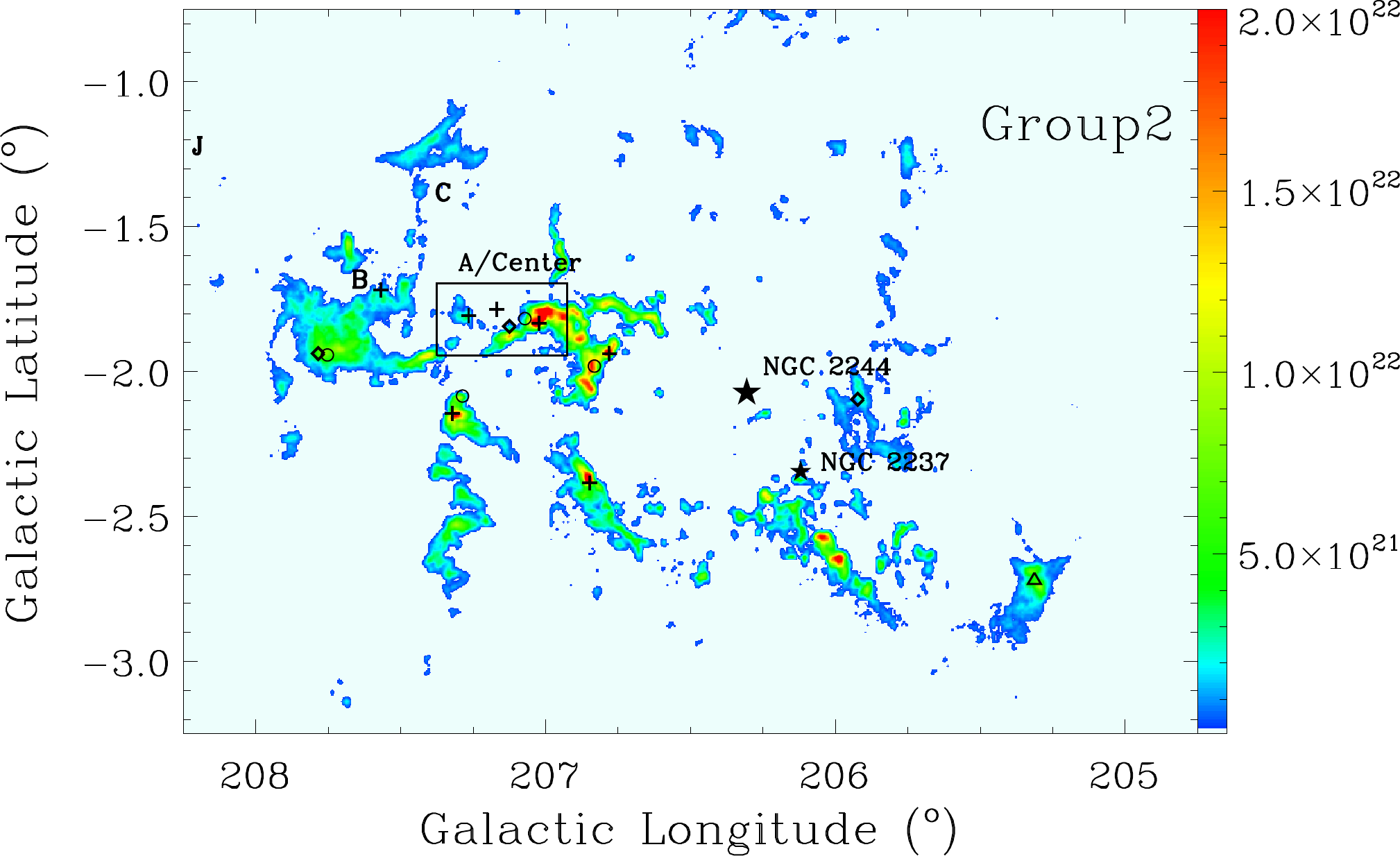}
\caption{Left: column density distribution of group 1 cloud (C3). Right: column density distribution of group 2 clouds (C1-C2 and C4-C9).}
\label{fig18}
\end{figure}

From above discussion it appears that cloud C3 is different from other clouds in the RMC complex in view of its velocity, its temperature, and velocity dispersion distribution. Therefore, we propose that the molecular clouds in the RMC complex can be generally divided into two groups, with cloud C3 as one group (group 1) that shows little impacts from the NGC 2244 and NGC 2237 clusters while other clouds, C1-C2 and C4-C9, as another group (group 2) that shows apparent influence from the NGC 2244 and/or NGC 2237 clusters. We present the column density distribution for group 1 and group 2 clouds separately in Figure \ref{fig18}.

We have estimated the distances to the group 1 and group 2 clouds using stellar extinction data. Based on 5-band grizy Pan-STARRS 1 (PS1) and 3-band 2MASS photometry, \citet{2015ApJ...810...25G} trace the extinction on 7$\arcmin$ scales and have obtained a three-dimensional map of interstellar dust reddening. We select 3 regions from group 1 and 4 regions from group 2 with a radius of 7$\arcmin$. There is only one velocity component in these regions. From the map we find a rapid increase in extinction centered at the distance modulus of 11 (1.4 kpc) for both groups of the clouds. Therefore, the both groups of the clouds are located at the same distance as the Rosette Nebula.

The age of the NGC 2244 cluster is estimated to be approximately $4 \times 10^6$ yr while that of the Rosette Nebula is about one order of magnitude younger \citep{1981PASJ...33..149O}. The age of the NGC 2237 cluster is 2 $\times$ 10$^6$ yr \citep{2010ApJ...716..474W}. The nearest known supernova remnant (SNR) to the RMC complex is SNR G205.5+0.5 (Monoceros Nebula) which is located about 2.6$\arcdeg$ to the northwest and has an angular diameter of 220$\arcmin$ \citep{2014BASI...42...47G}. With $^{12}$CO and $^{13}$CO data, \citet{2017ApJ...836..211S} identified six positions, denoted as a-f in their Figure \ref{fig1}, where they confirmed that SNR G205.5+0.5 is interacting with the surrounding molecular clouds. However, these positions are well separated from the RMC complex. Therefore, we propose that the stellar feedback to the RMC complex is dominated by the effects of photoionisation and winds from the OB clusters NGC 2244 and NGC 2237, although the nearby supernova remnant G205.5+0.5 may also play a role.

\begin{figure}[h]
\centering
\includegraphics[width=0.4\textwidth]{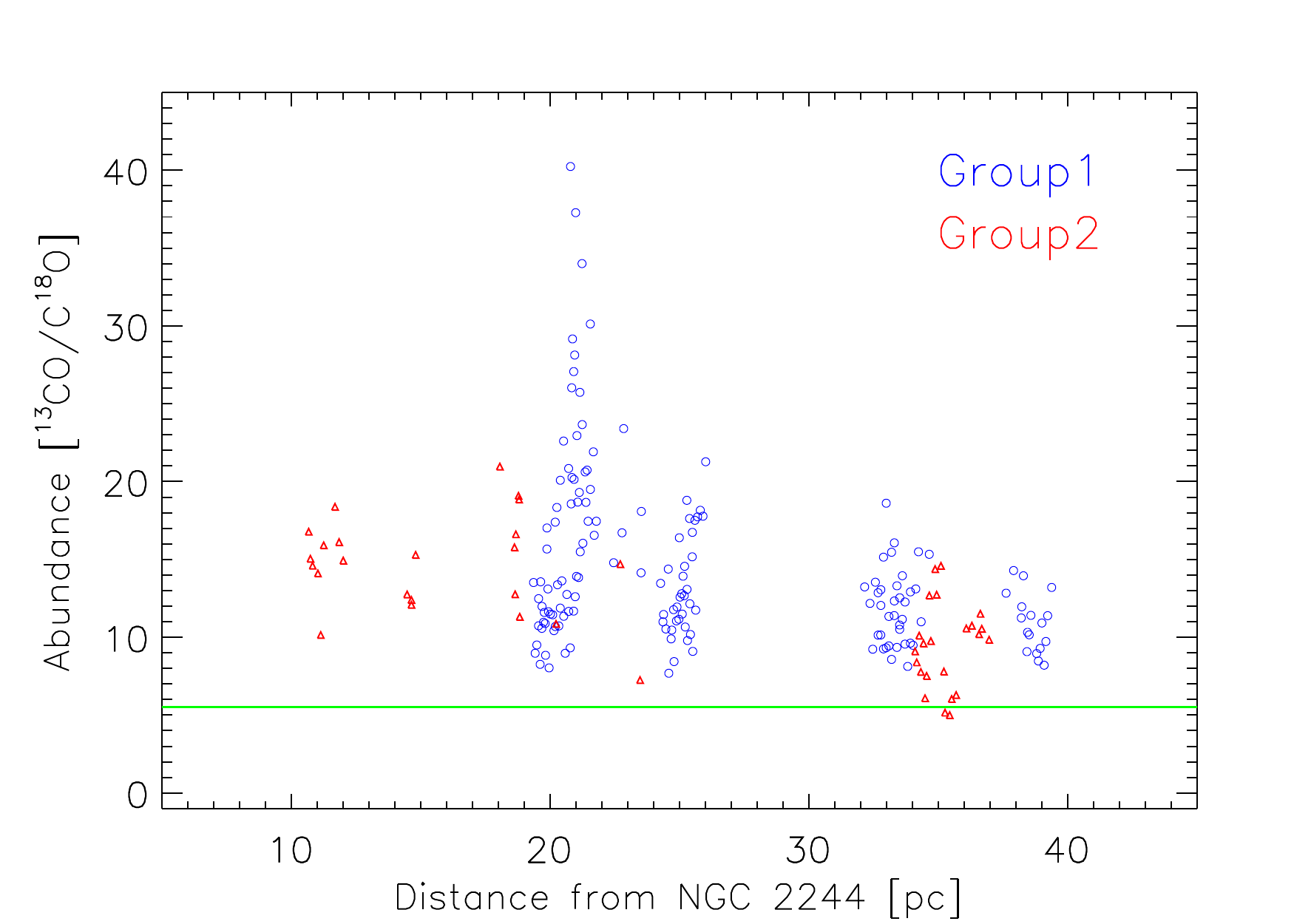}
\caption{Abundance ratio of $^{13}$CO to C$^{18}$O as a function of the projection distance from the NGC 2244 OB cluster for clouds in group 1 (blue circles) and clouds in group 2 (red triangles). The green line indicates the solar system value (5.5).}
\label{fig19}
\end{figure}

\subsection{Isotopologue abundance ratio }

Ultraviolet (UV) photons can dissociate CO and its isotopologues, therefore, have strong influence on the chemistry of molecular clouds. Due to the large difference in optical depth, $^{13}$CO and C$^{18}$O molecules are selectively dissociated by UV photons, with the $^{13}$CO abundance being less affected than the C$^{18}$O abundance. The abundance ratio between $^{13}$CO and C$^{18}$O, R$_{13,18}$, for molecular clouds that are irradiated by massive stars is expected to exceed the values for molecular clouds without strong UV irradiation. To investigate the effects of UV photons from the NGC 2244 and NGC 2237 OB clusters on the RMC complex, we calculated R$_{13,18}$ for clouds in both group 1 and group 2, and the results are presented in Figure \ref{fig19}. To ensure a high reliability of C$^{18}$O detection, only pixels with C$^{18}$O integrated intensity larger than three times the integrated noise are used in the R$_{13,18}$ calculation. Totally 161 pixels in group 1 clouds and 46 pixels in group 2 clouds satisfy this detection criterion. The R$_{13,18}$ values for group 1 clouds lie in the range 7.7-40.2 with a mean of 14.3. For group 2 clouds, the values lie in the range 5.0-21.0 with a mean of 12.0. No significant difference exists between the two groups in the R$_{13,18}$ value. We can see that clouds in both groups possess $^{13}$CO to C$^{18}$O abundance ratios higher than the solar system value (5.5) by a factor of 2.2-2.6, showing that the UV photons from the NGC 2244 and NGC 2237 OB clusters have strong influence on the chemistry of the clouds in the region, although clouds in group 1, as discussed in Section 4.1, exhibit no dynamical effects from the nearby OB clusters. We note that our estimates of $^{13}$CO to C$^{18}$O abundance ratio for the RMC complex are similar to the results by \citep{2014A&A...564A..68S} for the photon-dominated regions (PDRs) in the Orion A giant molecular clouds where they find values lying in the range of 5.7-33.0 with a mean of 16.5. From Figure \ref{fig19} we can see that there is no trend for the abundance ratio with the projection distance from the NGC 2244 OB cluster. This may reflect that the projection distance does not well represent the true distance. We note that all pixels with the $^{13}$CO to C$^{18}$O abundance ratio larger than 25 belong to the group 1 clouds and they are all located around the embedded clusters REFL9 and PL5. We attribute the high isotopologue ratios in that region to the presence of embedded clusters.

\section{Summary}

We have conducted a large-scale simultaneous survey of $^{12}$CO, $^{13}$CO, and C$^{18}$O J=1-0 emission toward the RMC region covering $3.5\arcdeg \times 2.5\arcdeg$. The survey has a sensitivity of 0.5 K for $^{12}$CO emission and 0.3 K for $^{13}$CO, and C$^{18}$O emission at a velocity resolution of 0.16 km s$^{-1}$. The majority emission comes from the Rosette molecular cloud complex with velocities in the range from -2 km s$^{-1}$ to 20.5 km s$^{-1}$ while 73 molecular clumps are identified to lie behind the Rosette molecular cloud complex with velocity extending to 58 km s$^{-1}$. The maps of column density, excitation temperature, and velocity dispersion for the Rosette molecular cloud complex are presented. Based on the spatial and velocity distribution, nine individual clouds, C1-C9, can be identified for the Rosette molecular cloud complex. It appears that the C3 cloud is different from other clouds in the RMC complex in view of its velocity, excitation temperature, and velocity dispersion distribution. We propose that the C3 cloud may have not been influenced by the NGC 2244 and NGC 2237 clusters while other clouds in the region, C1-C2 and C4-C9, show apparent impacts by the NGC 2244 and NGC 2237 clusters. Most of the young stellar clusters in the region are located in positions of high excitation temperature and high column density, showing that stellar clusters prefer to form in high density regions and would impose feedbacks to the environments in their early phase of formation. Seven new molecular filaments with physical lengths of around 10-40 pc are discovered in the RMC complex. Evidence for cloud-cloud collision is found in the REFL9 and PouF region. We propose that the REFL9 and PouF young stellar clusters probably result from this cloud-cloud collision. The abundance ratio of $^{13}$CO to C$^{18}$O for the RMC complex lies in the range of 5.0-40.2, with a mean value of 13.7 which is 2.5 times larger than the solar system value, showing that UV photons from the nearby OB clusters have heavily influenced the chemistry of clouds in the RMC complex.

\section{Acknowledgements}

We would like to thank the PMO-13.7m telescope staffs for their supports during the observation and the anonymous referee for constructive suggestions. This work is supported by the National Key R\&D Program of China (NO. 2017YFA0402701). We acknowledge the support by NSFC grants 11233007, 11127903, and 11503086.  C. Li acknowledges supports by NSFC grant 11503087 and by the Natural Science Foundation of Jiangsu Province of China (Grant No. BK20141046). This publication makes use of data products from the Wide-field Infrared Survey Explorer, which is a joint project of the University of California, Los Angeles, and the Jet Propulsion Lab-oratory/California Institute of Technology, funded by the National Aeronautics and Space Administration. This paper makes use of data obtained as part of the INT Photometric H$_{\alpha}$ Survey of the Northern Galactic Plane (IPHAS, www.iphas.org) carried out at the Isaac Newton Telescope (INT). The INT is operated on the island of La Palma by the Isaac Newton Group in the Spanish Observatorio del Roque de los Muchachos of the Instituto de Astrofisica de Canarias. All IPHAS data are processed by the Cambridge Astronomical Survey Unit, at the Institute of Astronomy in Cambridge. The bandmerged DR2 catalogue was assembled at the Centre for Astrophysics Research, University of Hertfordshire, supported by STFC grant ST/J001333/1. This work has also made use of the SIMBAD database, operated at CDS, Strasbourg, France.

\software{GILDAS \url{http://www.iram.fr/IRAMFR/GILDAS}}

\appendix

\begin{figure}[h!]
\centering
\includegraphics[width=0.6\textwidth]{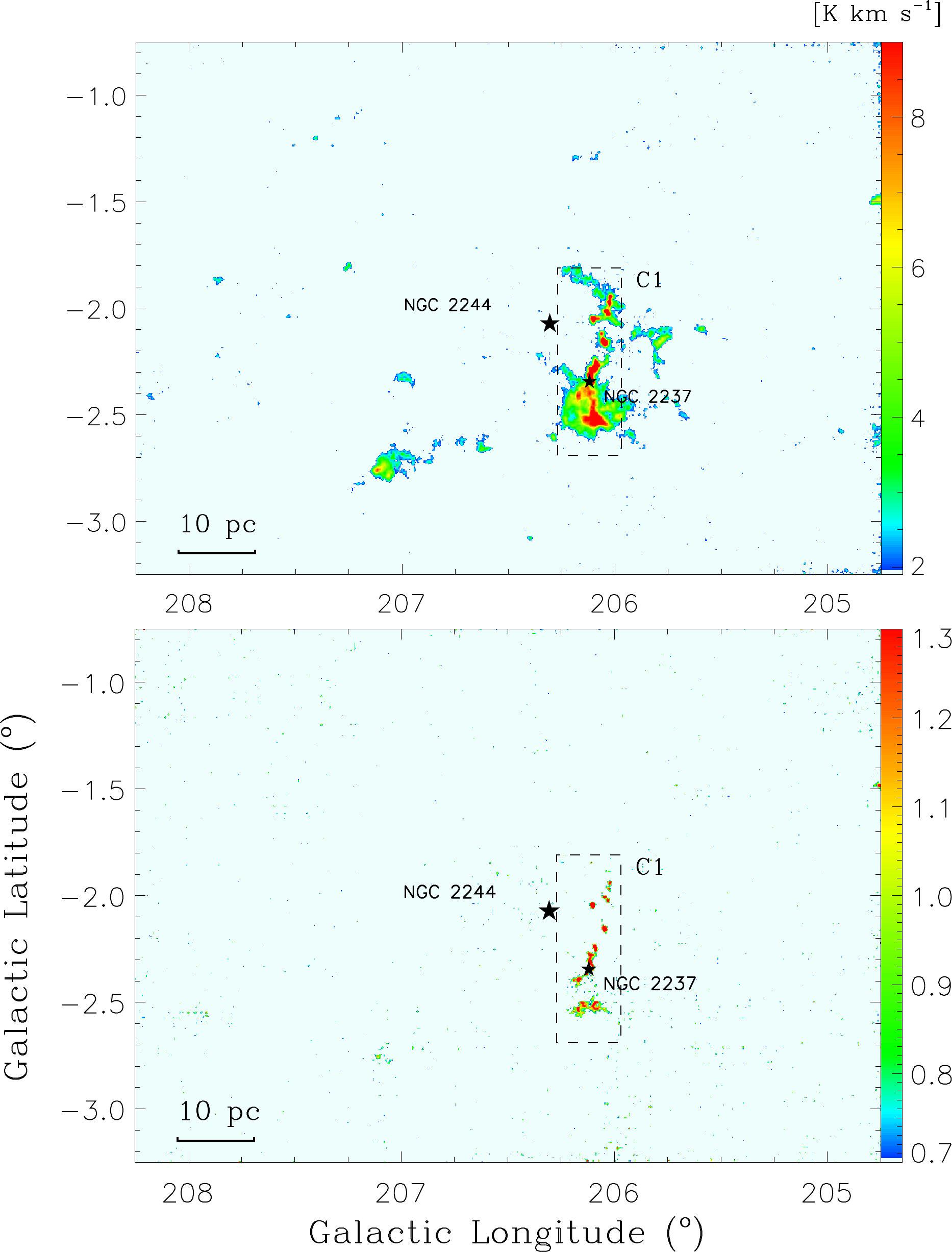}
\caption{Top: the integrated intensity map of $^{12}$CO emission of the RMC region in the velocity range from -2 km s$^{-1}$ to 3 km s$^{-1}$. The spatial extent of the C1 cloud is outlined with a dashed rectangle. The two asterisks mark the centers of the OB clusters NGC 2244 and NGC 2237. Bottom: the integrated intensity map of $^{13}$CO emission in the velocity range from -2 km s$^{-1}$ to 3 km s$^{-1}$.}
\label{fig20}
\end{figure}
\clearpage

\begin{figure}[b!]
\centering
\includegraphics[width=0.65\textwidth, trim = 0cm 0cm 0cm -3cm]{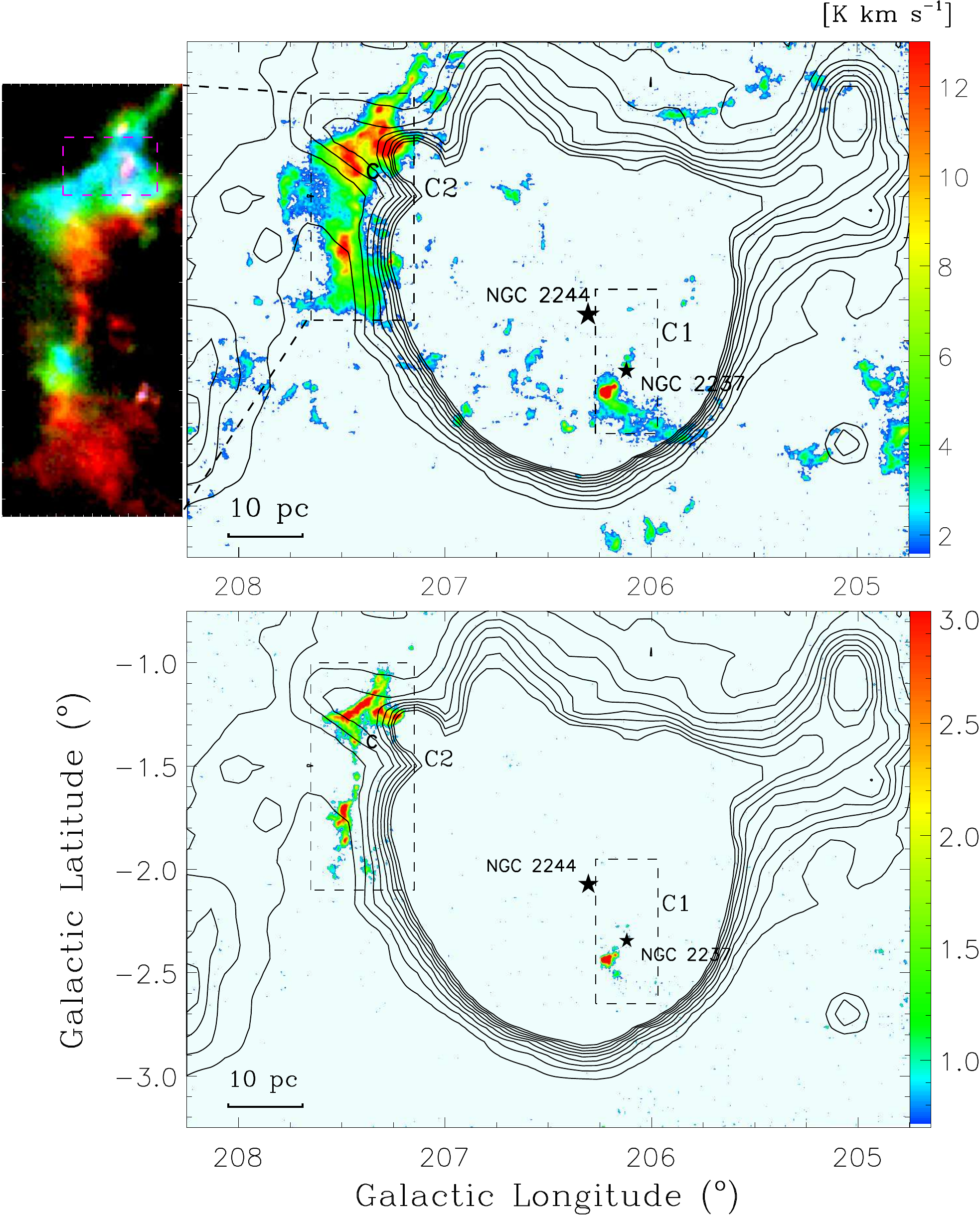}
\caption{Top: integrated intensity map of $^{12}$CO emission of the RMC region. The integrated velocity range is from 3 km s$^{-1}$ to 7 km s$^{-1}$. The cloud identified in this velocity range, C2, is outlined with a dashed rectangle. The colour-coded image shows the velocity distribution of the $^{12}$CO emission of cloud C2, where the blue represents the integrated intensity range from 2.7 to 4.2, the green from 4.2 to 5.7, and the red from 5.7 to 7.2 km s$^{-1}$. The region shown in Figure \ref{fig7} is indicated with a purple dashed rectangle. Bottom: integrated intensity map of the $^{13}$CO emission of the region from 3 km s$^{-1}$ to 7 km s$^{-1}$. All the others are the same as in Figure \ref{fig3}.}
\label{fig21}
\end{figure}
\clearpage

\begin{figure}[b!]
\centering
\includegraphics[width=0.6\textwidth]{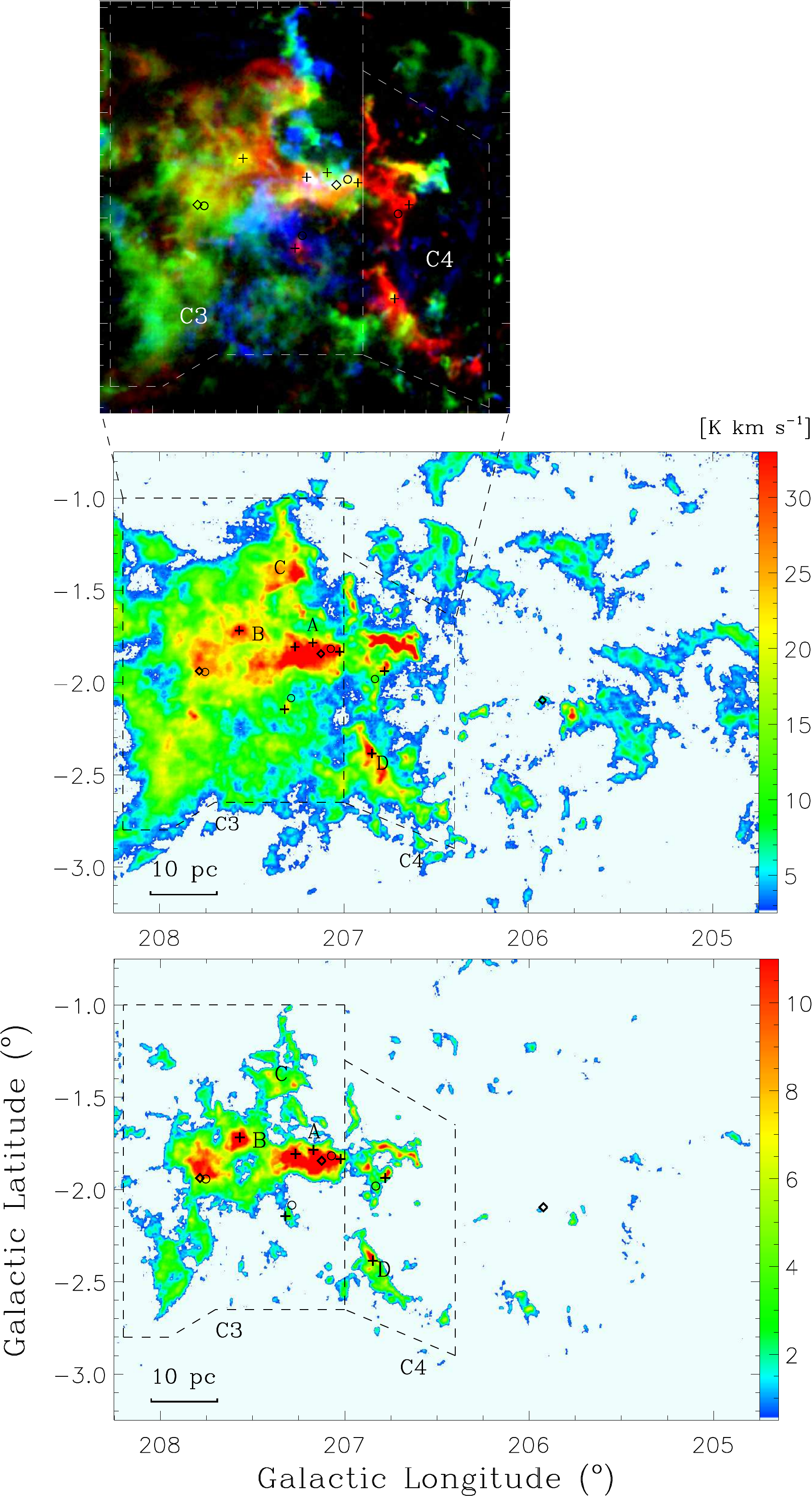}
\caption{Top: colour-coded image of the velocity distribution of $^{12}$CO emission, where the blue represents the integrated intensity in the velocity range from 6.6 to 9.2 km s$^{-1}$, the green from 9.2 to 11.8 km s$^{-1}$, and the red from 11.8 to 14.4 km s$^{-1}$. Middle: integrated intensity map of $^{12}$CO emission from 7 km s$^{-1}$ to 14 km s$^{-1}$. Bottom: integrated intensity map of $^{13}$CO emission from 7 km s$^{-1}$ to 14 km s$^{-1}$. The pluses, diamonds, and circles indicate the embedded clusters from \citet{1997ApJ...477..176P}, \citet{2008ApJ...672..861R}, and \citet{2008MNRAS.384.1249P}, respectively. The boundaries of clouds C3 and C4 are outlined with dashed lines. All the others are the same as in Figure \ref{fig3}.}
\label{fig22}
\end{figure}
\clearpage

\begin{figure}[b!]
\centering
\includegraphics[width=0.6\textwidth]{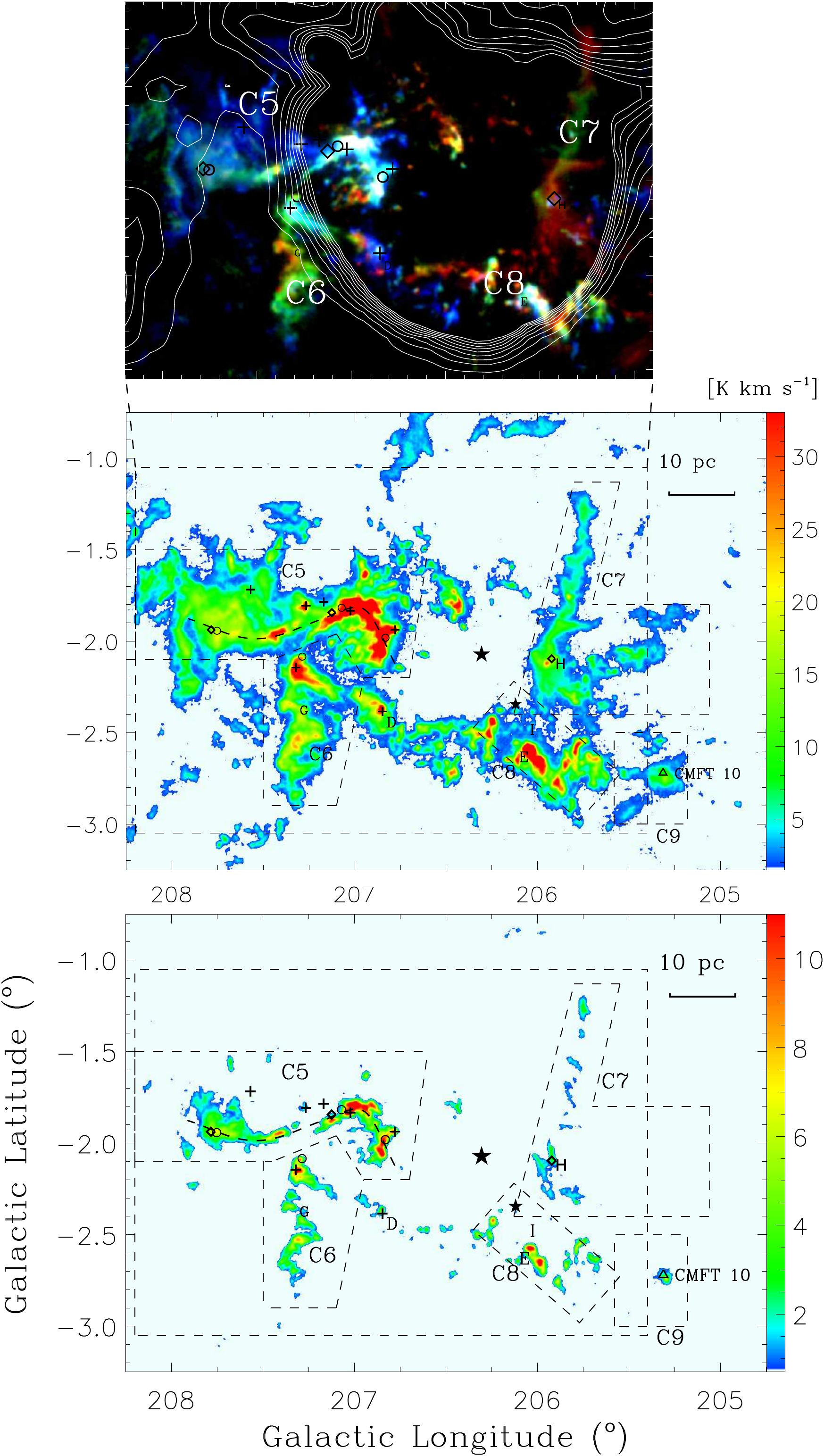}
\caption{Top: colour-coded image of the velocity distribution of $^{12}$CO emission, where the blue represents integrated intensity in the velocity range from 13.8 to 15.1 km s$^{-1}$, the green from 15.1 to 16.4 km s$^{-1}$, and the red from 16.4 to 17.7 km s$^{-1}$. The contours are the 21 cm radio continuum emission. Middle: the integrated intensity map of $^{12}$CO emission from 14 km s$^{-1}$ to 17.5 km s$^{-1}$. Bottom: the integrated intensity map of $^{13}$CO emission from 14 km s$^{-1}$ to 17.5 km s$^{-1}$. All the others are the same as in Figure \ref{fig3}.}
\label{fig23}
\end{figure}
\clearpage

\begin{figure}[b!]
\centering
\includegraphics[width=0.6\textwidth]{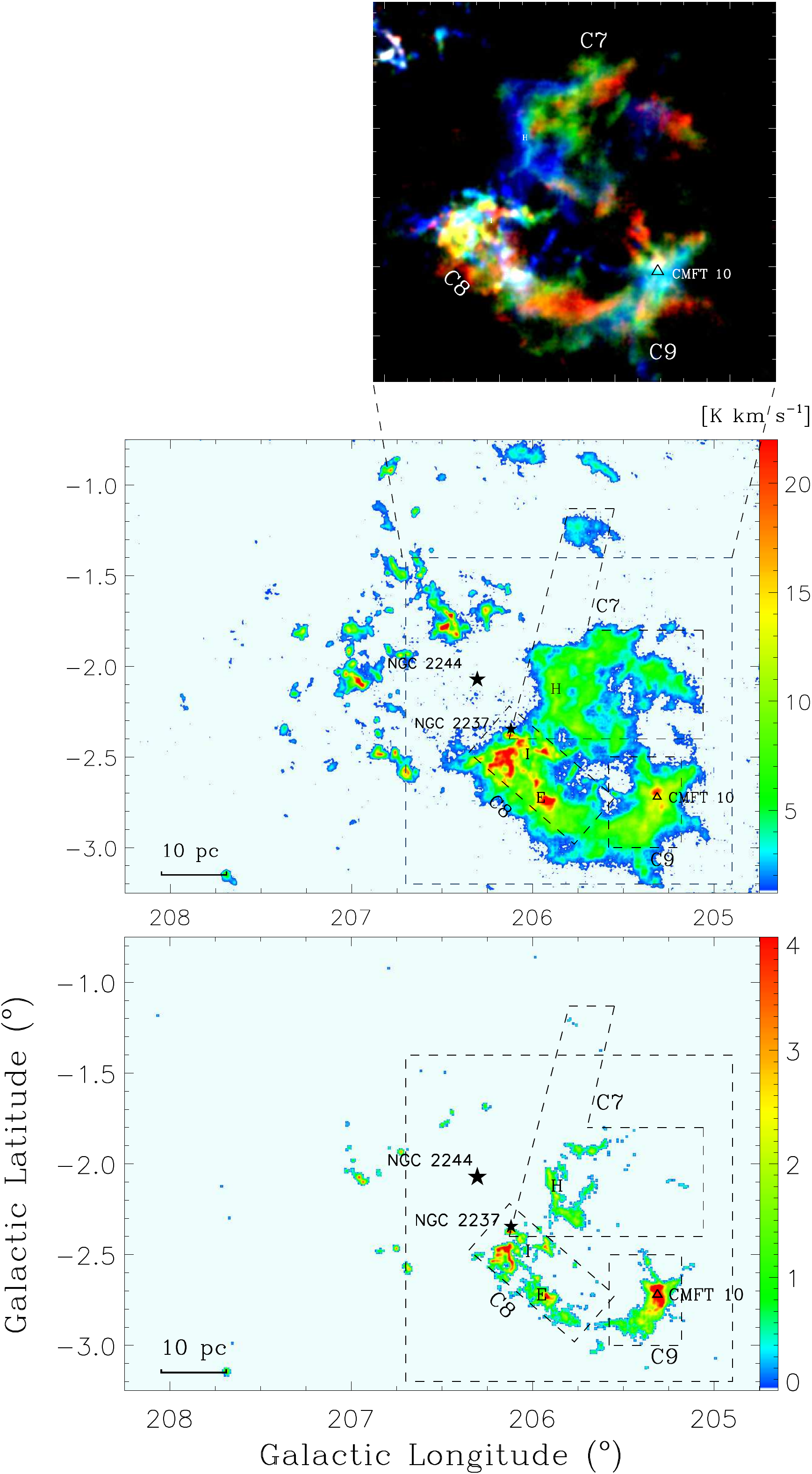}
\caption{Top: colour-coded image of the velocity distribution of $^{12}$CO emission, where the blue represents integrated intensity from 17.3 to 18.4, green from 18.4 to 19.5, and the red from 19.5 to 20.6 km s$^{-1}$. Middle: the integrated intensity of $^{12}$CO from 17.5 km s$^{-1}$ to 20.5 km s$^{-1}$. Bottom: the integrated intensity map of $^{13}$CO from 17.5 km s$^{-1}$ to 20.5 km s$^{-1}$. All the others are the same as in Figure \ref{fig3}.}
\label{fig24}
\end{figure}
\clearpage

\begin{figure}[t!]
\centering
\includegraphics[scale=.4]{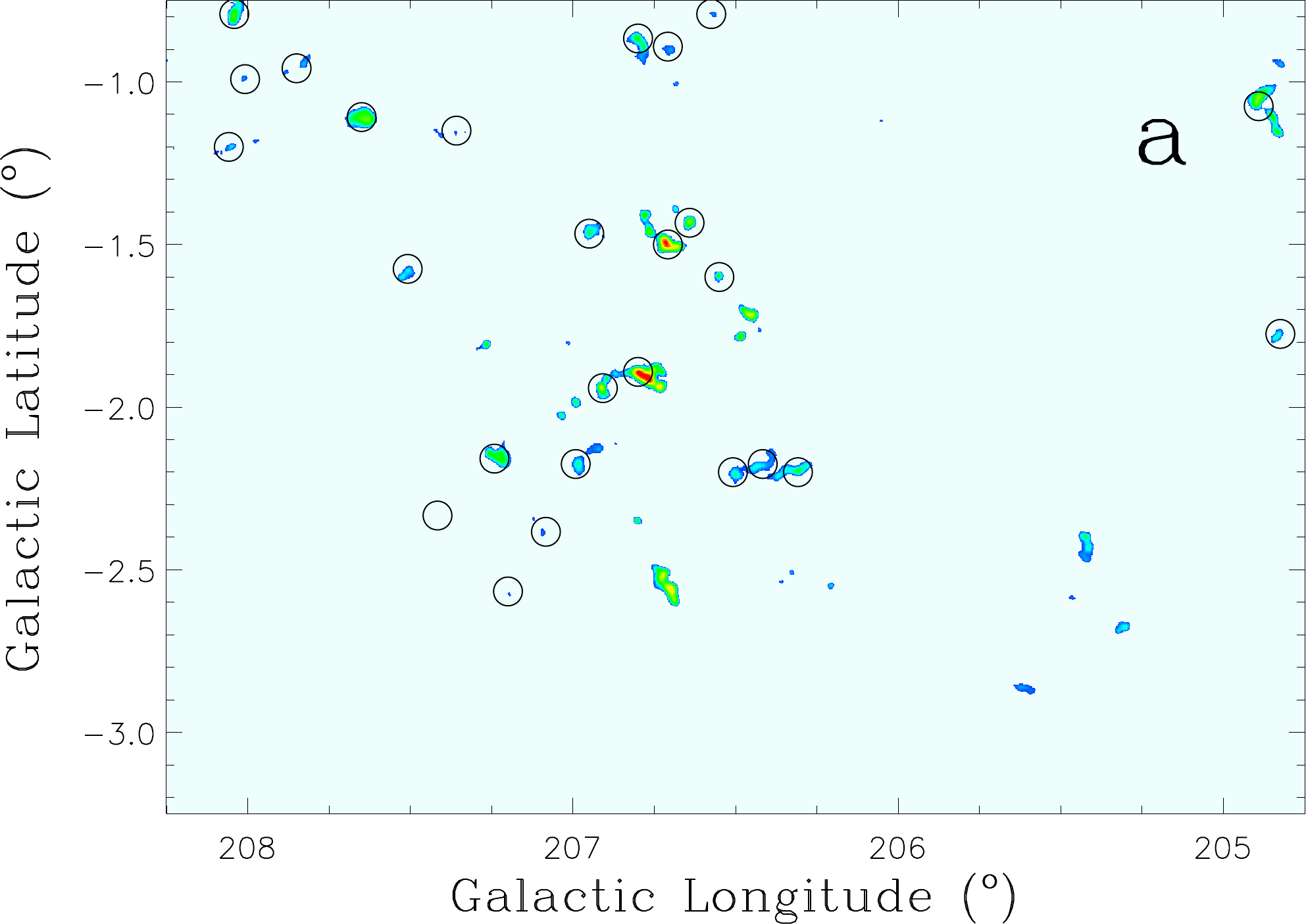}
\includegraphics[scale=.4]{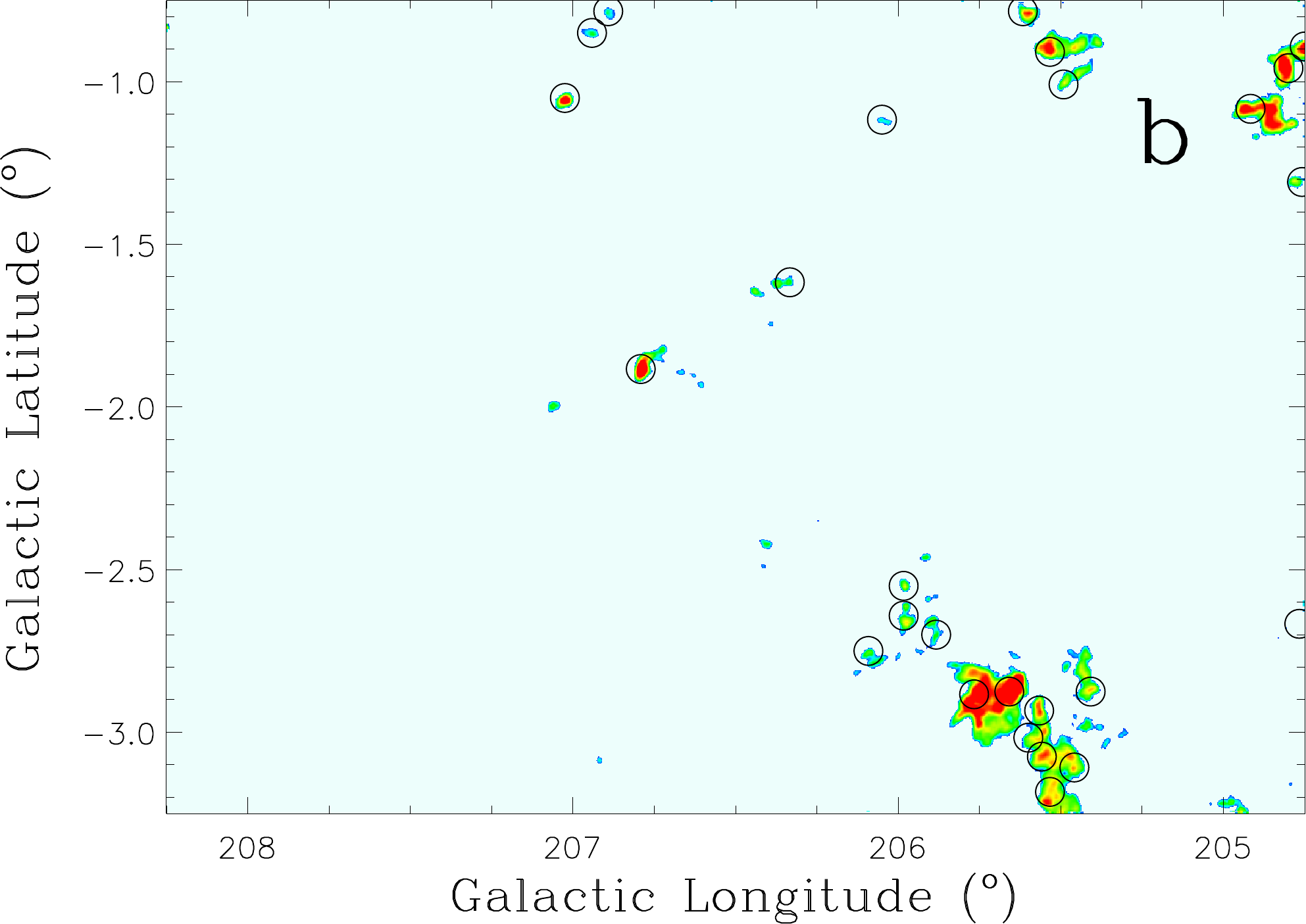}
\includegraphics[scale=.4]{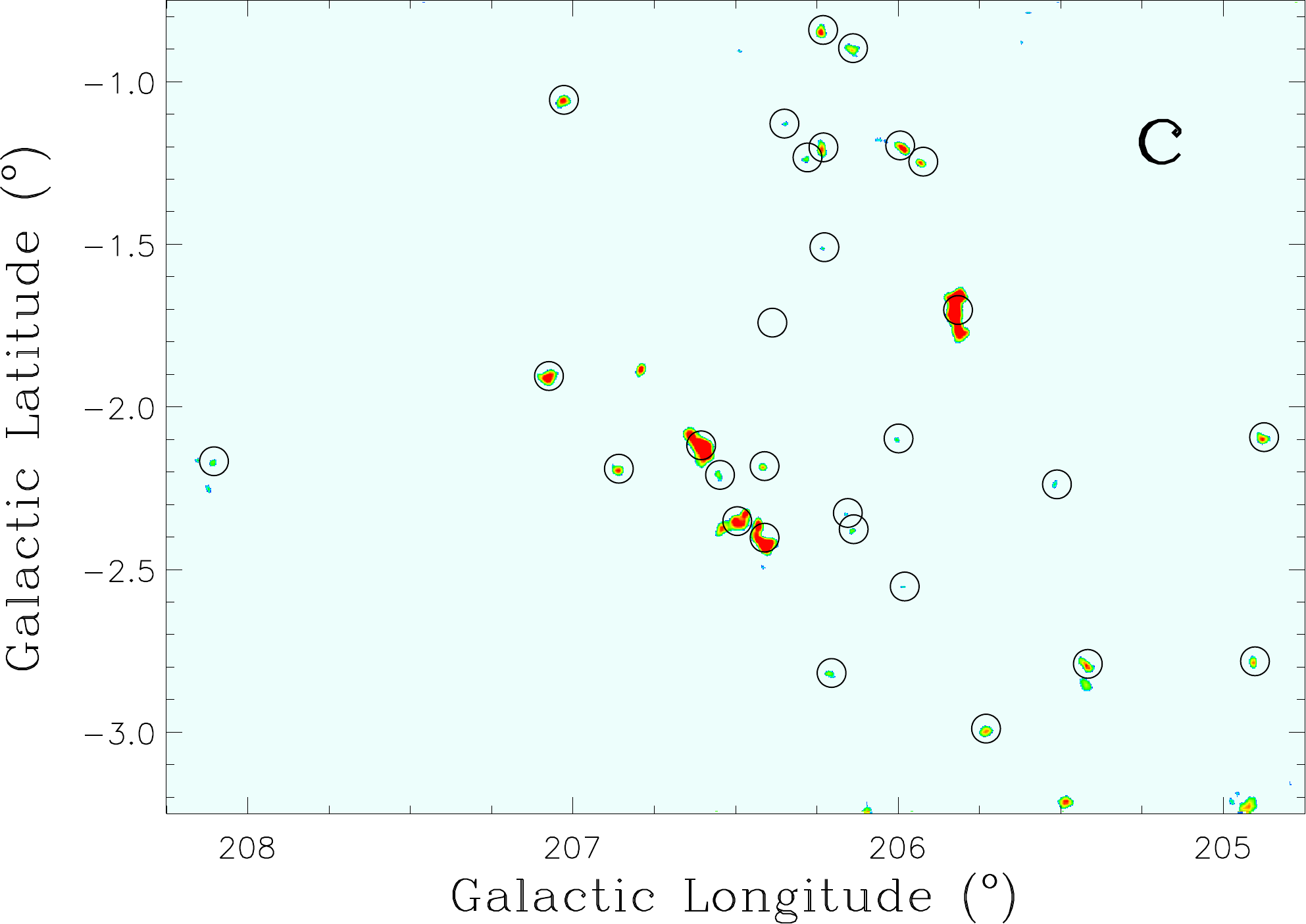}
\includegraphics[scale=.4]{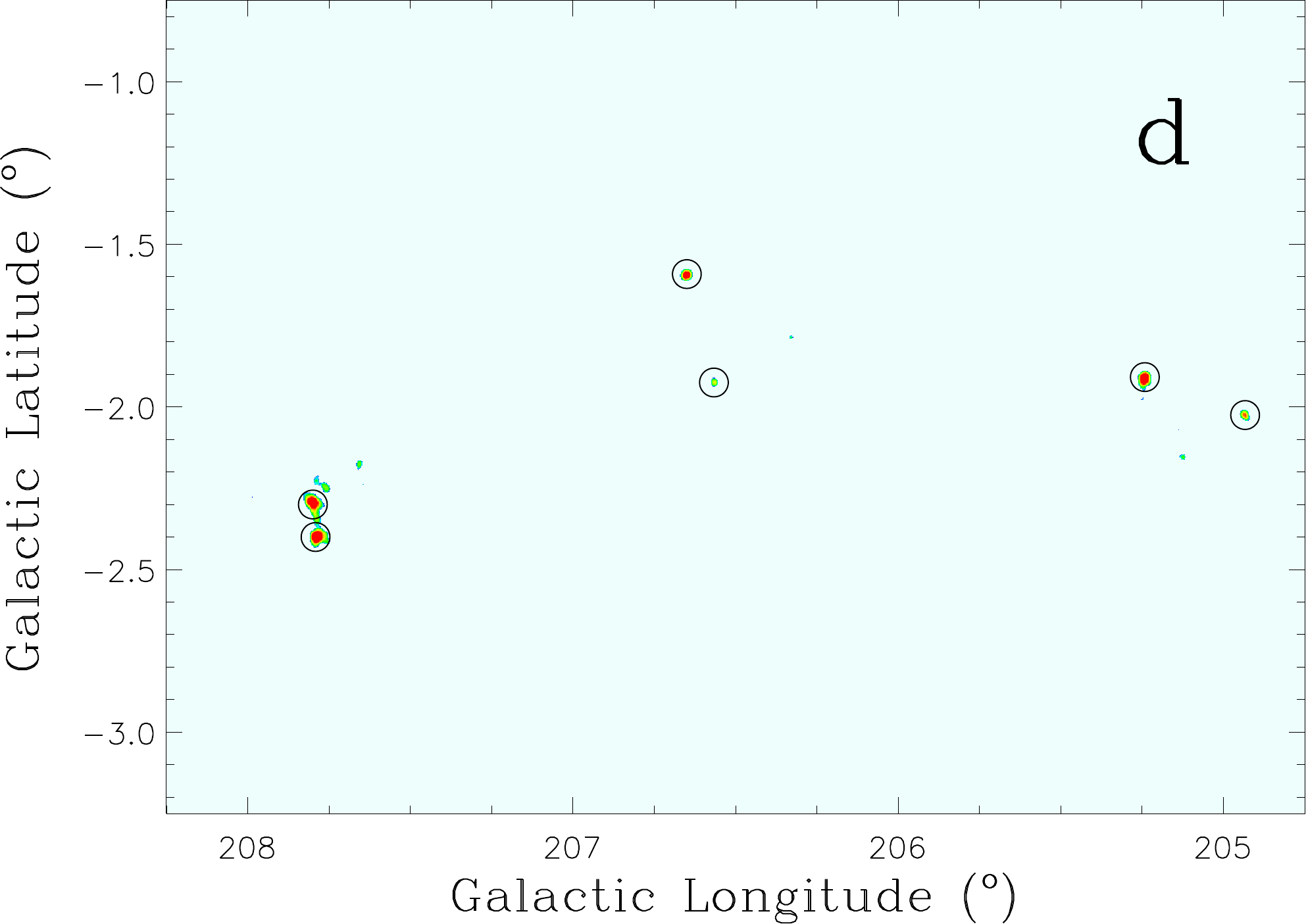}
\caption{Integrated intensity maps of $^{12}$CO emission in the velocity range of 21 to 30 km s$^{-1}$ (a), 30 to 40 km s$^{-1}$ (b), 40 to 50 km s$^{-1}$ (c), and 50 to 60 km s$^{-1}$ (d). The clumps identified by the Fellwalker algorithm are indicated with circles.}
\label{fig25}
\end{figure}

\bibliographystyle{apj}
\bibliography{my_ref_all}


\end{document}